\documentclass[11pt,english]{article}
\usepackage[latin9]{inputenc}
\usepackage{amsmath}
\usepackage{amssymb}
\usepackage{graphicx}
\usepackage{setspace}
\makeatletter

\providecommand{\tabularnewline}{\\}

\numberwithin{equation}{section}

\@ifundefined{date}{}{\date{}}
\usepackage{esint}
\usepackage{hyperref}
\usepackage{latexsym}
\usepackage{graphicx}\usepackage{bm}\usepackage{longtable}

\usepackage{xcolor}

\@addtoreset{equation}{section}

\makeatother

\usepackage{babel}
\begin{document}
\title{Chaos in the holographic matrix models for meson and baryon}
\maketitle
\begin{center}
Si-wen Li$^{a,}$\footnote{Email: siwenli@dlmu.edu.cn},
Xun Chen$^{b,c}$\footnote{Email: chenxun@usc.edu.cn}, 
\par\end{center}

\vspace{2mm}

\begin{center}
$^{a}$\emph{School of Science, Dalian Maritime University, Dalian
116026, China}\\
$^{b}$\emph{INFN---Istituto Nazionale di Fisica Nucleare---Sezione di Bari, Via Orabona 4, 70125 Bari, Italy}\\
$^{c}$\emph{School of Nuclear Science and Technology, University
of South China, Hengyang 421001, China}\\
\par\end{center}

\vspace{8mm}

\begin{abstract}
In recent years, the investigation of chaos has become a bridge connecting gravity theory and quantum field theory, especially within the framework of gauge-gravity duality. In this work, we study holographically the chaos in the matrix models for meson and baryon, which are derived from the $\mathrm{D4}/\mathrm{D6}/\overline{\mathrm{D6}}$ approach
as a top-down holographic model for QCD. Since these matrix
models can be simplified into coupled oscillator models with
special parameters, we analyze the chaos in the resultant
coupled oscillators. In the analysis of the classical chaos, we calculate
numerically the orbits on the Poincar\'e section, the Lyapunov exponent as a function of the total energy and derive the large $N_{c}$ behavior
analytically, then discuss the possible phase structure both in the
mesonic and baryonic matrix models. These analyses suggest that chaos might serve as an order parameter to detect the gauge theory with spontaneous
breaking or restoration of symmetry. Besides, in the analysis of the
quantum chaos, we demonstrate the numerical calculation of the OTOCs
and analytically derive their large $N_{c}$ behavior by using the
perturbation method in quantum mechanics. The numerical calculation
illustrates there is a critical temperature, as a critical energy
scale, that the OTOC begins to saturate, which covers qualitatively the classical analysis of the Lyapunov exponent. And the large $N_{c}$
analytics indicates the OTOCs are suppressed by the growth of $N_{c}$.
Overall, the investigation of chaos in this work may be helpful to identify common features shared by the matrix models, hadronic physics, gauge theory, quantum mechanics, and gravity theory.
\end{abstract}
\newpage{}

\tableofcontents{}

\section{Introduction}

\subsection{Chaos in classical physics}

Chaos is the conception originated from the investigation of the classical
nonlinear dynamics, as nonlinear dynamical systems are ubiquitous in physics. 
Such systems usually exhibits highly complex trajectories, that cannot be solved analytically, often leading to chaotic behavior. Thus, chaos is considered as a key characteristic for describing the classical properties of the nonlinearly dynamical system. While defining exactly the chaos may remain to be an open question, there are two popular methods to measure the chaos in the classically mechanical system: 1) use the Poincar\'e section, 2) compute the Lyapunov exponent
\cite{key-1,key-2,key-3,key-4}. Below, we briefly introduce some features
about these methods for measuring classical chaos.

Poincar\'e section: The Poincar\'e section is a surface in phase space (defined by canonical coordinates and momenta) that transversally intersects the system's flow. All possible orbits cross this section repeatedly. For a classical solution of the system, an orbit appears on the Poincar\'e section as discrete points which are the intersections between the orbit and the section. By changing the initial
conditions, the periodic trajectory comes back to a point, and quasi-periodic
orbits are also projected as dots on closed contours while it becomes
dense. Periodic and quasi-periodic orbits on the Poincar\'e section
illustrate that the system is in a regular or quasi-regular phase.
For the chaotic phase, the trajectory on Poincar\'e section is randomly
scattered dots. Therefore, using Poincar\'e section can help us to see
the chaos in the system. 

Lyapunov exponent: The Lyapunov exponent quantifies the sensitivity of a classical system's evolution to initial conditions. Consider $\mathbf{x}\left(t\right)$ is a solution of a given
classical system, the Lyapunov exponent is defined as,

\begin{equation}
e^{2Lt}=\left[\frac{\delta\mathbf{x}\left(t\right)}{\delta\mathbf{x}\left(0\right)}\right]^{2},\label{eq:1.1}
\end{equation}
where $t$ refers to time and $L$ is the classical Lyapunov coefficient.
For a classically chaotic system, the Lyapunov coefficient is positive
which illustrates the evolution is sensitive to the initial condition.
In addition, the system is regular or non-chaotic if $L\leq0$. Thus,
computing the Lyapunov exponent by using (\ref{eq:1.1}) is another
criterion to measure the classical chaos.

\subsection{Quantum chaos and OTOC}

In quantum theory, the quantum chaos can be measured by calculating
the out-of-time-order correlator (OTOC) $C_{T}\left(t\right)=e^{2Lt}$,
now $L$ refers to the quantum Lyapunov coefficient. The exact formulas
of $C_{T}\left(t\right)$ is conjectured based on the definition of
the classical chaos in (\ref{eq:1.1}). Let us rewrite (\ref{eq:1.1})
by using the Poisson bracket $\left\{ ...\right\} _{\mathrm{P.B.}}$
as,

\begin{equation}
\left[\frac{\delta\mathbf{x}\left(t\right)}{\delta\mathbf{x}\left(0\right)}\right]^{2}=\left\{ \mathbf{x}\left(t\right),\mathbf{p}\left(0\right)\right\} _{\mathrm{P.B.}}^{2},\label{eq:1.2}
\end{equation}
therefore it implies the quantum version of (\ref{eq:1.2}) can be
obtained by replacing the Poisson bracket $\left\{ ...\right\} _{\mathrm{P.B.}}$
with the quantum commutator $-i\left[...\right]$. Hence it is conjectured
that the quantum chaos can be measured by OTOC which is defined as
\cite{key-5,key-6,key-7},

\begin{equation}
C_{T}\left(t\right)=-\left\langle \left[x\left(t\right),p\left(0\right)\right]^{2}\right\rangle _{T},
\end{equation}
where $\left\langle ...\right\rangle _{T}$ refers to the thermal
average of the statistics. And as the quantum operators, $x\left(t\right),p\left(t\right)$
are in the Heisenberg picture. In fact, OTOC can take the general
form for arbitrary operators $W,V$ given in Heisenberg picture as

\begin{equation}
C_{T}=-\left\langle \left[W\left(t\right),V\left(0\right)\right]^{2}\right\rangle _{T},
\end{equation}
and OTOC has become a measure of the magnitude of quantum chaos \cite{key-8,key-9,key-10,key-11}
in recent years while it was first introduced to analyze the vertex
correction of a current for the superconductor \cite{key-12}. 

\subsection{Chaos in matrix models and gauge-gravity duality}

The gauge-gravity duality and AdS/CFT illustrate the equivalence of
the gravity theory and quantum field theory in holography, which have
been proposed for nearly 30 years \cite{key-13,key-14}. In history,
the BFSS (Banks-Fischler-Shenker-Susskind) matrix model was suggested
to be the holographically equivalent description of the 11-dimensional
M-theory \cite{key-15,key-16}. The BFSS matrix model provides a more
simple way to study the quantum gravity since it is nothing but a
quantum mechanical system with non-Abelian gauge symmetry. Due to the attention to the chaos in recent years, people find analyzing chaos may be a bridge to connect holographically the gravity theory and quantum field theory in the framework of gauge-gravity duality, and it is also constructive to quantum gravity.  Consequently, there are several studies explored the chaotic behavior in the matrix models, aiming to investigate the manifestations of chaos in quantum gravity through the framework of gauge-gravity duality \cite{key-17,key-18,key-19,key-20,key-21,key-22,key-22b1,key-22b2}.
For a remarkable instance, the quantum Lyapunov coefficient of the
SYK (Sachdev-Ye-Kitaev) model \cite{key-23,key-24,key-25,key-26}
saturates the upper bound $2\pi T$ which is originally found in the
context of quantum information on the black hole horizon \cite{key-27,key-28,key-29,key-30,key-31,key-32,key-32a}.
This conclusion strongly implies the SYK model (also as a quantum
mechanical matrix model) can somehow describe a quantum black hole
\cite{key-33}, as an equivalent theory of quantum gravity in holography. 

Motivated by these, in this work, we study the chaos in the matrix
models for meson and baryon derived from the $\mathrm{D4}/\mathrm{D6}/\overline{\mathrm{D6}}$
approach \cite{key-34} which is a top-down holographic system for
QCD constructed by the framework of string theory. The background
geometry of this model is produced by $N_{c}$ coincident D4-branes
compactified on a circle $S^{1}$ \cite{key-35,key-36,key-37} which
is the soliton solution of the IIA supergravity. At low-energy, the
supersymmetric fermion on the compactified D4-branes is decoupled
since they acquire mass due to the anti-periodic boundary condition
along $S^{1}$. Therefore the low-energy theory on the $N_{c}$ D4-branes
including only the massless gauge bosons is pure Yang-Mills theory
as the color sector (i.e. the dynamics of gluon) of the holographic
QCD, and it exhibits confinement due to the behavior of the Wilson
loop in the background geometry. Besides, the flavors can be further
introduced to the D4-brane background by embedding intersectionally
$N_{f}$ pairs of the $\mathrm{D6}/\overline{\mathrm{D6}}$-branes.
Meson is identified to the open string on the $\mathrm{D6}/\overline{\mathrm{D6}}$-branes
by analyzing the symmetries and spectrum in this system, its dynamics
is described by the action of the D6-brane. In particular, the transverse
excitations on the $\mathrm{D6}/\overline{\mathrm{D6}}$-branes correspond
to the scalar and pseudo-scalar mesons in this model, which can be
described by a matrix model due to the form of the D-brane action,
hence we name this matrix model as mesonic matrix model in this paper.
Moreover, it is possible to define the baryon vertex by introducing
the D4-branes wrapped on $S^{4}$ in the D4-brane background according
to the classical literatures \cite{key-38,key-39}, and the effective
dynamics of the baryon vertex can be effectively described by a matrix
model \cite{key-40,key-41,key-42} through the T-duality rules in
string theory \cite{key-16}. In this sense, we name this matrix model
as baryonic matrix model in this work. 

Altogether, we can summarize the motivation to study the chaos in
the $\mathrm{D4}/\mathrm{D6}/\overline{\mathrm{D6}}$ approach as
follows. First, the $\mathrm{D4}/\mathrm{D6}/\overline{\mathrm{D6}}$
approach is a top-down holographic model for QCD which includes the
dynamics of meson and baryon. In the traditional theories of nuclear
physics, it is very challenging to analyze analytically the dynamics
of meson and baryon by using the methods in quantum field theory since
meson and baryon live in the strongly coupled region of QCD. Second,
the dynamics of meson and baryon in the $\mathrm{D4}/\mathrm{D6}/\overline{\mathrm{D6}}$
approach can be effectively described by the matrix models. Third,
as we will see in the following sections, our concerned matrix models
can be simplified to be various coupled oscillators with special parameters
which provides a simple way to investigate the chaos. Accordingly,
the investigation of the chaos in the various matrix models may give
some extraordinary results to understand the properties of gauge-gravity
duality and the behaviors of hadrons, even to somehow understand
quantum gravity.

\subsection{Gauge theory as coupled oscillator}

The concerns in this work are the chaos in the coupled oscillator whose
Hamiltonian takes the generic form in $D$-dimensional Euclidean space
as,

\begin{equation}
H=\sum_{a=1}^{D}\left(\frac{1}{2}p_{a}^{2}+\frac{1}{2}m_{a}x_{a}^{2}\right)+\frac{1}{2}g\sum_{a,b=1}^{D}\left|\epsilon^{ab}\right|x_{a}^{2}x_{b}^{2},\label{eq:1.5}
\end{equation}
where $a,b$ run from $1$ to $D$, $x_{a},p_{a}$ refers to the canonical
coordinates and the associated canonical momenta. $\epsilon^{ab}$
is the Levi-Civita symbol, $m_{a}$ is the mass term for $x_{a}$,
and $g$ refers to the coupling constant. We note here if all the
mass $m_{a}$'s are equal, the Hamiltonian (\ref{eq:1.5}) takes the
homogeneous mass term; otherwise it has inhomogeneous mass term.
We will focus on the case of $D=2,3$ in this paper. The reasons that
we focus on the model (\ref{eq:1.5}) to study the chaos are as follows.
First, in the case of $D=2$ and $m_{1}=m_{2}$, the model (\ref{eq:1.5})
reduces to a two-dimensional coupled oscillator with homogeneous mass. This system
has attracted interests for long years due to its classical chaos
\cite{key-1,key-43,key-44,key-45}. This is one of the most popular
models for studying the chaos, since its classical Lyapunov exponent and
quantum level statistics are well-established \cite{key-46,key-47}.
Second, the model (\ref{eq:1.5}) can be obtained by reducing the
$SU\left(N\right)$ Yang-Mills theory with spontaneous breaking of
symmetry which is a fundamental mechanism in the Standard Model of particle
physics. Third, the model (\ref{eq:1.5}) emerges as a simplification of our mesonic and baryonic matrix models with special values of $m_{a}$ and $g$. Thus, analyzing chaos in model (\ref{eq:1.5}) may reveal generic features of chaos in gauge theories, matrix models, and even quantum gravity.

As we will see that our holographic matrix models can be simplified
to the model (\ref{eq:1.5}) in the following sections, here we specify
in general how the $SU\left(N\right)$ Yang-Mills theory with spontaneous
breaking of symmetry can be reduced to the model (\ref{eq:1.5}).
As a simple example, we consider the case of $SU\left(2\right)$ Yang-Mills
theory in the flat spacetime parameterized by $\left\{ x^{\mu}\right\} ,\mu=0,1,2,3$.
The Yang-Mills Lagrangian with a Higgs scalar is given by

\begin{align}
L & =-\frac{1}{2}\mathrm{Tr}F_{\mu\nu}F^{\mu\nu}+\left(D_{\mu}\phi\right)^{\dagger}\left(D_{\mu}\phi\right)-V\left(\phi\right),\nonumber \\
V\left(\phi\right) & =-\chi^{2}\left|\phi\right|^{2}+\frac{\lambda}{2}\left|\phi\right|^{4},\ \chi^{2}>0\label{eq:1.6}
\end{align}
where the Higgs scalar $\phi$ is in the fundamental representation
of $SU\left(2\right)$, the covariant derivative is given as $D_{\mu}\phi=\partial_{\mu}\phi-iA_{\mu}\phi$
and the gauge field strength is $F_{\mu\nu}=\partial_{\mu}A_{\nu}-\partial_{\nu}A_{\mu}-i\left[A_{\mu},A_{\nu}\right]$.
So the potential $V\left(\phi\right)$ is minimized at $\left\langle \phi\right\rangle =\left(\chi^{2}/\lambda\right)^{1/2}\xi$
which means $\phi$ acquires the vacuum expectation value (vev) and
the $SU\left(2\right)$ symmetry is spontaneously broken in the vacuum.
Here $\xi$ is a constant spinor satisfying the normalization condition
$\xi^{\dagger}\xi=1$. Therefore, by decomposing the Higgs scalar
$\phi$ as $\phi=\left\langle \phi\right\rangle +\varphi,$ then substituting
it for (\ref{eq:1.6}), we can obtain a mass term $m_{A}=\left(\frac{\chi^{2}}{\lambda}\right)^{1/2}$
for the gauge potential $A_{\mu}$ from the kinetic term of $\phi$
in the Lagrangian (\ref{eq:1.6}) as,

\begin{equation}
L=-\frac{1}{2}\mathrm{Tr}F_{\mu\nu}F^{\mu\nu}-m_{A}^{2}\xi^{\dagger}A_{\mu}A^{\mu}\xi+...\label{eq:1.9}
\end{equation}
which illustrates that the $SU\left(2\right)$ symmetry are totally
broken down. If we further assume the fields are spatially homogeneous,
then turn on 
\begin{equation}
A_{1}=\frac{1}{\sqrt{2}}x\left(t\right)\sigma^{1},A_{2}=\frac{1}{\sqrt{2}}y\left(t\right)\sigma^{2},
\end{equation}
(where $\sigma^{1,2}$ refer to the Pauli matrices), the (\ref{eq:1.9})
leads to the following Hamiltonian ($p_{x,y}=\dot{x},\dot{y}$)

\begin{equation}
H=\frac{1}{2}\left(p_{x}^{2}+p_{y}^{2}\right)+\frac{1}{2}m^{2}\left(x^{2}+y^{2}\right)+gx^{2}y^{2},
\end{equation}
which describes a two-dimensional coupled oscillator with homogeneous
mass and corresponds to the model (\ref{eq:1.5}) in the case of $D=2,m_{1}=m_{2}=m$. 

Following the above discussion, we can in addition reach to the Hamiltonian
for the coupled oscillator with inhomogeneous mass by incorporating the
Higgs scalar into the vector representation of $SU\left(2\right)$.
In this case, the covariant derivative operator for the Higgs scalar
$\phi$ becomes $\left(D_{\mu}\phi\right)^{a}=\left(\partial_{\mu}\phi-i\left[A_{\mu},\phi\right]\right)^{a}=\partial_{\mu}\phi^{a}+f^{abc}A_{\mu}^{b}\phi^{c}$,
where $a,b,c$ runs 1,2,3 in the vector representation of $SU\left(2\right)$.
Then the Higgs scalar $\phi$ can be decomposed as $\phi^{a}=\left\langle \phi^{a}\right\rangle +\varphi^{a}.$
Further, by substituting the decomposition for (\ref{eq:1.6}), we obtain
the mass term for $A_{\mu}^{a}$:

\begin{equation}
L=-\frac{1}{2}\mathrm{Tr}F_{\mu\nu}F^{\mu\nu}-\frac{1}{2}f^{abc}f^{dec}A_{\mu}^{a}A_{\mu}^{d}\left\langle \phi^{e}\right\rangle \left\langle \phi^{b}\right\rangle +...\label{eq:1.13}
\end{equation}
Consider the case where $SU\left(2\right)$ symmetry is partially broken,
e.g., the rotation symmetry in the third direction of vector space remains,
this leads to $\left\langle \phi^{a}\right\rangle =m_{A}\delta^{3a}.$
Imposing this into (\ref{eq:1.13}), we can obtain the Lagrangian of $A_{\mu}$
as

\begin{equation}
L=-\frac{1}{4}F_{\mu\nu}^{a}F^{a\mu\nu}-\frac{1}{2}m_{A}^{2}\left[\left(A_{\mu}^{1}\right)^{2}+\left(A_{\mu}^{2}\right)^{2}\right]+...\label{eq:1.15}
\end{equation}
which indicates that $A_{\mu}^{1,2}$ are massive while $A_{\mu}^{3}$
remains to be massless. Finally, we assume the fields are spatially
homogeneous and turn on

\begin{equation}
A_{1}^{1}=x\left(t\right),A_{2}^{2}=y\left(t\right),A_{3}^{3}=w\left(t\right),
\end{equation}
the Lagrangian (\ref{eq:1.15}) leads to the Hamiltonian,

\begin{equation}
H=\frac{1}{2}\left(p_{x}^{2}+p_{y}^{2}+p_{w}^{2}\right)+\frac{1}{2}m^{2}\left(x^{2}+y^{2}\right)+g\left(x^{2}y^{2}+x^{2}w^{2}+y^{2}w^{2}\right).
\end{equation}
corresponding to the model (\ref{eq:1.5}) in the case of $D=3$ with
inhomogeneous mass term ($m_{1}=m_{2}=m,m_{3}=0$). Altogether, we
can see the model (\ref{eq:1.5}) can be obtained by reducing the
Yang-Mills theory with spontaneous breaking of symmetry. In particular,
if the gauge symmetry is completely broken, the resultant Hamiltonian
describes the coupled oscillator with homogeneous mass. If the gauge
symmetry is partially broken, we arrive at the Hamiltonian describing
the coupled oscillator with homogeneous mass. In this sense, investigating
the chaos in model (\ref{eq:1.5}) may reveal generic features of
the chaotic behavior in gauge theory and it may also be significant
for the gauge-gravity duality.

\subsection{The outline}

In this work, we focus on chaos in the mesonic and baryonic matrix
models obtained from the $\mathrm{D4}/\mathrm{D6}/\overline{\mathrm{D6}}$
approach. For actual calculations, we study the chaos in the model
(\ref{eq:1.5}), since the mesonic and baryonic matrix models can
be simplified to be the model (\ref{eq:1.5}) with special parameters.
The manuscript is organized as follows: In Section 2, the essential
parts of the $\mathrm{D4}/\mathrm{D6}/\overline{\mathrm{D6}}$ system
are collected, including the D4-brane soliton background, the
antipodal embedding of the $\mathrm{D6}/\overline{\mathrm{D6}}$-branes, and the baryon vertex. In Section 3, we demonstrate how to obtain
the mesonic and baryonic matrix models from the $\mathrm{D4}/\mathrm{D6}/\overline{\mathrm{D6}}$
approach and how to simplify these matrix models to the special case
of model (\ref{eq:1.5}). In Section 4, the analysis of the classical
chaos is specified with respect to the mesonic and baryonic matrix
models. It includes a numerical evaluation of the orbits in
the Poincar\'e section and the classical Lyapunov exponent. To illustrate
exactly the large $N_{c}$ behavior, we further use the perturbative
method to analytically solve the classical equations of motion by
perturbing the coupling constant. In Section 5, we turn to the analysis
of the quantum chaos by computing numerically the OTOCs with respect
to the quantum version of the mesonic and baryonic matrix models.
In particular, the large $N_{c}$ behavior is also evaluated analytically
by using the perturbative method in quantum mechanics since the coupling
constants in the associated Hamiltonian become perturbative in the
large $N_{c}$ limit. A summary and discussion are given in Section
6. In addition, we append the full bosonic action for a D-brane, the
details for deriving the baryonic matrix model, the calculations of
the OTOC and Lyapunov coefficient for 1d harmonic oscillator to the
end of this manuscript which are very useful in this work.

\section{The holographic setup of the $\mathrm{D4}/\mathrm{D6}/\overline{\mathrm{D6}}$
approach}

In this section, we summarize the essential features of the D4-brane soliton background at large \(N_c\) (color sector) and the \(\mathrm{D6}/\overline{\mathrm{D6}}\)-brane embeddings (flavor sector) in our holographic QCD setup. To model baryonic states, we further introduce the baryon vertex which is a D4-brane wrapped on \(S^4\). All D-brane configurations are described with explicit geometric interpretations.

\subsection{The D4-brane soliton background}

The D4-brane soliton background is produced by $N_{c}$ coincident
D4-branes compactified on a circle $S^{1}$ whose low-energy dynamics
is described by IIA supergravity. The background metric $g$, dilaton
$\phi$ and the Romand-Romand $C_{3}$ form take the solution as \cite{key-34,key-35,key-36,key-37},

\begin{align}
ds^{2} & =\left(\frac{U}{R}\right)^{3/2}\left[\eta_{\mu\nu}dx^{\mu}dx^{\nu}+f\left(U\right)d\tau^{2}\right]+\left(\frac{R}{U}\right)^{3/2}\frac{dU^{2}}{f\left(U\right)}+R^{3/2}U^{1/2}d\Omega_{4},\nonumber \\
e^{\phi} & =\left(\frac{U}{R}\right)^{3/4},\ F_{4}=dC_{3}=\frac{N_{c}}{\Omega_{4}}\epsilon_{4},\ f\left(U\right)=1-\frac{U_{KK}^{3}}{U^{3}}.\label{eq:2.1}
\end{align}
The indices $\mu,\nu$ run over 0,1,2,3 parametrizing the four non-compact
directions along the D4-branes. The direction $\tau$ is the 4th spacial
direction of the D4-branes compactified on the circle $S^{1}$, the
size of $S^{1}$ is denoted as $\delta\tau$ satisfying

\begin{equation}
M_{KK}=\frac{2\pi}{\delta\tau}=\frac{3U_{KK}^{1/2}}{2R^{3/2}},
\end{equation}
in order to avoid the conical singularity. $d\Omega_{4}^{2}$ and
$\epsilon_{4}$ refers respectively to the line element and volume
form on a unit four-sphere $S^{4}$, $\Omega_{4}=8\pi^{2}/3$ is the
volume of a unit $S^{4}$. The coordinate $U\in\left[U_{KK},+\infty\right)$
refers to the holographic radius in the 56789-directions and the holographic
boundary is defined at $U\rightarrow\infty$. The gravity-gauge
theory correspondence relates the variables as follows:

\begin{equation}
R^{3}=\frac{1}{2}\frac{\lambda l_{s}^{2}}{M_{KK}},g_{s}=\frac{1}{2\pi}\frac{\lambda}{M_{KK}l_{s}N_{c}},U_{KK}=\frac{2}{9}\lambda M_{KK}l_{s}^{2},\lambda=g_{\mathrm{YM}}^{2}N_{c}.
\end{equation}
Here, \( R \) denotes the bulk radius, and \( g_s \) is the string coupling constant. The parameters \( \lambda \) and \( g_{\text{YM}} \) represent the 't Hooft and Yang-Mills coupling constants, respectively. The supersymmetry on the D4-branes breaks down below the Kaluza-Klein (KK) mass scale due to the anti-periodical boundary condition for the fermion along $S^{1}$. The behavior of the Wilson loop in this geometry satisfies the area law exhibiting confinement in the dual field theory \cite{key-16,key-35}. Altogether, the dual theory corresponding to the background (\ref{eq:2.1}) below the energy scalar $M_{KK}$ is the confining $U\left(N_{c}\right)$ Yang-Mills theory.

\subsection{The intersection embedding of the $\mathrm{D6}/\overline{\mathrm{D6}}$-branes}

We consider the embedding of the $N_{f}$ coincident $\mathrm{D6}/\overline{\mathrm{D6}}$-branes
as the flavor branes in the D4-brane soliton background following \cite{key-34}.
By imposing the coordinate transformation

\begin{align}
U\left(\rho\right) & =A\left(\rho\right)U_{KK},\nonumber \\
A\left(\rho\right) & =\left(\rho^{3/2}+\frac{1}{4\rho^{3/2}}\right)^{2/3},\nonumber \\
K\left(\rho\right) & =\frac{R^{3/2}U_{KK}^{1/2}A^{1/2}}{\rho^{2}}.
\end{align}
The metric (\ref{eq:2.1}) can be written as

\begin{equation}
ds^{2}=\left(\frac{U}{R}\right)^{3/2}\left(\eta_{\mu\nu}dx^{\mu}dx^{\nu}+fd\tau^{2}\right)+K\left(d\rho^{2}+\rho^{2}d\Omega_{4}\right).\label{eq:2.4}
\end{equation}
For the intersectional embedding, the D-brane configuration is illustrated
in Table \ref{tab:1}. 
\begin{table}
\begin{centering}
\begin{tabular}{|c|c|c|c|c|c|c|c|c|c|c|}
\hline 
 & $0$ & $1$ & $2$ & $3$ & $4\left(\tau\right)$ & $5\left(U\right)$ & $6$ & $7$ & $8$ & $9$\tabularnewline
\hline 
\hline 
D4-branes & - & - & - & - & - &  &  &  &  & \tabularnewline
\hline 
$\mathrm{D6}/\overline{\mathrm{D6}}$-branes & - & - & - & - &  & - & - & - &  & \tabularnewline
\hline 
Baryon vertex & - &  &  &  &  &  & - & - & - & -\tabularnewline
\hline 
\end{tabular}
\par\end{centering}
\caption{\label{tab:1}The D-brane configuration in the $\mathrm{D4/D6/\overline{D6}}$
approach.}

\end{table}
So, further decompose (\ref{eq:2.4}), we obtain

\begin{align}
ds^{2} & =\left(\frac{U}{R}\right)^{3/2}\left(\eta_{\mu\nu}dx^{\mu}dx^{\nu}+fd\tau^{2}\right)+K\left(d\xi^{2}+\xi^{2}d\Omega_{2}+dX_{8}^{2}+dX_{9}^{2}\right)\nonumber \\
 & =\left(\frac{U}{R}\right)^{3/2}\left(\eta_{\mu\nu}dx^{\mu}dx^{\nu}+fd\tau^{2}\right)+K\left(d\xi^{2}+\xi^{2}d\Omega_{2}+dr^{2}+r^{2}d\phi^{2}\right),
\end{align}
where $X_{8}=r\cos\phi,X_{9}=r\sin\phi$ denotes the 89-direction
and $\rho^{2}=\xi^{2}+r^{2}$. In this sense, $\mathrm{D6}/\overline{\mathrm{D6}}$-branes
can be embedded at $\phi=\phi_{0}$ where $\phi_{0}$ is a constant,
thus the induced metric on the $\mathrm{D6}/\overline{\mathrm{D6}}$-branes
is 

\begin{align}
ds_{\mathrm{D6}/\overline{\mathrm{D6}}}^{2} & =\left(\frac{U}{R}\right)^{3/2}\eta_{\mu\nu}dx^{\mu}dx^{\nu}+\left[\left(\frac{U}{R}\right)^{3/2}f\tau^{\prime2}+Kr^{\prime2}+K\right]d\xi^{2}+K\xi^{2}d\Omega_{2}^{2}.\label{eq:2.7}
\end{align}
$"\prime"$ refers to the derivative with respect to $\xi$. The equations of motion for the embedding functions $\tau\left(\xi\right),r\left(\xi\right)$
can be obtained by varying the DBI (Dirac-Born-Infeld) action of the
D6-brane given in the Appendix A as,

\begin{equation}
S_{\mathrm{D6}}=T_{\mathrm{D6}}\int d^{7}xe^{-\phi}\sqrt{-g_{\mathrm{D6}}}.
\end{equation}
Resultantly, it leads respectively to the solutions $\tau\left(U\right)=\pm\frac{\pi}{2}$
and $r\left(\xi\right)=0$ \cite{key-34} which denote that the $\mathrm{D6}/\overline{\mathrm{D6}}$-branes
are located at the antipodal position of $S^{1}$ and at $\left(X_{8},X_{9}\right)=\left(0,0\right)$.
With this solution, the induced metric on $\mathrm{D6}/\overline{\mathrm{D6}}$-branes
(\ref{eq:2.7}) becomes,

\begin{equation}
ds_{\mathrm{D6}/\overline{\mathrm{D6}}}^{2}=\left(\frac{U}{R}\right)^{3/2}\eta_{\mu\nu}dx^{\mu}dx^{\nu}+K\left(d\xi^{2}+\xi^{2}d\Omega_{2}^{2}\right).\label{eq:2.9}
\end{equation}
We will use the induced metric (\ref{eq:2.9}) as the intersection
embedding of the $\mathrm{D6}/\overline{\mathrm{D6}}$-branes.

\subsection{The baryon vertex}

The baryon vertex (b.v.) in the D4-brane background is identified
as a baryonic D4-brane wrapped on $S^{4}$, occupying the directions
of the spherical coordinates presented in (\ref{eq:2.1}) \cite{key-36,key-37,key-38,key-39}
with $N_{c}$ open strings ending on it. The configuration of the
baryon vertex is illustrated in Table \ref{tab:1}. The reason that
the wrapped D4-brane is a baryon vertex is given as follows. Since
there is a $U\left(1\right)$ chemical potential $A_{0}$ on the wrapped
D4-brane, it couples to the CS (Chern-Simons) action of the D4-brane
as,

\begin{align}
S_{\mathrm{CS}} & =-g_{s}T_{\mathrm{D4}}\left(2\pi\alpha^{\prime}\right)\int_{\mathbb{R}\times S^{4}}F\wedge C_{3}\nonumber \\
 & =-g_{s}T_{\mathrm{D4}}\left(2\pi\alpha^{\prime}\right)\int_{\mathbb{R}\times S^{4}}A\wedge F_{4}\nonumber \\
 & =N_{c}\int_{\mathbb{R}}dtA_{0},
\end{align}
where we have used the solution (\ref{eq:2.1}) for $F_{4}$. Accordingly,
$F_{4}$ contributes $N_{c}$ units to the charge carried by $A_{0}$
which in this sense is expected to be a baryonic chemical potential.
On the other hand, the $N_{c}$ open strings (with opposite orientation
to the wrapped D4-brane) ending on the wrapped D4-brane take $-N_{c}$
units charges which cancel exactly $N_{c}$ charges contributed from
$F_{4}$. So the total charge vanishes which implies the baryon number
is conserved. Therefore the wrapped D4-brane is a baryon vertex (or
an anti-baryon vertex dependent on the orientation).

Since the stable location of the baryon vertex is at $U=U_{KK}$,
the DBI action for the wrapped D4-brane can be written as,

\begin{equation}
S_{\mathrm{b.v.}}=-T_{\mathrm{D4}}\int d^{5}xe^{-\phi}\sqrt{-g}=-\frac{\lambda M_{KK}N_{c}}{27\pi}\int_{\mathbb{R}}dx^{0},\label{eq:2.10}
\end{equation}
where the background solution (\ref{eq:2.1}) at $U=U_{KK}$ has been
imposed. And (\ref{eq:2.10}) illustrates the baryon mass is $m_{b}=\frac{\lambda M_{KK}N_{c}}{27\pi}$
proportional to $N_{c}$ as it is expected in the large $N_{c}$ QCD
\cite{key-48}. 

\section{The matrix models}

The scalar mesons on the flavor $\mathrm{D6}/\overline{\mathrm{D6}}$-branes
can be described by a matrix model obtained from the D-brane action,
and the baryon vertex can also be rewritten as a matrix model if we
impose the rules of T-duality in the string theory. In this section,
we will introduce briefly these matrix models as the mesonic and baryonic
matrix models, then simplify them to be the models of the coupled
harmonic oscillators.

\subsection{The mesonic matrix model}

In our setup of $\mathrm{D4}/\mathrm{D6}/\overline{\mathrm{D6}}$,
let us take into account the fluctuations of $X_{8,9}$ as $w^{8}=\delta X_{8},w^{9}=\delta X_{9}$
which correspond to the scalar and pseudo-scalar mesons excited on
the $\mathrm{D6}/\overline{\mathrm{D6}}$-branes in this model \cite{key-34}.
Note that $w^{8,9}$ are transverse modes of the $\mathrm{D6}/\overline{\mathrm{D6}}$-branes,
and on the other hand, under the T-duality, $w^{8,9}$ are matrices
in the adjoint representation of $U\left(N_{f}\right)$ group. Therefore,
by turning off the gauge field, we can write down a matrix model for
$\left(w_{8},w_{9}\right)$ on the $\mathrm{D6}/\overline{\mathrm{D6}}$-branes
by following the DBI action given in Appendix A as,

\begin{equation}
S_{\mathrm{D6}}=-T_{\mathrm{D6}}\int d^{7}xe^{-\phi}\sqrt{-g}\mathrm{Tr}\left\{ \frac{1}{4}g^{ab}\partial_{a}w^{m}\partial_{b}w^{n}g_{mn}-\frac{1}{4\left(2\pi\alpha^{\prime}\right)^{2}}g_{mr}g_{ns}\left[w^{m},w^{n}\right]\left[w^{r},w^{s}\right]\right\} ,\label{eq:3.1}
\end{equation}
where $w^{m}$ is chosen as Hermitian matrix and $m,n=8,9$. If we
consider the case of $N_{f}=1$, the matrix $w^{m}$ is Abelian hence
the commutators presented in (\ref{eq:3.1}) vanishes leading to a
trivially non-chaotic solution for $w^{m}$. In this sense, we consider
the two-flavor case $N_{f}=2$ as the most simple but non-trivial
case, since, on the other hand, meson is usually consisted of two
flavored quarks. So (\ref{eq:3.1}) can be further simplified by setting

\begin{equation}
w^{8}=\mathcal{N}P\left(\xi\right)\frac{x\left(t\right)}{2\pi\alpha^{\prime}}\sigma^{1},w^{9}=\mathcal{N}P\left(\xi\right)\frac{y\left(t\right)}{2\pi\alpha^{\prime}}\sigma^{2},P\left(\xi\right)=A^{-1}\rho^{-1/2},
\end{equation}
leading to an action for the coupled harmonic oscillator with homogeneous
mass term as

\begin{equation}
S_{\mathrm{D6}}=V_{3}\int dt\left[\frac{1}{2}\left(\dot{x}^{2}+\dot{y}^{2}\right)-\frac{1}{2}m^{2}\left(x^{2}+y^{2}\right)-gx^{2}y^{2}\right].\label{eq:3.3}
\end{equation}
where $\mathcal{N},m,g$ are constants given as,

\begin{align}
\frac{M_{KK}^{2}N_{c}\lambda^{2}\mathcal{N}^{2}}{2^{4}3^{4}}\int d\xi\frac{A^{2}\xi^{2}P^{2}}{\rho^{4}} & =\frac{1}{4},\nonumber \\
\frac{M_{KK}^{4}N_{c}\lambda^{2}\mathcal{N}^{2}}{2\times3^{6}}\int d\xi\frac{A^{3}\xi^{2}}{\rho^{2}}P^{\prime2} & =m^{2}=\frac{4}{27}M_{KK}^{2}\left[7+11\ _{2}F_{1}\left(1,1,\frac{1}{3},-1\right)\right]\simeq M_{KK}^{2},\nonumber \\
\frac{M_{KK}^{4}N_{c}\lambda^{4}\mathcal{N}^{4}}{2^{2}3^{8}\pi^{2}}\int d\xi\frac{A^{4}\xi^{2}}{\rho^{6}}P^{4} & =g=\frac{16\times2^{2/3}}{5\pi^{2}N_{c}}\simeq0.55N_{c}^{-1}.\label{eq:3.4}
\end{align}
Therefore, action (\ref{eq:3.3}) is the case of $D=2$, taking the
homogeneous mass, with respect to the model (\ref{eq:1.5}), and we
will investigate the chaos in the holographic mesonic system by using
the action (\ref{eq:3.3}) for simplification.

\subsection{The baryonic matrix model}

The matrix model for baryon describes effectively the dynamics of
the baryon vertex which is a D4-brane wrapped on $S^{4}$ in our $\mathrm{D4}/\mathrm{D6}/\overline{\mathrm{D6}}$
setup. Here, we give the conjectured action for the baryonic matrix
model and leave its derivation in the Appendix B. The conjectured
action for the baryonic matrix model is,

\begin{align}
S= & \frac{2}{27\pi}\lambda M_{KK}N_{c}\int dt\mathrm{Tr}\left\{ \frac{1}{2}D_{0}X^{M}D_{0}X^{M}-\frac{2}{3}M_{KK}^{2}\left(X^{4}\right)^{2}+\frac{2}{3^{6}\pi^{2}}\lambda^{2}M_{KK}^{4}\left[X^{M},X^{N}\right]^{2}\right\} \nonumber \\
 & +N_{c}\int dt\mathrm{Tr}A_{0},\label{eq:3.5}
\end{align}
where the covariant derivative is defined as $D_{0}X^{M}=\partial_{0}X^{M}-i\left[A_{0},X^{M}\right],M,N=1,2,3,4$.
The matrices $X^{M}$ and $A_{0}$ are in the adjoint representation
of $U\left(n_{b}\right)$ group where $n_{b}$ refers to the number
of the baryon vertex. This model is a simplified version of the baryonic
matrix model proposed in \cite{key-40,key-42}, and it describes the
dynamics of the matrices $X^{M}$ under the external baryonic potential
field $A_{0}$. The action (\ref{eq:3.5}) of the baryonic matrix
model can be obtained by imposing the T-duality rules on a baryon
vertex i.e. the wrapped D4-brane on $S^{4}$ in the background (\ref{eq:2.1}).
Since $A_{0}$ refers to the baryonic chemical potential, its Abelian
part is mostly relevant in the action. On the other hand, if we consider
the case of $n_{b}=1$, all the matrices $X^{M}$ becomes Abelian
and the action becomes harmonic oscillator as,

\begin{equation}
S=\frac{2}{27\pi}\lambda M_{KK}N_{c}\int dt\left[\frac{1}{2}\partial_{0}X^{M}\partial_{0}X^{M}-\frac{2}{3}M_{KK}^{2}\left(X^{4}\right)^{2}\right]+N_{c}\int dt\mathrm{Tr}A_{0},\label{eq:3.6}
\end{equation}
which is not really chaotic according to the analysis in the Appendix
C. Therefore, to investigate the chaos, we consider the case of $n_{b}=2$
and chose the following ansatz for the matrix model as,

\begin{equation}
X^{1}=\frac{1}{\mathcal{N}}\frac{\sigma^{1}}{\sqrt{2}}x\left(t\right),X^{2}=\frac{1}{\mathcal{N}}\frac{\sigma^{2}}{\sqrt{2}}y\left(t\right),X^{3}=\frac{1}{\mathcal{N}}\frac{h\left(t\right)}{\sqrt{2}}\mathbf{1}_{2\times2},X^{4}=\frac{1}{\mathcal{N}}\frac{\sigma^{3}}{\sqrt{2}}w\left(t\right),A_{0}=\frac{\hat{A}_{0}\left(t\right)}{\sqrt{2}}\mathbf{1}_{2\times2},
\end{equation}
hence the action (\ref{eq:3.5}) is simplified as a coupled harmonic
oscillator with inhomogeneous mass term as

\begin{align}
S & =S_{\mathrm{ocs}}+S_{h}+S_{A},\nonumber \\
S_{\mathrm{ocs}} & =\int dt\left[\frac{1}{2}\left(\dot{x}^{2}+\dot{y}^{2}+\dot{w}^{2}\right)-\frac{1}{2}m^{2}w^{2}-g\left(x^{2}y^{2}+x^{2}w^{2}+y^{2}w^{2}\right)\right],\nonumber \\
S_{h} & =\frac{1}{2}\int dt\dot{h}^{2},\ S_{A}=N_{c}\int dt\hat{A}_{0}.\label{eq:3.8}
\end{align}
where $\mathcal{N},m,g$ are constants given as

\begin{equation}
m=\frac{2\sqrt{3}}{3}M_{KK},g=\frac{4M_{KK}^{3}}{27\pi N_{c}},\mathcal{N}=\left(\frac{2}{27\pi}\lambda M_{KK}N_{c}\right)^{1/2}.
\end{equation}
Note that, the case of baryon number $n_{b}=2$ corresponds to the
baryon system with two-body interaction in the large $N_{c}$ expansion,
e.g. deuteron as dibaryon could be the typical system. Besides, the classical
action (\ref{eq:3.8}) implies that $h$ can be solved as a constant,
hence we can omit $S_{h},S_{A}$, just focus on $S_{\mathrm{ocs}}$
which describes a three-dimensional coupled harmonic oscillator with
inhomogeneous mass term. Therefore the baryonic matrix model can also
be simplified to be the model (\ref{eq:1.5}) with special mass terms
and coupling constant. We will investigate the chaos in the baryonic
matrix model with respect to the action $S_{\mathrm{ocs}}$ in (\ref{eq:3.8}).

\section{Analysis of the classical chaos}

In this section, we will discuss the chaos in the mesonic and baryonic
matrix models with respect to the classical version of action (\ref{eq:3.3})
and action (\ref{eq:3.8}). Our concerns would be the classical orbits
on the Poincar\'e section, the relation between the classical Lyapunov
coefficient and the total energy. We will work in the unit of $M_{KK}=1$
throughout this manuscript.

\subsection{The Poincar\'e section}

\begin{figure}
\begin{centering}
\includegraphics[scale=0.25]{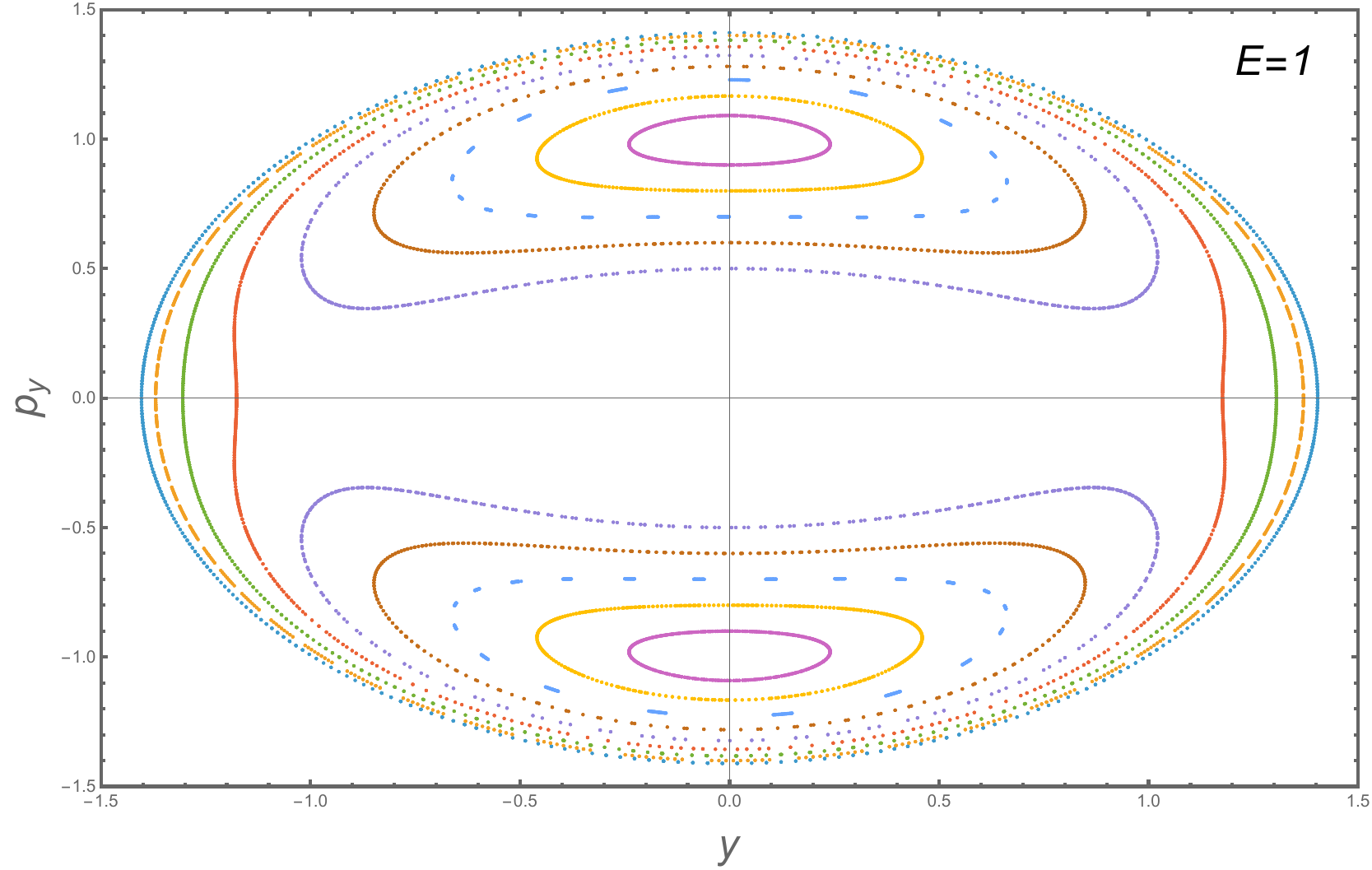}\includegraphics[scale=0.25]{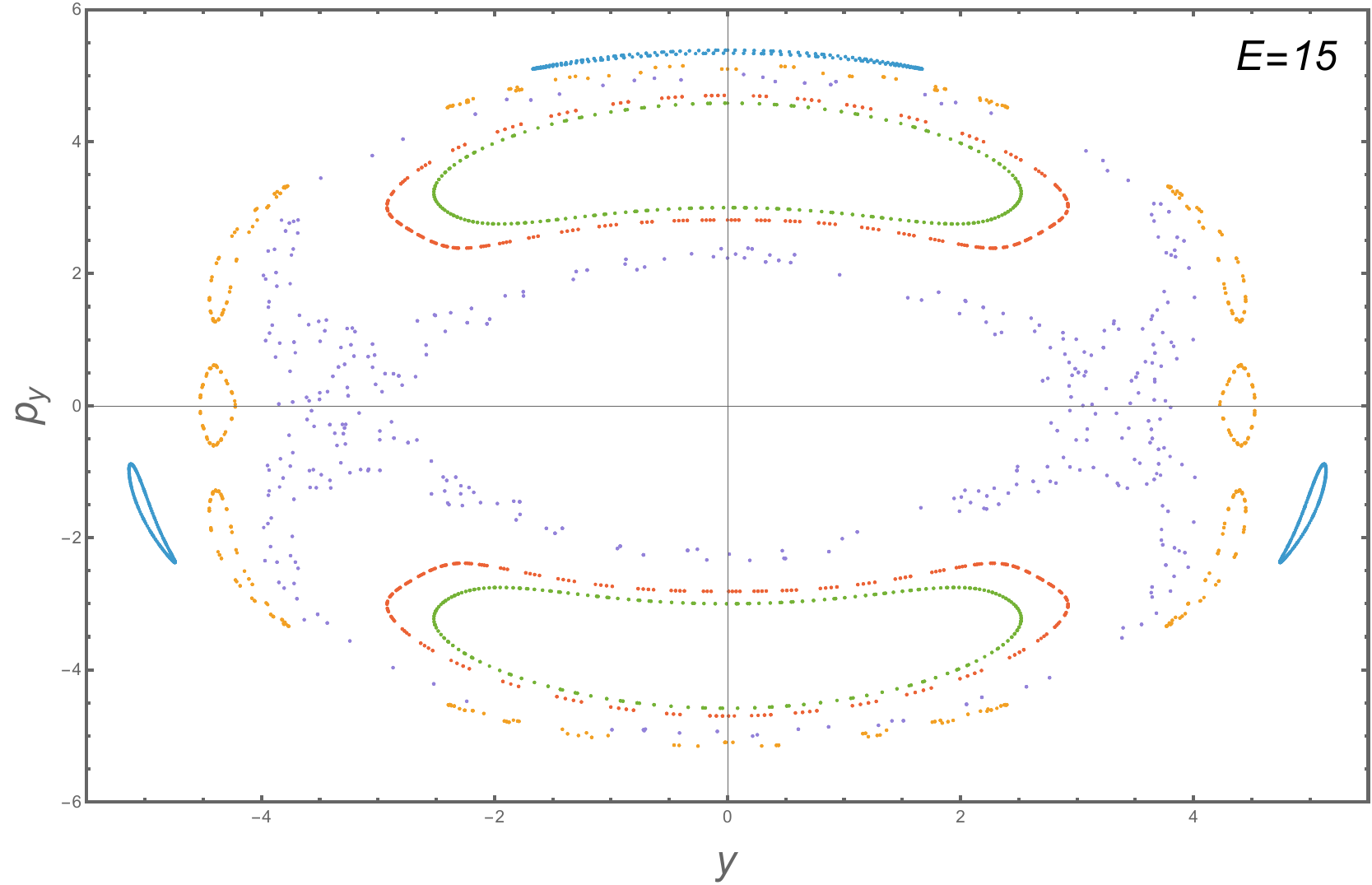}
\par\end{centering}
\begin{centering}
\includegraphics[scale=0.25]{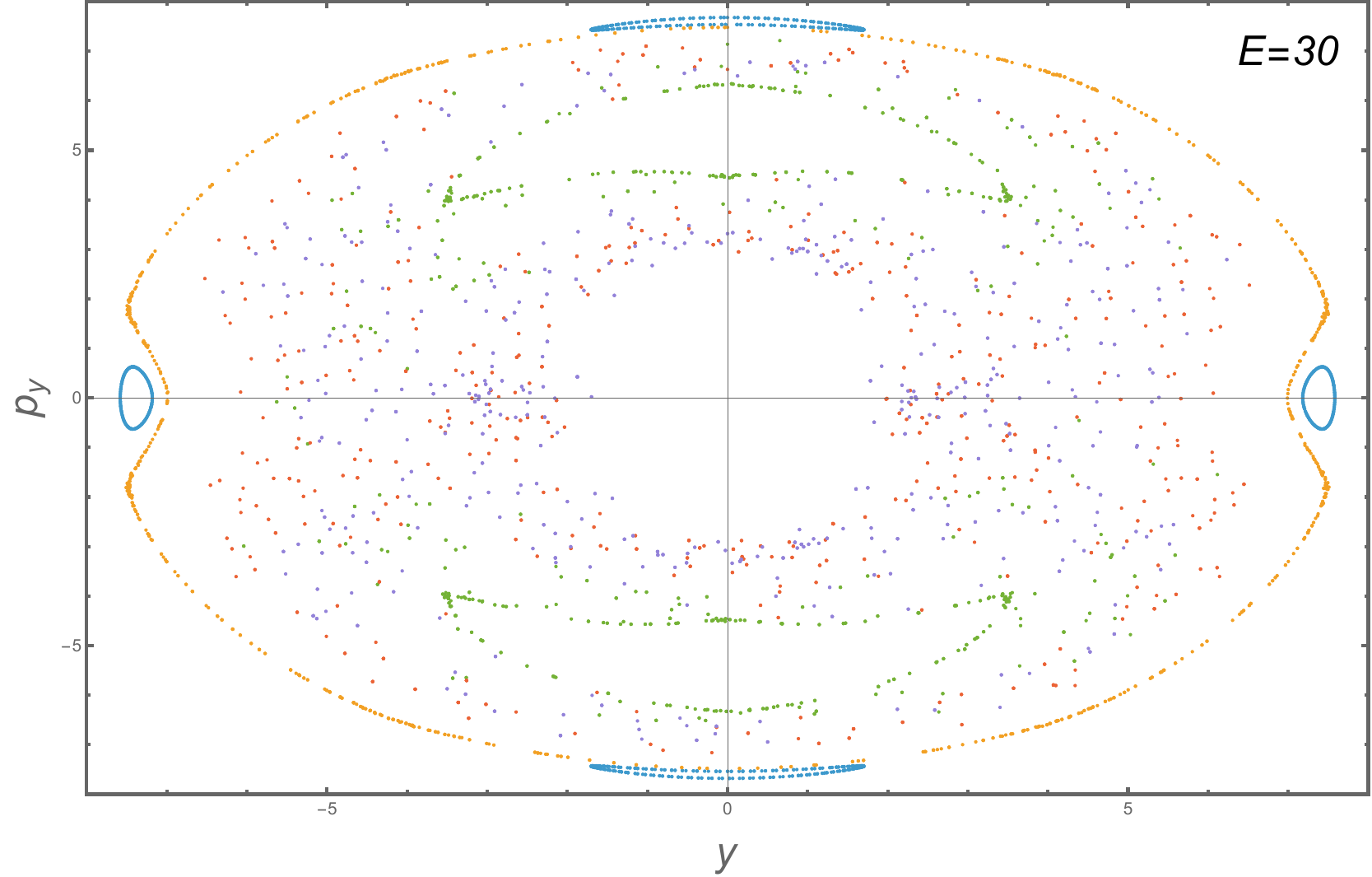}\includegraphics[scale=0.25]{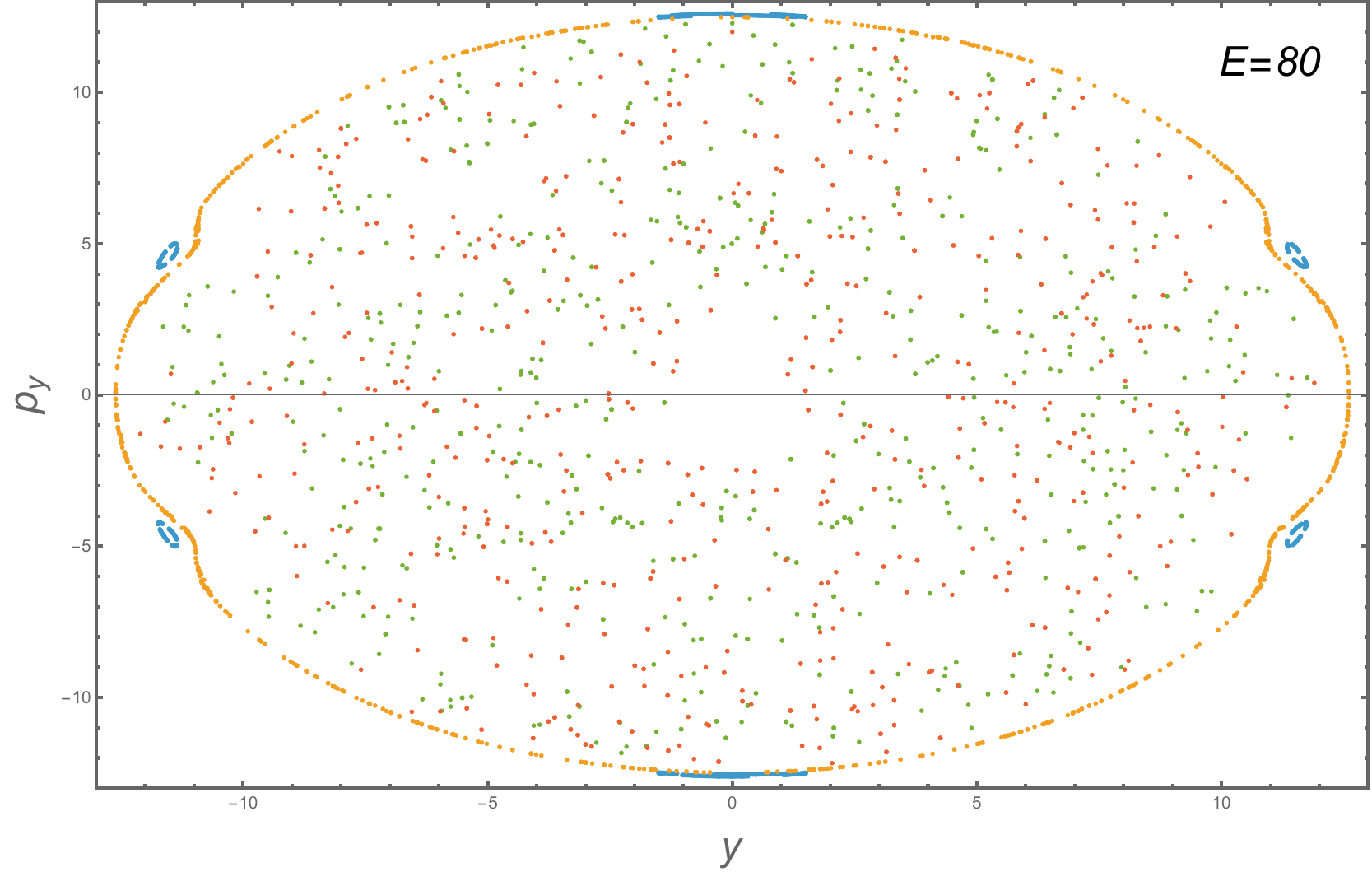}
\par\end{centering}
\caption{\label{fig:1}The classical trajectory on the Poincare section of
the mesonic matrix model at $x\left(t\right)=0$ with various energy
$E=1,15,30,80$. The horizontal axis is $y\left(t\right)$, the vertical
axis is $p_{y}=\dot{y}\left(t\right)$.}
\end{figure}
\begin{figure}
\begin{centering}
\includegraphics[scale=0.25]{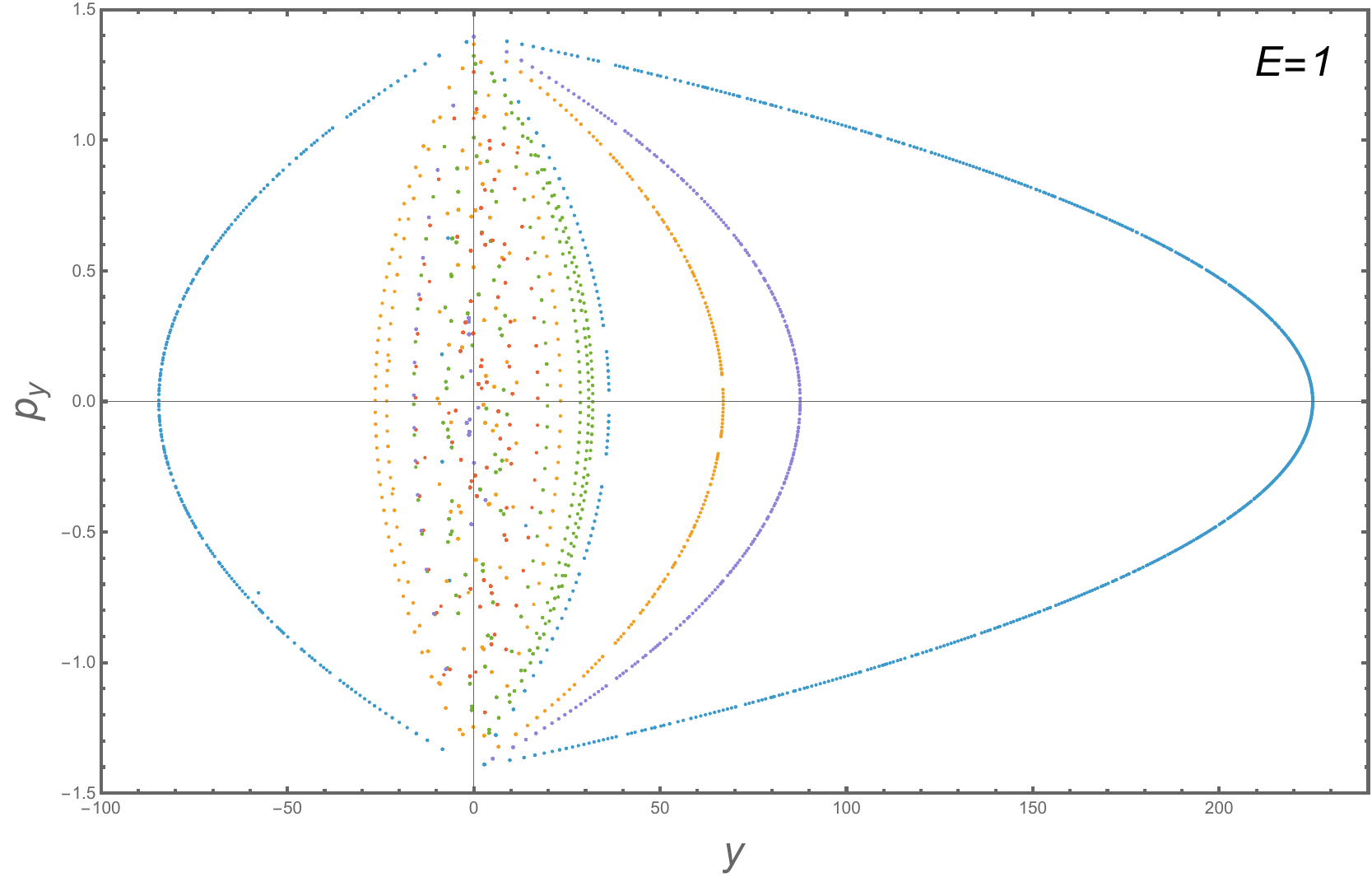}\includegraphics[scale=0.25]{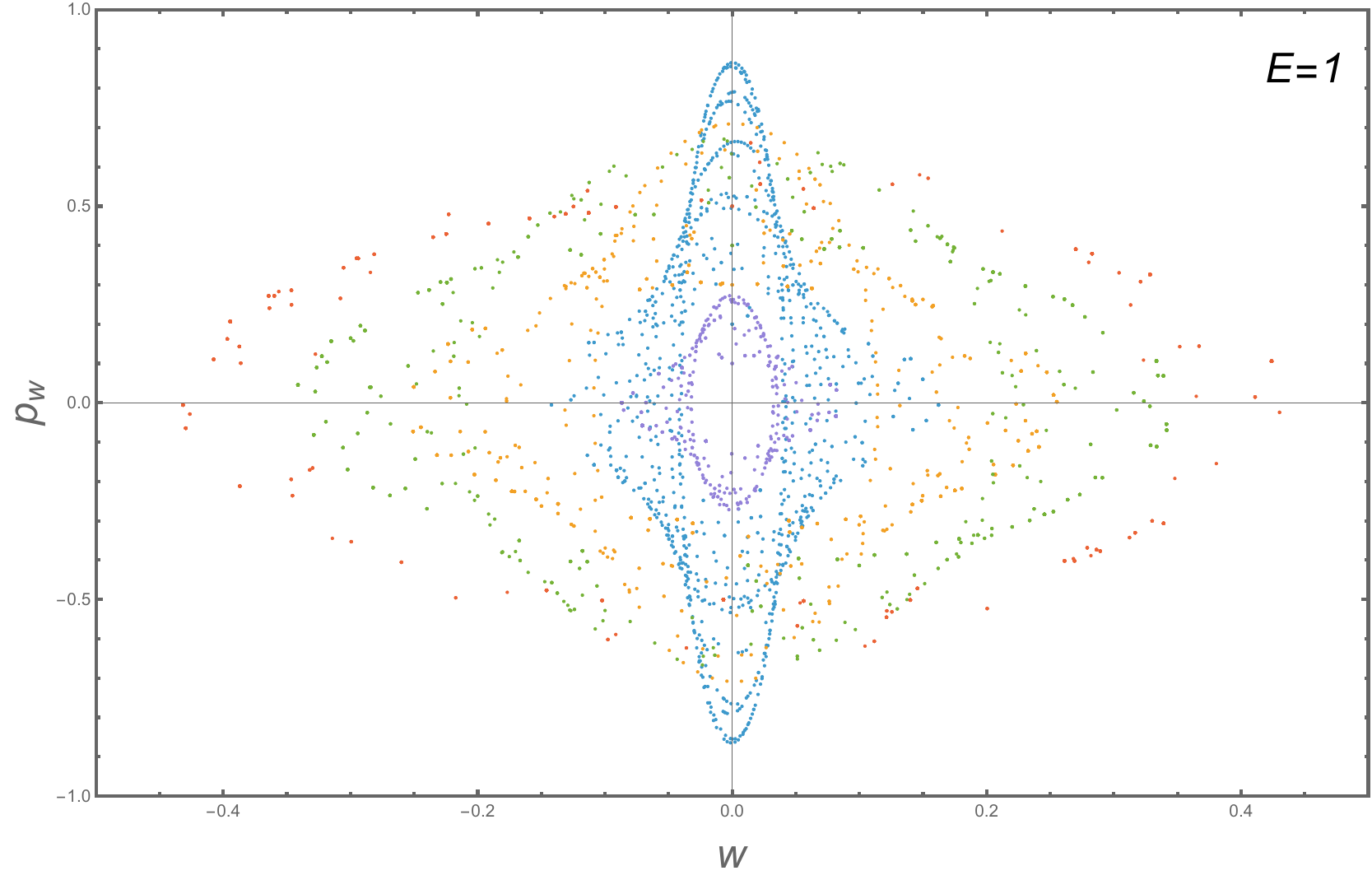}
\par\end{centering}
\begin{centering}
\includegraphics[scale=0.25]{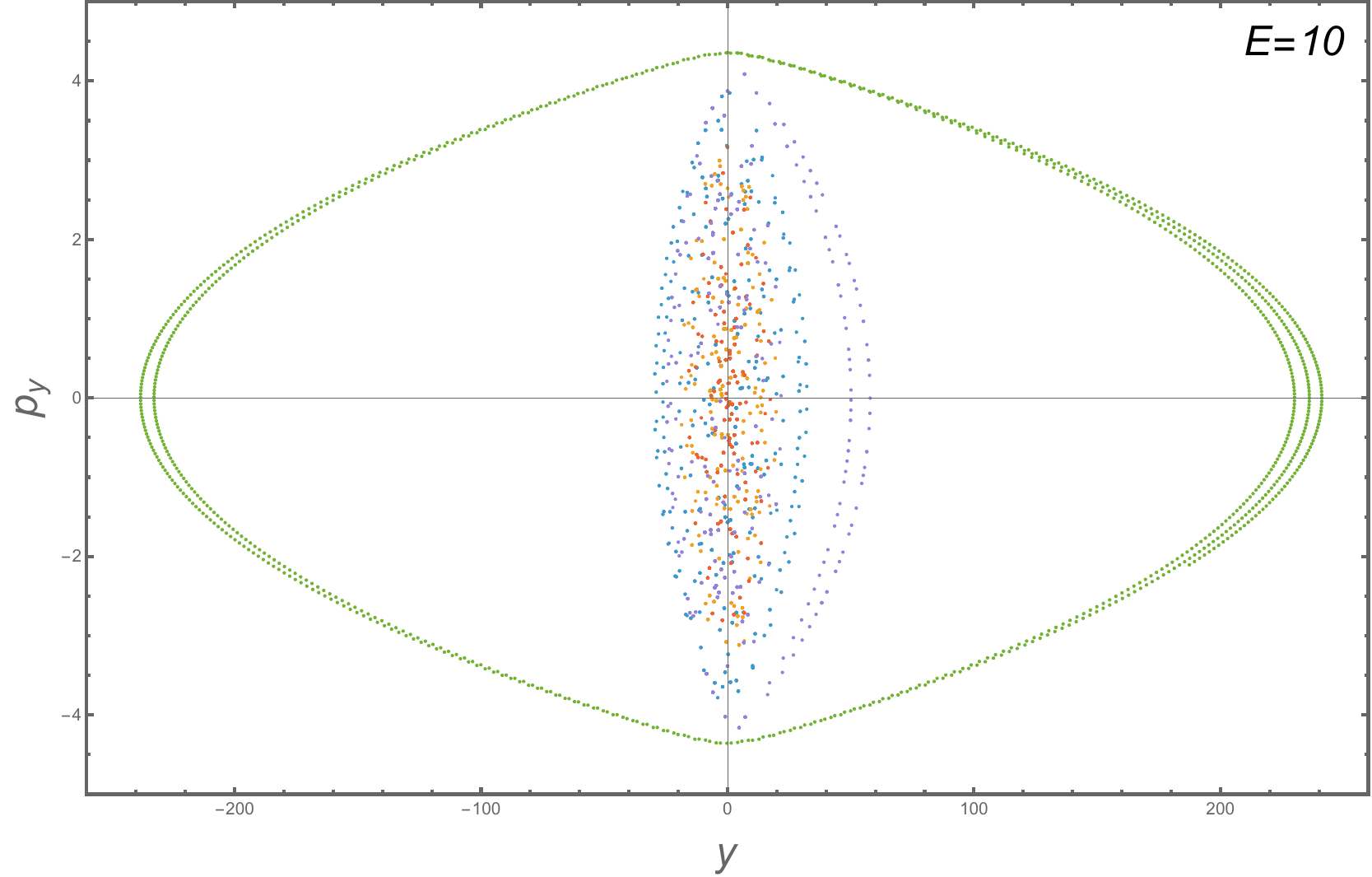}\includegraphics[scale=0.25]{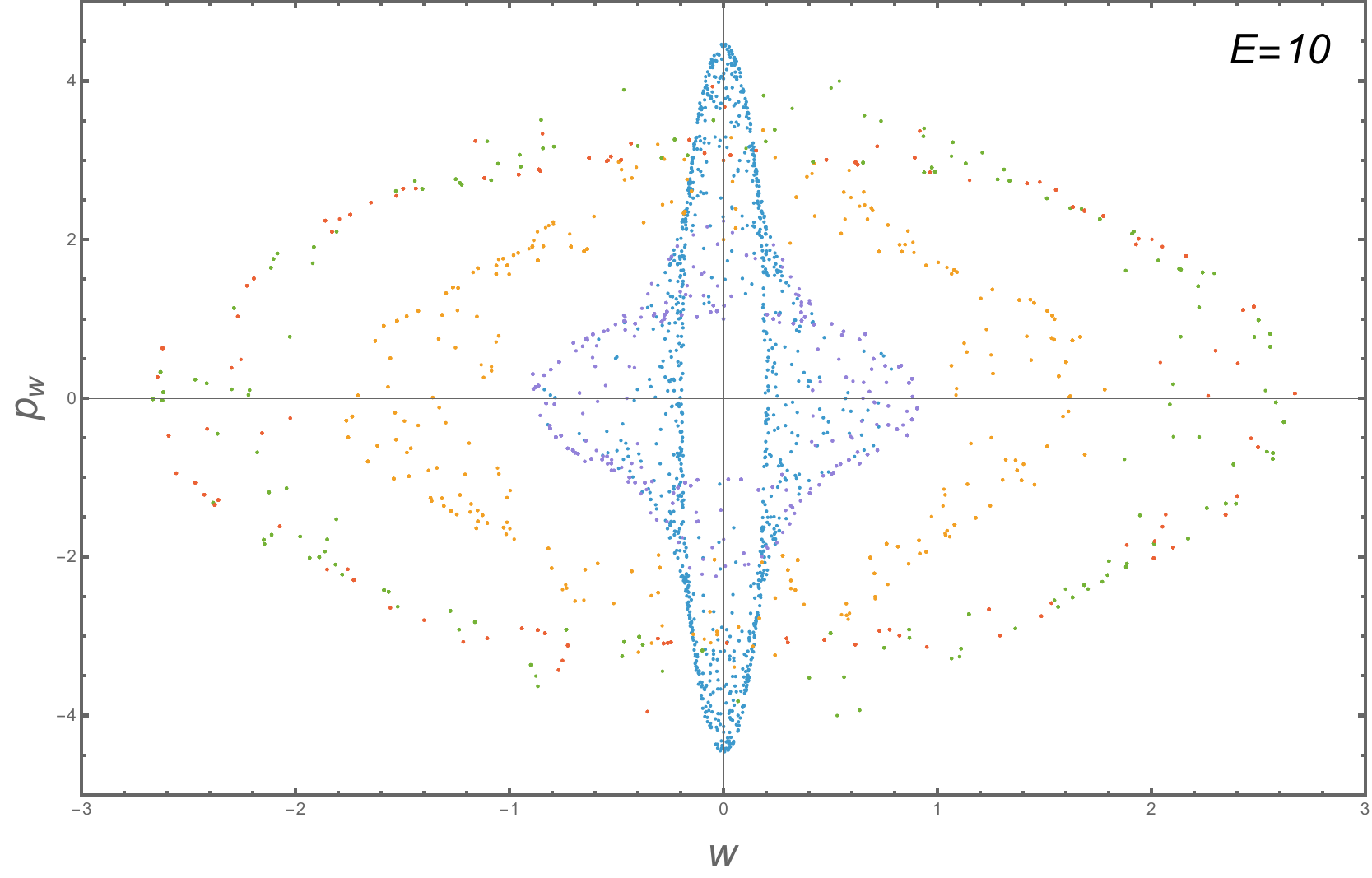}
\par\end{centering}
\caption{\label{fig:2}The classical trajectory on the Poincare section of
the baryonic matrix model at $x\left(t\right)=0$ with various energy
$E=1,10$. The horizontal axis is $y\left(t\right)$ or $w\left(t\right)$,
the vertical axis is $p_{y}=\dot{y}\left(t\right)$ or $p_{w}=\dot{w}\left(t\right)$.}
\end{figure}

Working out the Poincar\'e sections is one of the popular method to
find qualitatively the classical chaos. According to \cite{key-1,key-44},
the system is in the regular phase if the Poincar\'e section displays
specific orbits, otherwise the system is in the chaotic phase and
the Poincar\'e section is filled. Keeping these in mind, we plot out
the Poincar\'e sections at $x\left(t\right)=0$ at several chosen values
of the total energy $E$ with respect to the mesonic matrix model
in Figure \ref{fig:1} and with respect to the baryonic matrix model
in Figure \ref{fig:2}. 
For the mesonic matrix model, we see clearly the orbits in the Poincar\'e
section at very low energy while the orbits are destroyed at high
energy. So the system is in the regular phase and goes to the chaotic
phase when the total energy increases. Hence there is a phase transition
between the regular and chaotic phase due to the classical analysis
of the mesonic matrix model. These conclusions agree basically with
the discussion in \cite{key-6,key-7}. However, there is not a definite
phase transition in the baryonic matrix model since its Poincar\'e section
aways contains the random scattered plots. It is believed the baryonic
matrix model is always in the chaotic phase due to its inhomogeneous
mass terms in the action (\ref{eq:3.8}).

\subsection{The classical Lyapunov exponent}

\begin{figure}
\begin{centering}
\includegraphics[scale=0.25]{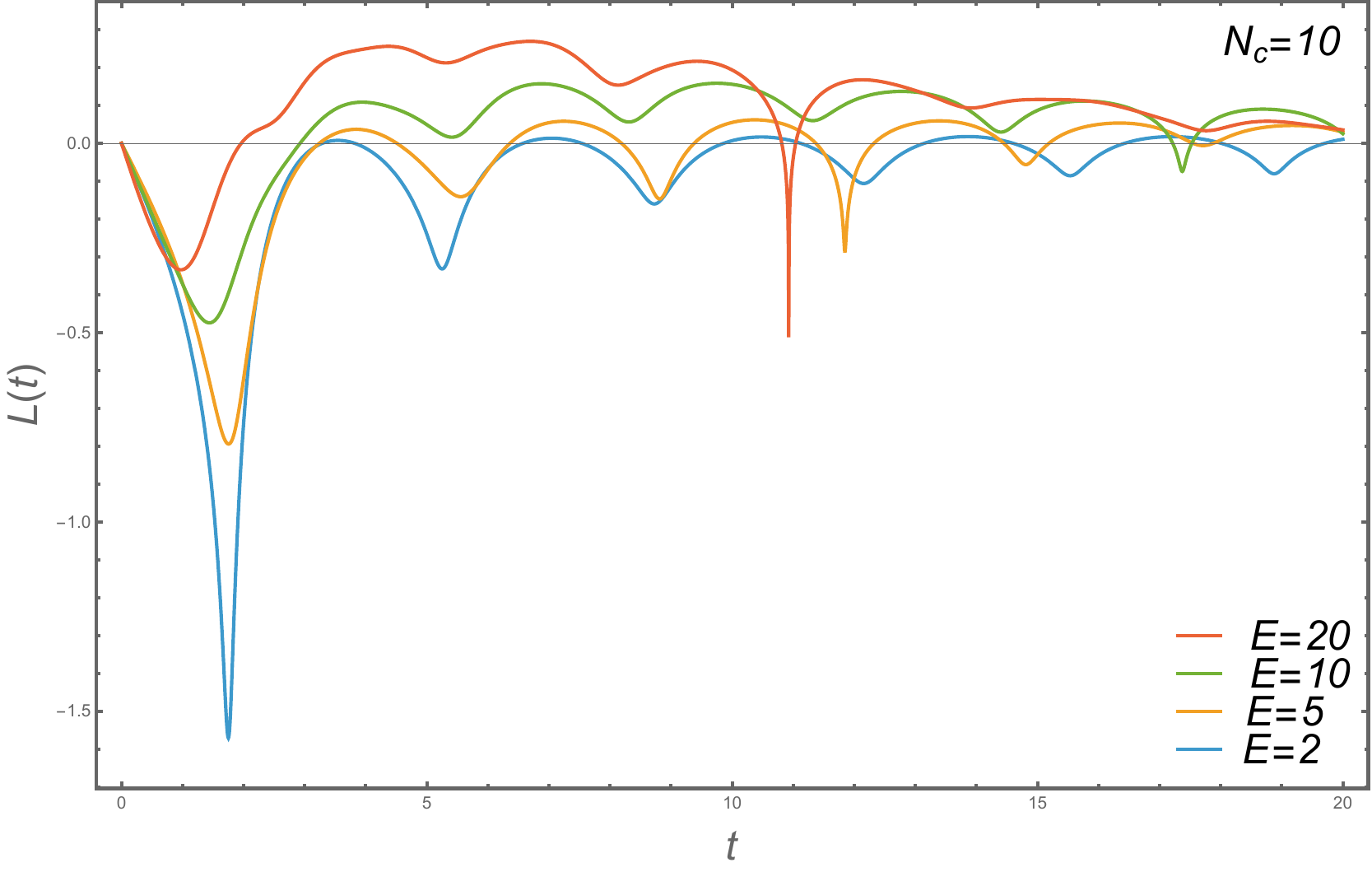}\includegraphics[scale=0.25]{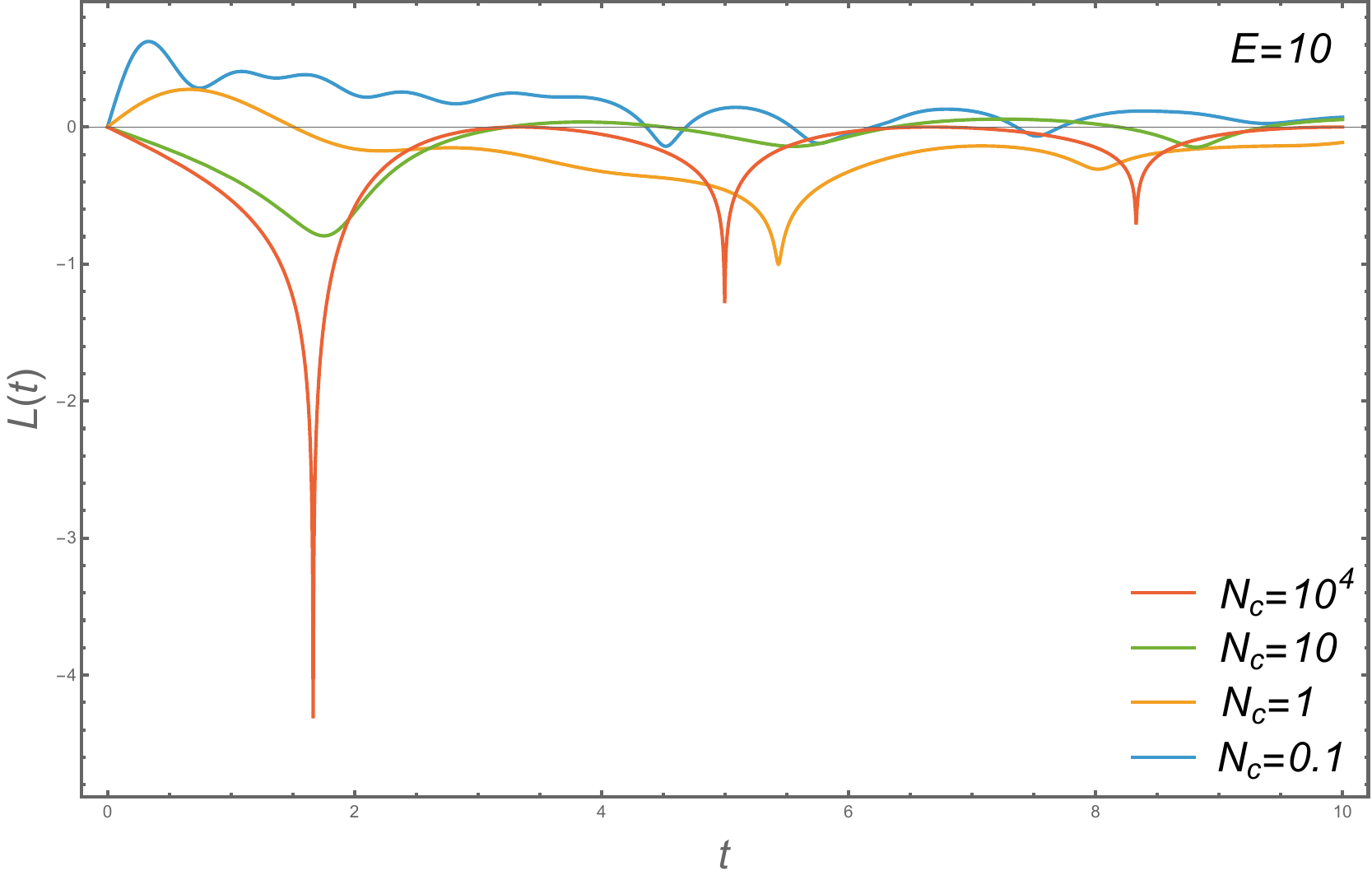}
\par\end{centering}
\begin{centering}
\includegraphics[scale=0.25]{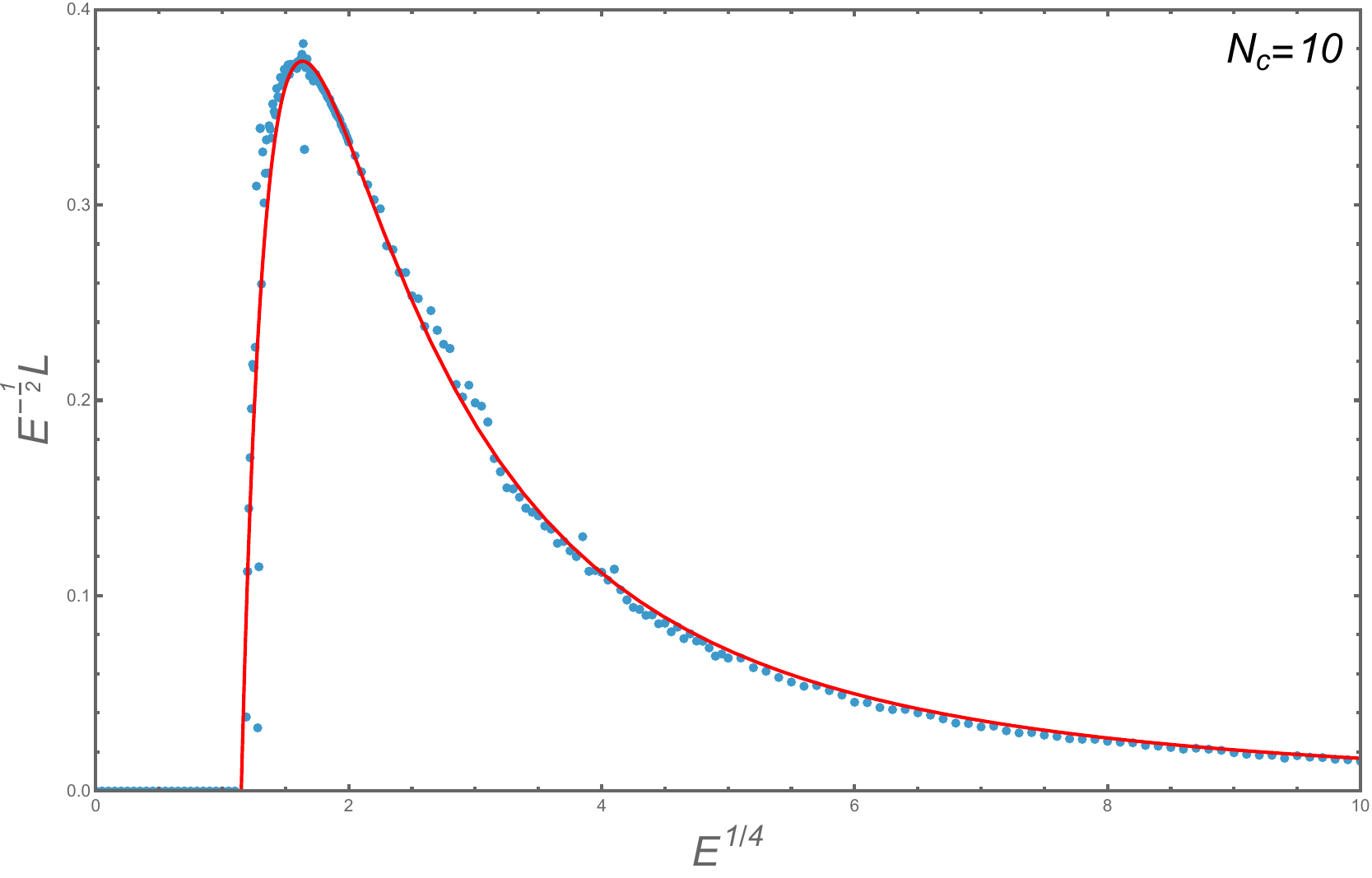}\includegraphics[scale=0.25]{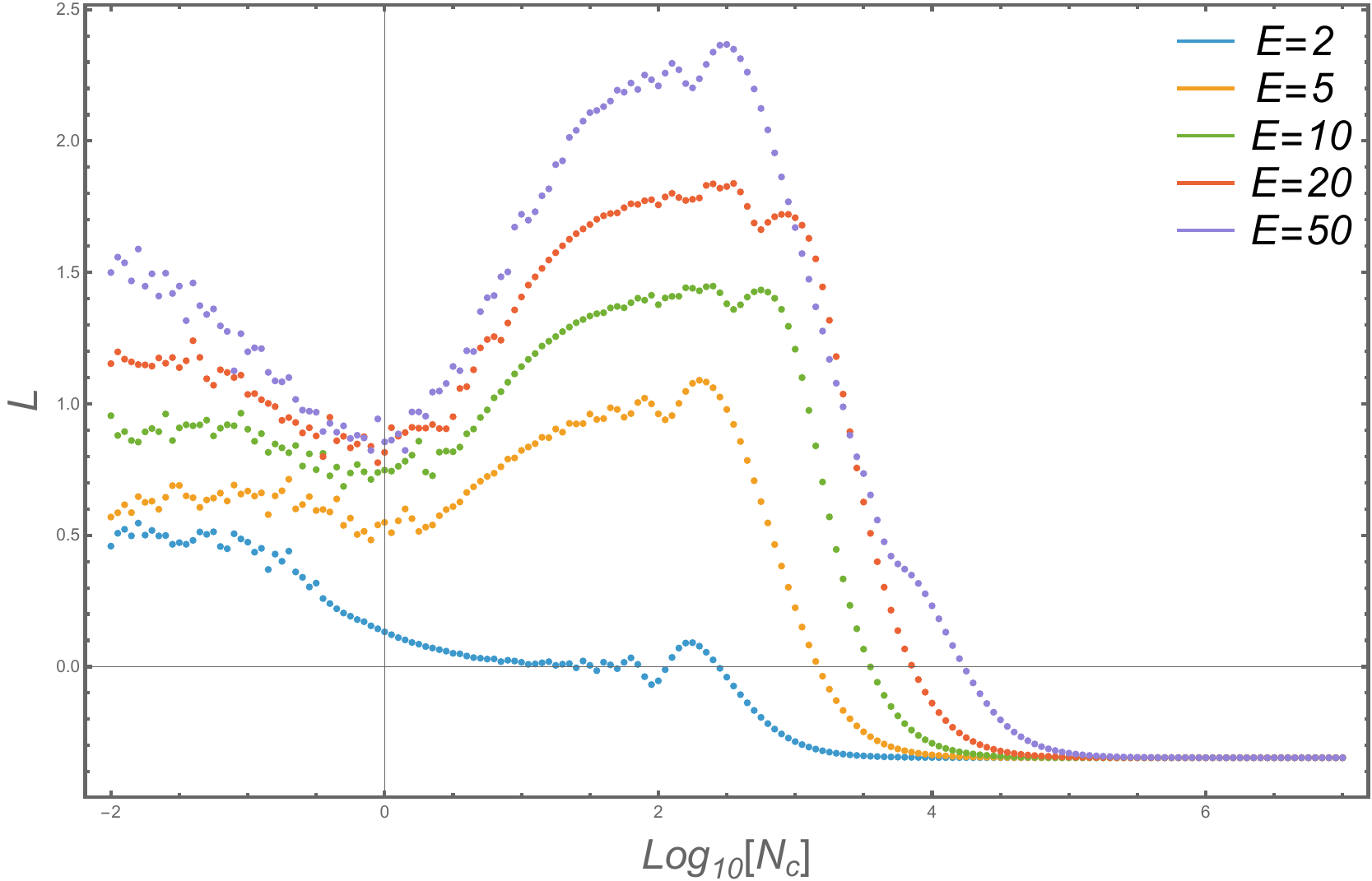}
\par\end{centering}
\caption{\label{fig:3}Lyapunov coefficient of the mesonic matrix model . \textbf{Upper:
}Lyapunov coefficient as a function of time $t$. \textbf{Lower:}
Lyapunov coefficient as a function of $E$ and $N_{c}$.}

\end{figure}

\begin{figure}
\begin{centering}
\includegraphics[scale=0.25]{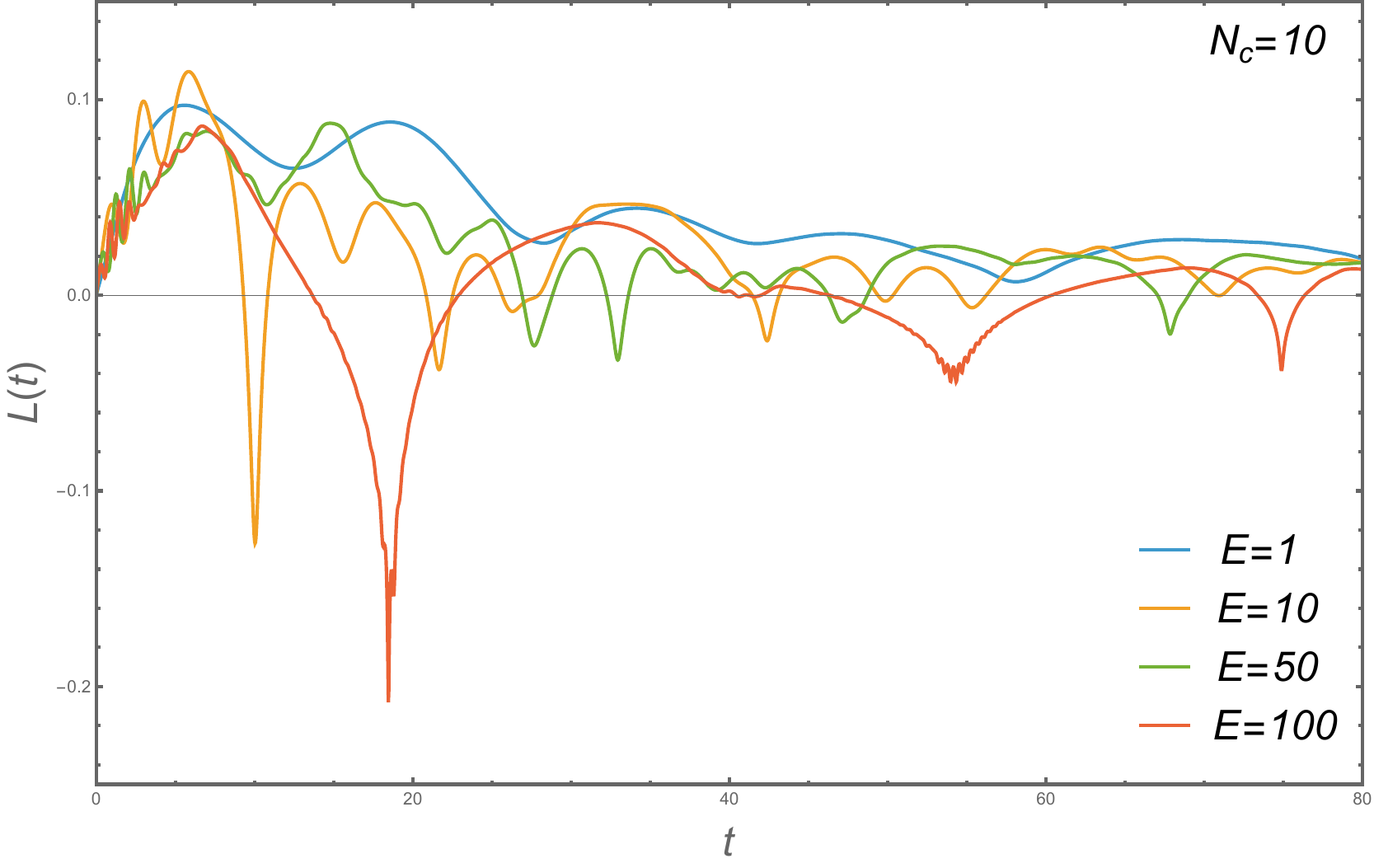}\includegraphics[scale=0.25]{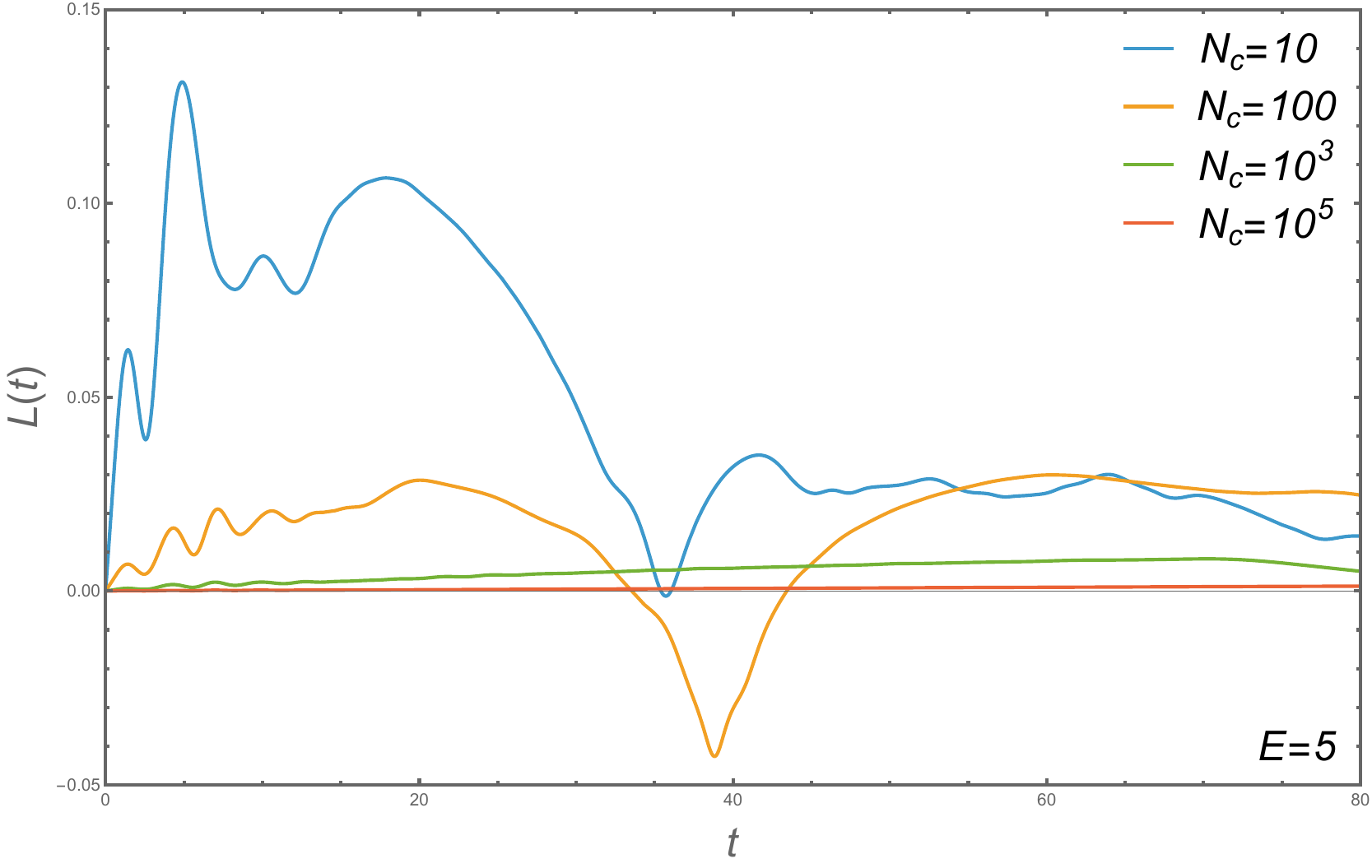}
\par\end{centering}
\begin{centering}
\includegraphics[scale=0.25]{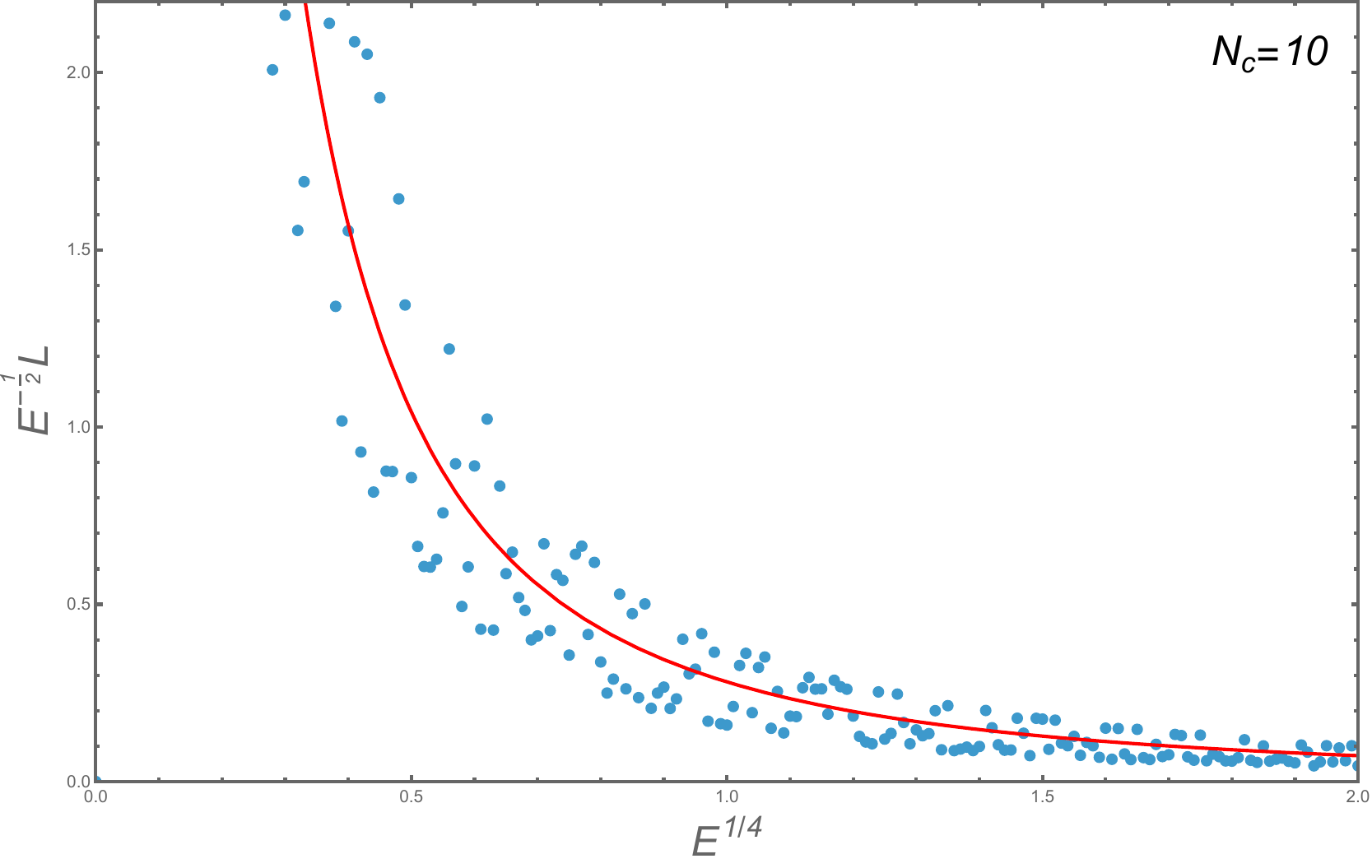}\includegraphics[scale=0.25]{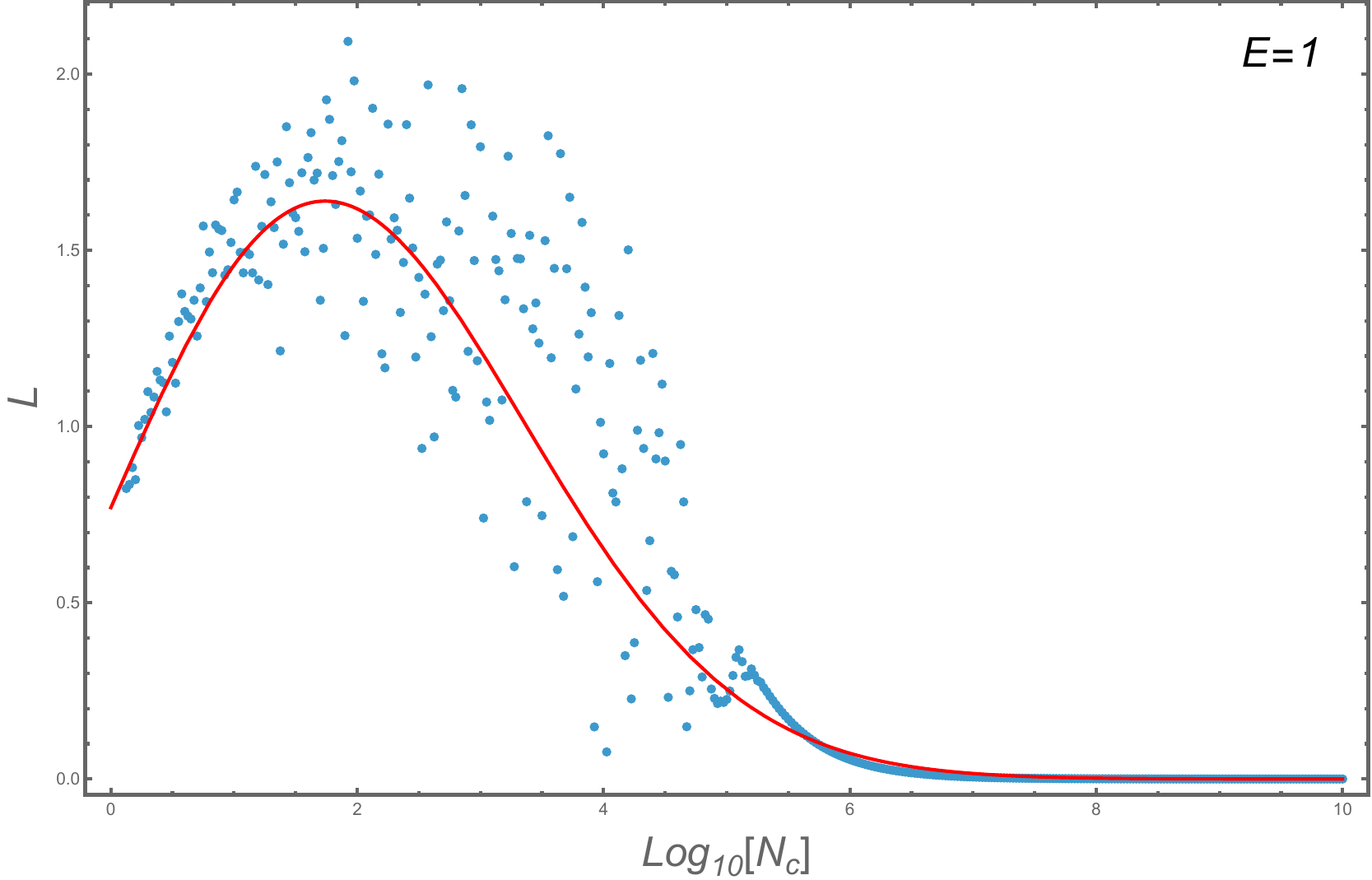}
\par\end{centering}
\caption{\label{fig:4}Lyapunov coefficient of the baryonic matrix model .
\textbf{Upper: }Lyapunov coefficient as a function of time $t$. \textbf{Lower:}
Lyapunov coefficient as a function of $E$ and $N_{c}$.}

\end{figure}

To compare with the analysis of the Poincar\'e section, we consider
another popular method to measure quantitatively the chaos of the
system which is to evaluate the classical Lyapunov exponent. In the
classical analysis, the Lyapunov coefficient $L\left(t\right)$ is
defined as \cite{key-4}

\begin{equation}
C\left(t\right)=e^{2Lt}=\left[\frac{\delta\mathbf{x}\left(t\right)}{\delta\mathbf{x}\left(0\right)}\right]^{2},\label{eq:4.1}
\end{equation}
where $\delta\mathbf{x}\left(t\right)$ refers to the distance between
two nearby orbits $\mathbf{x}\left(t\right)$. Hence (\ref{eq:4.1})
can be rewritten as,

\begin{equation}
C\left(t\right)=e^{2Lt}=\lim_{\mathbf{x}_{1}\rightarrow\mathbf{x}_{2}}\left[\frac{\mathbf{x}_{1}\left(t\right)-\mathbf{x}_{2}\left(t\right)}{\mathbf{x}_{1}\left(0\right)-\mathbf{x}_{2}\left(0\right)}\right]^{2},
\end{equation}
where $\mathbf{x}_{1,2}\left(t\right)$ are two solutions of the system
with different initial conditions $\mathbf{x}_{1,2}\left(0\right)$.
So the function $C\left(t\right)$ illustrates, for a given system,
how sensitive the dependence on the initial condition is, and Lyapunov
coefficient refers to the growth speed of the distance between two
nearby orbits. Accordingly, the time-independent Lyapunov coefficient
$L$ is usually defined by the late-time behavior of the function
$C\left(t\right)$ as \cite{key-4},

\begin{equation}
L=\lim_{t\rightarrow\infty}\frac{1}{2t}\log\left[\frac{\delta\mathbf{x}\left(t\right)}{\delta\mathbf{x}\left(0\right)}\right]^{2}=\lim_{t\rightarrow\infty}\left\{ \frac{1}{2t}\log\left[\lim_{\mathbf{x}_{1}\rightarrow\mathbf{x}_{2}}\left[\frac{\mathbf{x}_{1}\left(t\right)-\mathbf{x}_{2}\left(t\right)}{\mathbf{x}_{1}\left(0\right)-\mathbf{x}_{2}\left(0\right)}\right]^{2}\right]\right\} .\label{eq:4.3}
\end{equation}
However, using the definition (\ref{eq:4.3}) to evaluate the Lyapunov
coefficient is mostly out of reach in the actual numerical calculation.
Therefore, the definition (\ref{eq:4.3}) is usually modified slightly
in the actual calculations, e.g. consider the average value of the
function $C\left(t\right)$ with two close but different solutions
of the system as,

\begin{equation}
L=\frac{1}{2t_{max}}\log\left[\frac{1}{t_{max}}\int_{0}^{t_{max}}dt\left[\frac{\mathbf{x}_{1}\left(t\right)-\mathbf{x}_{2}\left(t\right)}{\mathbf{x}_{1}\left(0\right)-\mathbf{x}_{2}\left(0\right)}\right]^{2}\right],\label{eq:4.4}
\end{equation}
where $t_{max}\rightarrow\infty$ is a large upper bound of the time
we are taking into account. In this work, we adopt (\ref{eq:4.4})
as the simple definition for the average value of the Lyapunov coefficient
i.e. time-independent Lyapunov coefficient. Keeping these in hand,
we evaluate numerically the Lyapunov coefficient with various variables
both in the mesonic and baryonic matrix models and the associated
results are illustrated in Figure \ref{fig:3} and Figure \ref{fig:4}.

For the mesonic matrix model, we find the time-dependent Lyapunov
coefficient $L\left(t\right)$ is always negative at low energy which
means the system is in a regular phase. When the energy increases,
we find $L\left(t\right)$ becomes gradually positive which implies
a chaotic phase transition. The average value of the Lyapunov coefficient
$L$ as a function of the total energy $E$ can be fitted by a function

\begin{equation}
L\simeq\begin{cases}
\frac{2.83E^{1/2}}{\left(E^{1/4}-0.11\right)^{2.6}}\log\left(E^{1/4}-0.15\right), & E\geq E_{c}\\
0, & E<E_{c},
\end{cases}\label{eq:4.5}
\end{equation}
which also indicates a chaotic phase transition at the critical energy
$E_{c}\simeq1.2^{4}$. These behaviors of the Lyapunov coefficient
agree qualitatively with \cite{key-6,key-46,key-47} and with the
conclusion from the analysis of the Poincar\'e section. Furthermore,
we can see that $L\left(t\right)$ decreases basically when $N_{c}$
increases and its average $L$ converges to a negative value at $N_{c}\rightarrow\infty$.
Since at $N_{c}\rightarrow\infty$ the coupling constant in (\ref{eq:3.4})
goes to zero i.e. $g\rightarrow0$, the action (\ref{eq:3.3}) becomes
a 2-dimensional decoupled harmonic oscillator which is integrable.
As the average of Lyapunov coefficient for harmonic oscillator is
definitely negative according to the Appendix C, it signals the negative
convergence of $L$ in the large $N_{c}$ limit.

For the baryonic matrix model, we can see the time-dependent Lyapunov
coefficient $L\left(t\right)$ is dominantly positive which means
the system does not have a definitely regular phase. And its average
$L$ as a function of the total energy $E$ can be fitted by the following
function

\begin{equation}
L\simeq\frac{1.9E^{1/2}}{\left(2.5E^{1/4}+0.1\right)^{2}},\label{eq:4.6}
\end{equation}
which indicates $L$ grows rapidly by the increasing of the energy
$E$. This behavior of the Lyapunov coefficient is also consistent
with the analysis of the Poincar\'e section since the system is alway
in a chaotic phase for various energies. In addition, the numerical
results display that the time-dependent Lyapunov coefficient $L\left(t\right)$
is suppressed by the increasing of $N_{c}$ and its average $L$ converges
to zero at $N_{c}\rightarrow\infty$. Since the action (\ref{eq:3.8})
at $N_{c}\rightarrow\infty,g\rightarrow0$ returns to be integrable
i.e. it becomes to a harmonic oscillator of $w\left(t\right)$ with
free moving $x\left(t\right),y\left(t\right)$, the average $L$ of
the Lyapunov coefficient can be evaluated as,

\begin{align}
L & =\lim_{t\rightarrow\infty}\frac{1}{2t}\log\left\{ \frac{1}{3}\left(\left[\frac{\delta w\left(t\right)}{\delta w\left(0\right)}\right]^{2}+\left[\frac{\delta x\left(t\right)}{\delta x\left(0\right)}\right]^{2}+\left[\frac{\delta y\left(t\right)}{\delta y\left(0\right)}\right]^{2}\right)\right\} \nonumber \\
 & \simeq\lim_{t\rightarrow\infty}\frac{1}{2t}\log\left[\frac{1}{3}\left(\cos\frac{2\sqrt{3}}{3}t\right)^{2}+\frac{2}{3}\right]\rightarrow0,
\end{align}
signaling its large $N_{c}$ behavior.

\subsection{The large $N_{c}$ analytics}

In order to investigate quantitatively the dependence on $N_{c}$
at $N_{c}\rightarrow\infty$ (i.e. $g\rightarrow0$), let us solve
analytically the mesonic action (\ref{eq:3.3}) and baryonic action
(\ref{eq:3.8}) by using the method of perturbation. 

\subsubsection*{The mesonic matrix model}

For the mesonic action, we impose the following ansatz as a series
of coupling constant $g$,
\begin{equation}
x\left(t\right)=x_{0}\left(t\right)+gx_{1}\left(t\right)+\mathcal{O}\left(g^{2}\right),y\left(t\right)=y_{0}\left(t\right)+gy_{1}\left(t\right)+\mathcal{O}\left(g^{2}\right),
\end{equation}
to the equations of motion for $x\left(t\right),y\left(t\right)$
associated to the action (\ref{eq:3.3}). Here $x_{0}\left(t\right),y_{0}\left(t\right)$
are zero-th order solution which are nothing but the solution for
classical harmonic oscillator as it is given in (\ref{eq:C-2}). And
$x_{1}\left(t\right),y_{1}\left(t\right)$ are first order solutions
satisfying

\begin{equation}
\ddot{x}_{1}+m^{2}x_{1}+2x_{0}y_{0}=0,\ddot{y}_{1}+m^{2}y_{1}+2x_{0}y_{0}=0,
\end{equation}
which can be solved analytically. Therefore the leading order solution
is totally integrable. By fixing partly the initial condition $x_{1}\left(0\right)=y_{1}\left(0\right)=0,x\left(0\right)^{2}+y\left(0\right)^{2}=1$,
we can obtain a simply formula of $C\left(t\right)$ as

\begin{align}
C\left(t\right) & =\frac{1}{2}\left\{ \left[\frac{\delta x\left(t\right)}{\delta x\left(0\right)}\right]^{2}+\left[\frac{\delta y\left(t\right)}{\delta y\left(0\right)}\right]^{2}\right\} \nonumber \\
 & \simeq\cos^{2}t-\frac{g}{16}\sin2t\left(6t+\sin2t\right)+\mathcal{O}\left(g^{2}\right),
\end{align}
hence its average $\overline{C\left(t\right)}$ at large $t$ is evaluated
as,

\begin{equation}
\lim_{t\rightarrow\infty}\overline{C\left(t\right)}\simeq0.5-0.01N_{c}^{-1},
\end{equation}
leading to a negative Lyapunov coefficient $L\simeq-\log2$ at large
$N_{c}$ as it is expected for an integrable system. So this analytical
discussion specify the large $N_{c}$ behavior of the Lyapunov coefficient
in the mesonic matrix model presented in Figure \ref{fig:3}.

\subsubsection*{The baryonic matrix model}

For the baryonic matrix model, we impose the same ansatz as,

\begin{equation}
x\left(t\right)=x_{0}\left(t\right)+gx_{1}\left(t\right)+\mathcal{O}\left(g^{2}\right),y\left(t\right)=y_{0}\left(t\right)+gy_{1}\left(t\right)+\mathcal{O}\left(g^{2}\right),w\left(t\right)=w_{0}\left(t\right)+gw_{1}\left(t\right)+\mathcal{O}\left(g^{2}\right),
\end{equation}
to the equations of motion associated to action (\ref{eq:3.8}). The
zero-th order part $x_{0},y_{0},w_{0}$ is very simply to solve, resultantly
$x_{0},y_{0}$ are linear functions of $t$ and $w_{0}$ is the solution
for classical harmonic oscillator. Hence the first order functions
$x_{1},y_{1},w_{1}$ satisfy the equations of motion as,

\begin{align}
\ddot{x}_{1}+2x_{0}\left(w_{0}^{2}+y_{0}^{2}\right) & =0,\nonumber \\
\ddot{y}_{1}+2y_{0}\left(x_{0}^{2}+w_{0}^{2}\right) & =0,\\
\ddot{w}_{1}+m^{2}w_{1}+2w_{0}\left(x_{0}^{2}+y_{0}^{2}\right) & =0,\nonumber 
\end{align}
which can be solved analytically. By fixing partly the initial conditions
as $x_{1}\left(0\right)=y_{1}\left(0\right)=0,x\left(0\right)^{2}+y\left(0\right)^{2}=w\left(0\right)^{2}=1$,
we can obtain the function $C\left(t\right)$ as,

\begin{align}
C\left(t\right) & =\frac{1}{3}\left\{ \left[\frac{\delta w\left(t\right)}{\delta w\left(0\right)}\right]^{2}+\left[\frac{\delta x\left(t\right)}{\delta x\left(0\right)}\right]^{2}+\left[\frac{\delta y\left(t\right)}{\delta y\left(0\right)}\right]^{2}\right\} \nonumber \\
 & \simeq\frac{1}{6}\left[5+\cos\left(\frac{4t}{\sqrt{3}}\right)\right]-\frac{1}{12}g\left[3+16t^{2}-3\cos\left(\frac{4t}{\sqrt{3}}\right)+2\sqrt{3}\sin\left(\frac{4t}{\sqrt{3}}\right)\right]+\mathcal{O}\left(g^{2}\right),
\end{align}
so the average value of the function $C\left(t\right)$ at large $t$
is

\begin{equation}
\lim_{t\rightarrow t_{\infty}}\overline{C\left(t\right)}\simeq\frac{5}{6}-\frac{1}{81\pi N_{c}}\left(3+\frac{16}{3}t_{\infty}^{2}\right),
\end{equation}
leading to negative Lyapunov coefficient $L=\lim_{t\rightarrow\infty}\frac{1}{2t}\log C\left(t\right)\rightarrow0$
at large $N_{c}$. Accordingly, the leading order solution is integrable
and it specifies the large $N_{c}$ behavior of Lyapunov coefficient
in the baryonic matrix model presented in Figure \ref{fig:4}.

\section{Analysis of the quantum chaos}

In this section, we will focus on the analysis of the quantum chaos
in the matrix models with respect to the quantum version of the action
(\ref{eq:3.3}) and action (\ref{eq:3.8}).To measure quantitatively
the chaos of a quantum system, we follow the definition of OTOC in
\cite{key-5} and collect the formulas to calculate the OTOC. The
quantum operators in this section are in the Heisenberg picture. However
in general systems, the Heisenberg equation may not be analytically
solved, so we have to rely on a numerical evaluation.

\subsection{Formulas for the quantum OTOC}

Let us take into account a $d$ dimensional quantum mechanical system
described by a time-independent Hamiltonian as $H\left(x_{1},x_{2}...x_{d},p_{1},p_{2}...p_{d}\right)$.
In the Heisenberg picture, we denote $x_{1}\left(t\right)\equiv x\left(t\right),p_{1}=p\left(t\right)$
and $x\equiv x\left(0\right),p=p\left(0\right)$. Thus the operator
at time $t$ and at $t=0$ is connected by a unitary transformation
$x\left(t\right)=e^{iHt}xe^{-iHt}$. Keeping these in hand, let us
recall the definition of classical function $C\left(t\right)$ in
(\ref{eq:4.1}) and rewrite it by using the Poisson bracket as,
\begin{equation}
C\left(t\right)=\left[\frac{\delta x\left(t\right)}{\delta x}\right]^{2}=\left\{ x\left(t\right),p\right\} _{\mathrm{P.B.}}^{2}\label{eq:5.1}
\end{equation}
Thus the quantum version of (\ref{eq:5.1}) is conjectured by replacing
the Poisson bracket with the quantum commutator, i.e. $\left\{ ,\right\} _{\mathrm{P.B.}}\rightarrow\frac{1}{i}\left[,\right]$.
Therefore, we can define the microcanonical OTOC $c_{n}\left(t\right)$
as \cite{key-5},

\begin{equation}
c_{n}\left(t\right)=-\left\langle n\left|\left[x\left(t\right),p\right]^{2}\right|n\right\rangle ,\label{eq:5.2}
\end{equation}
where $\left|n\right\rangle ,E_{n}$ refers respectively to the eigenstates
and eigenvalues of the Hamiltonian satisfying $H\left|n\right\rangle =E_{n}\left|n\right\rangle $.
Furthermore, to take into account the quantum statistics, it is significant
to define the canonical thermal OTOC $C_{T}\left(t\right)$ as,

\begin{equation}
C_{T}\left(t\right)=-\left\langle \left[x\left(t\right),p\right]^{2}\right\rangle _{T}.
\end{equation}
Here the thermal average $\left\langle ...\right\rangle _{T}$ is
defined as,

\begin{equation}
\left\langle O\right\rangle _{T}=\frac{1}{\mathcal{Z}\left(T\right)}\mathrm{Tr}\left[e^{-\beta H}O\right],\mathcal{Z}\left(T\right)=\mathrm{Tr}\left[e^{-\beta H}\right],\beta=1/T,
\end{equation}
and $T$ refers to the temperature. Therefore, we have

\begin{align}
C_{T}\left(t\right) & =\frac{1}{\mathcal{Z}\left(T\right)}\sum_{n}c_{n\left(t\right)}e^{-\frac{E_{n}}{T}}=e^{2L\left(T\right)t},\nonumber \\
\mathcal{Z}\left(T\right) & =\sum_{n}e^{-\frac{E_{n}}{T}},\label{eq:5.5}
\end{align}
and $L\left(T\right)$ depends on the temperature named as the quantum
Lyapunov coefficient. Note that the quantum Lyapunov exponent is an
analogue of the classical Lyapunov exponent.

\subsection{The numerical method}

Here we outline how to calculate the OTOCs numerically by using the
above formulas, since, in general, the Heisenberg equation can not
be solved analytically. We first rewrite (\ref{eq:5.2}) by using
the completeness condition $\sum_{m}\left|m\right\rangle \left\langle m\right|=1$
as,

\begin{equation}
c_{n}\left(t\right)=-\sum_{m}\left\langle n\left|\left[x\left(t\right),p\right]\left|m\right\rangle \left\langle m\right|\left[x\left(t\right),p\right]\right|n\right\rangle \equiv\sum_{m}b_{nm}b_{nm}^{*},\label{eq:5.6}
\end{equation}
where 

\begin{equation}
b_{nm}\left(t\right)=-i\left\langle n\left|\left[x\left(t\right),p\right]\right|m\right\rangle .\label{eq:5.7}
\end{equation}
Then impose $x\left(t\right)=e^{iHt}xe^{-iHt}$ to (\ref{eq:5.7}),
we can obtain

\begin{equation}
b_{nm}\left(t\right)=-i\sum_{k}\left(e^{iE_{nk}t}x_{nk}p_{km}-e^{iE_{km}t}p_{nk}x_{km}\right),\label{eq:5.8}
\end{equation}
where $x_{nm},p_{nm}$ refer to the matrix elements as $\left\langle n\left|x\right|m\right\rangle =x_{nm},p_{nm}=\left\langle n\left|x\right|m\right\rangle $
and $E_{nm}=E_{n}-E_{m}$. Since the Hamiltonian in our concern takes
the form as,

\begin{equation}
H=\sum_{i=1}^{d}\frac{p_{i}^{2}}{2}+U\left(x_{1},x_{2}...x_{d}\right),
\end{equation}
we can further obtain $\frac{1}{i}\left[x,H\right]=p$, so it leads
to

\begin{equation}
p_{nm}=-i\left\langle n\left|xH-Hx\right|m\right\rangle =iE_{nm}x_{nm}.
\end{equation}
Hence (\ref{eq:5.8}) can be rewritten as,

\begin{equation}
b_{nm}\left(t\right)=\sum_{k}x_{nk}x_{km}\left(e^{iE_{nk}t}E_{km}-e^{iE_{km}t}E_{nk}\right).\label{eq:5.11}
\end{equation}
Therefore, we can see only the eigenstates and eigenvalues of the
Hamiltonian are the inputs of the numerical calculation. 

\begin{figure}[th]
\begin{centering}
\includegraphics[scale=0.25]{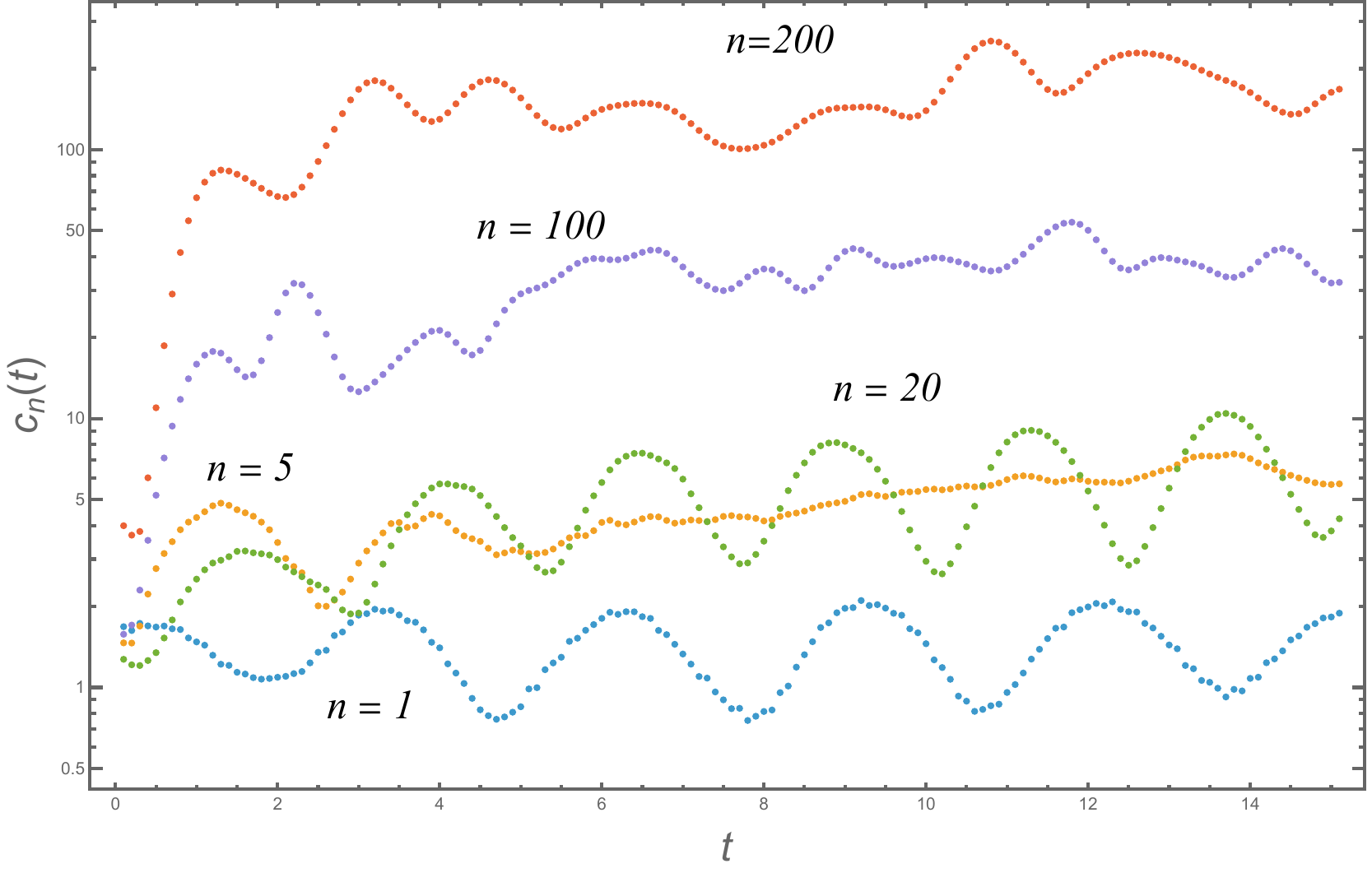}\includegraphics[scale=0.25]{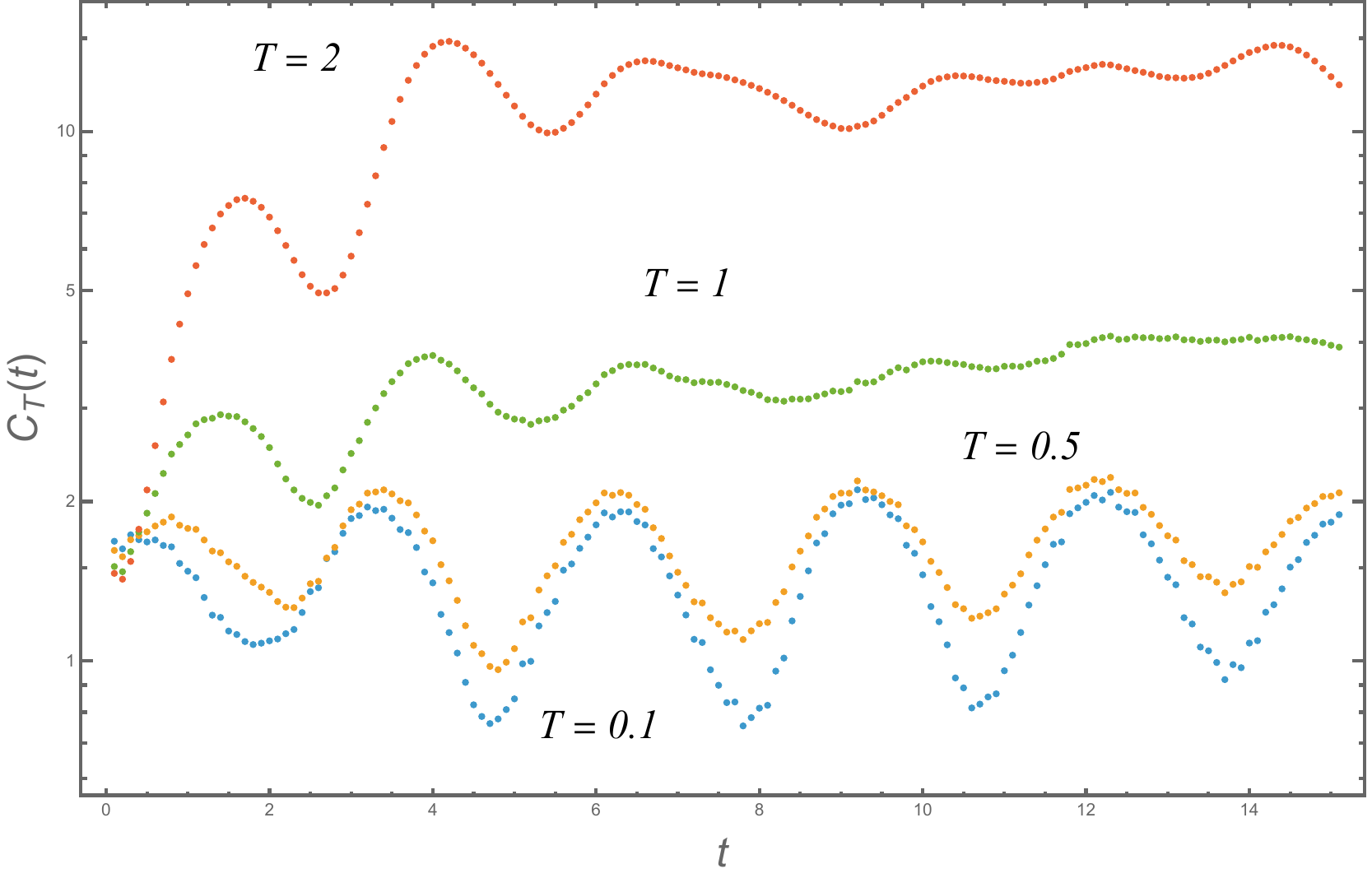}
\par\end{centering}
\begin{centering}
\includegraphics[scale=0.25]{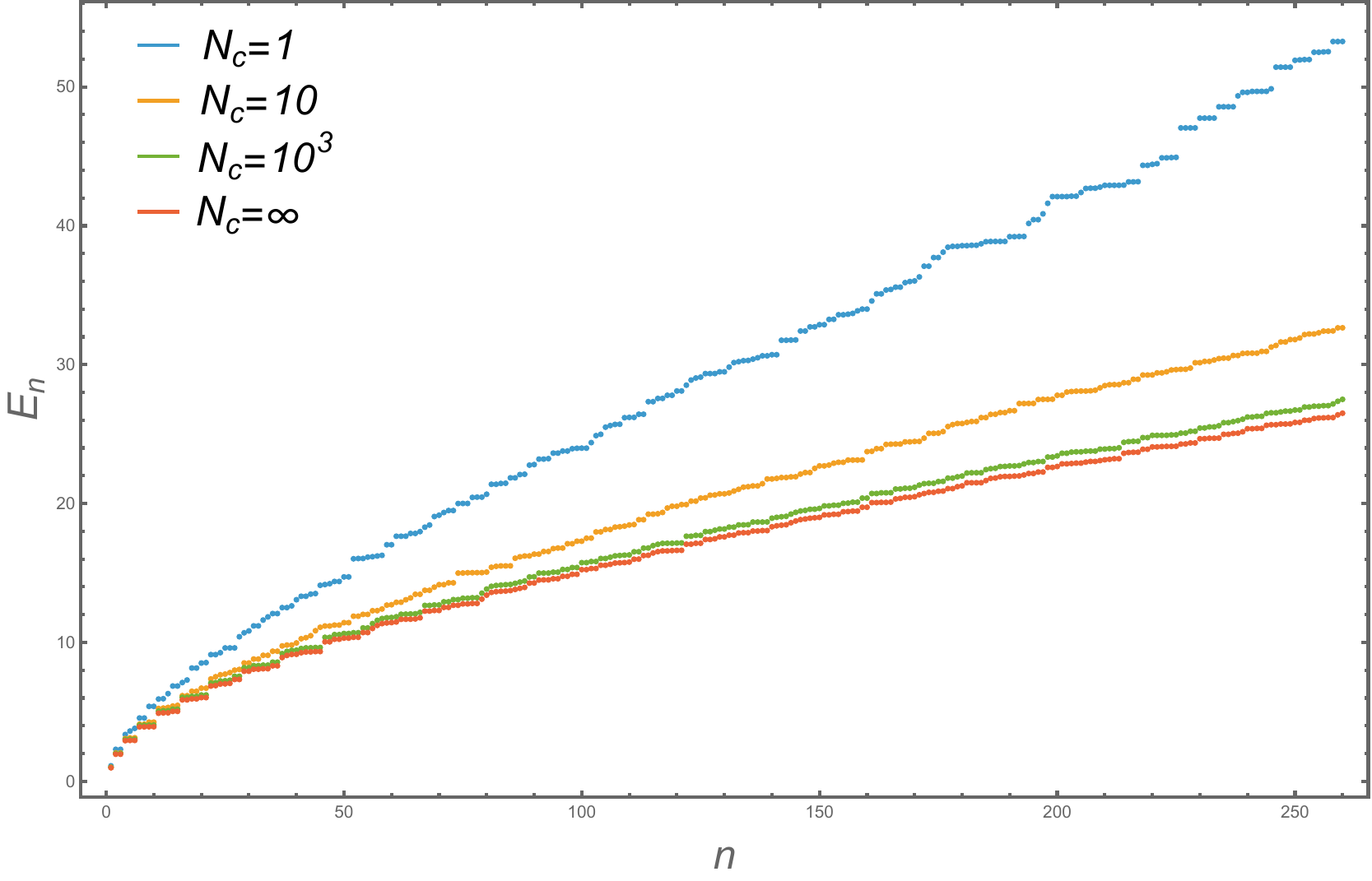}\includegraphics[scale=0.25]{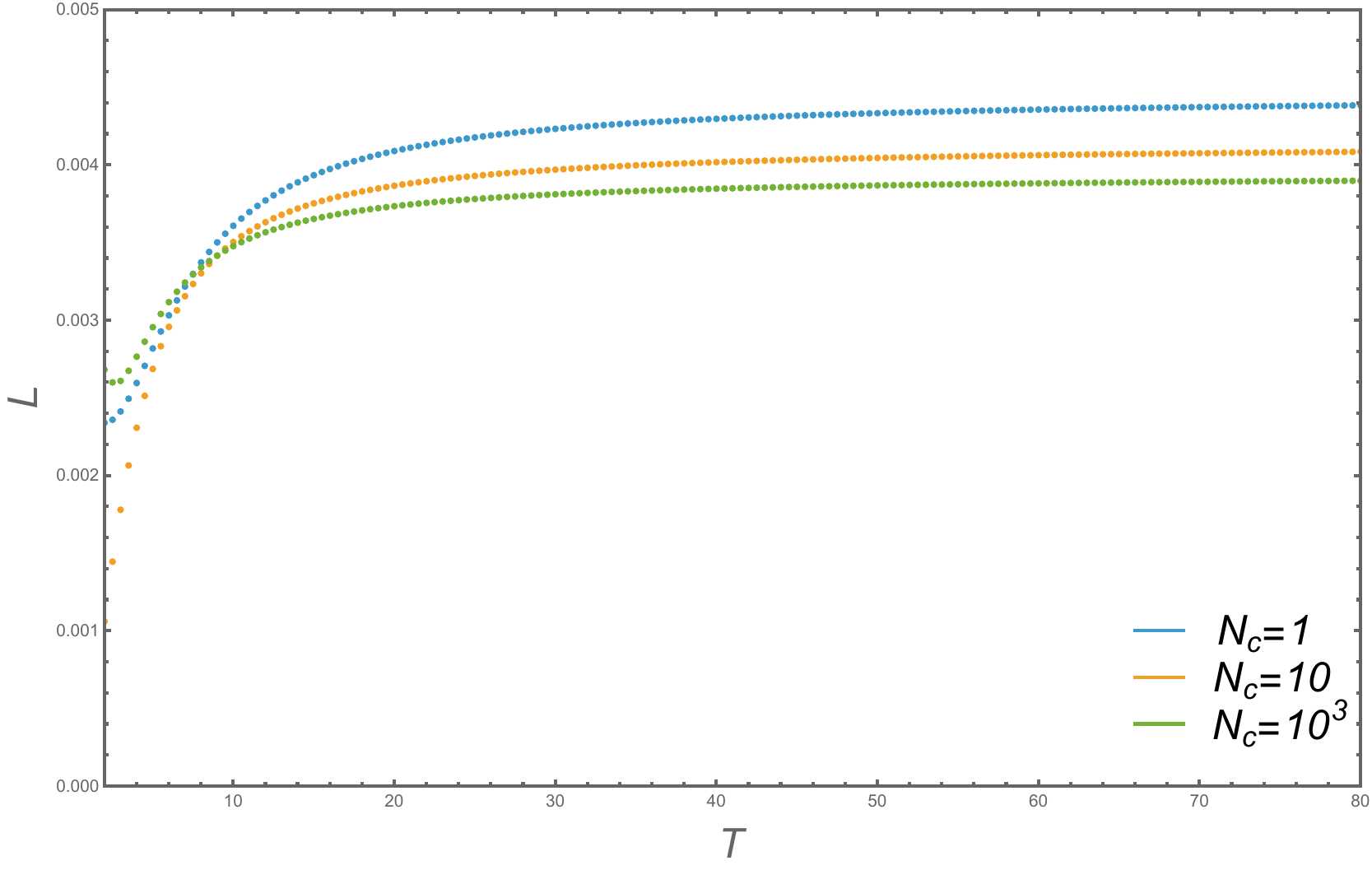}
\par\end{centering}
\caption{\label{fig:5}The quantum properties of the mesonic matrix model including
the microcanonical and thermal OTOCs, average of the quantum Lyapunov
coefficient and eigen energies.\textbf{ Upper:} The microcanonical
OTOC $c_{n}\left(t\right)$ and thermal OTOC $C_{T}\left(t\right)$
as functions of time $t$ at various temperatures with $N_{c}=10$.
\textbf{Lower:} Eigen energies $E_{n}$ as a function of quantum number
$n$ and the average of quantum Lyapunov coefficient $L$ as a function
of temperature $T$ with various $N_{c}$.}

\end{figure}

\begin{figure}
\begin{centering}
\includegraphics[scale=0.25]{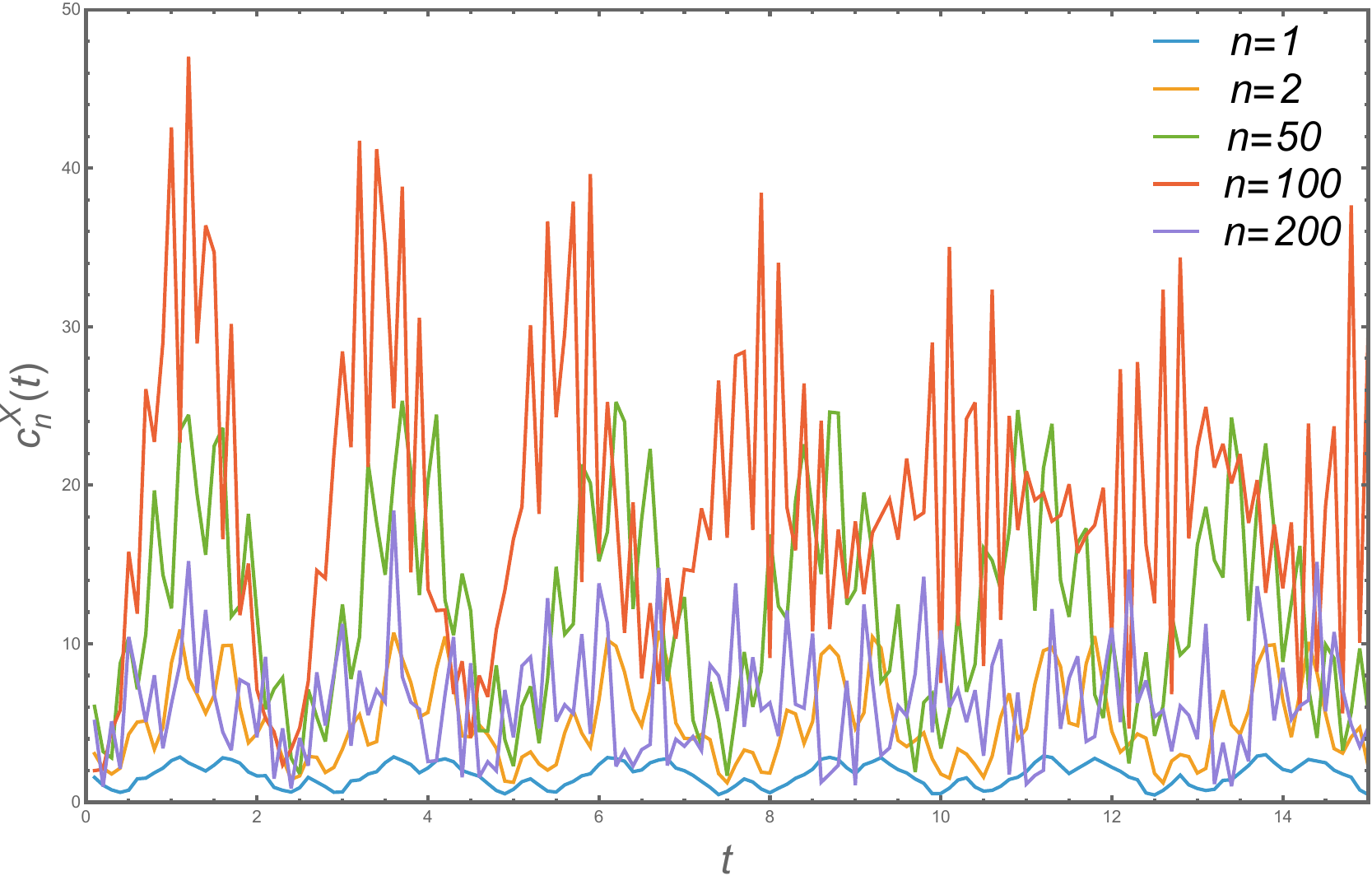}\includegraphics[scale=0.25]{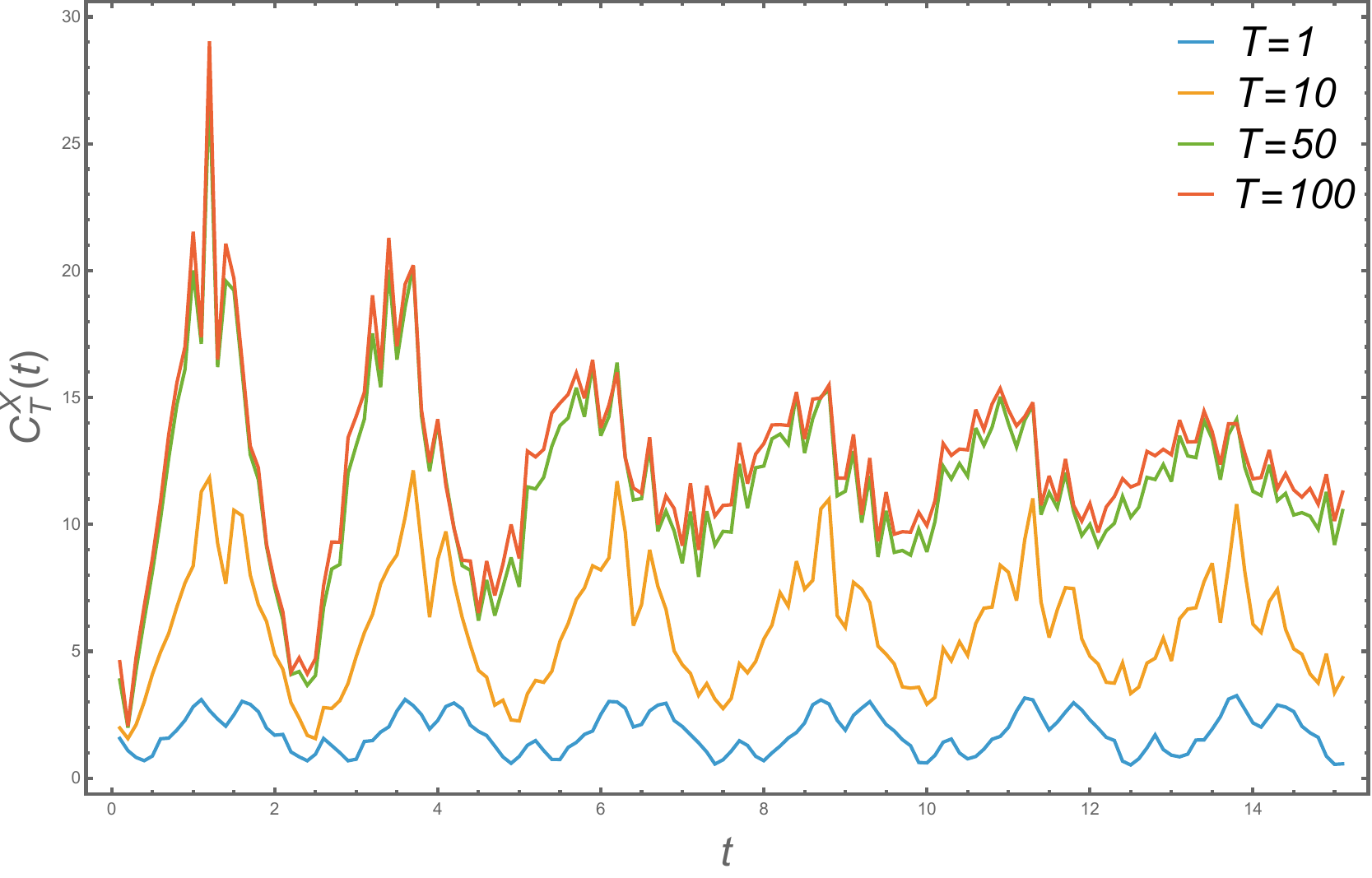}
\par\end{centering}
\begin{centering}
\includegraphics[scale=0.25]{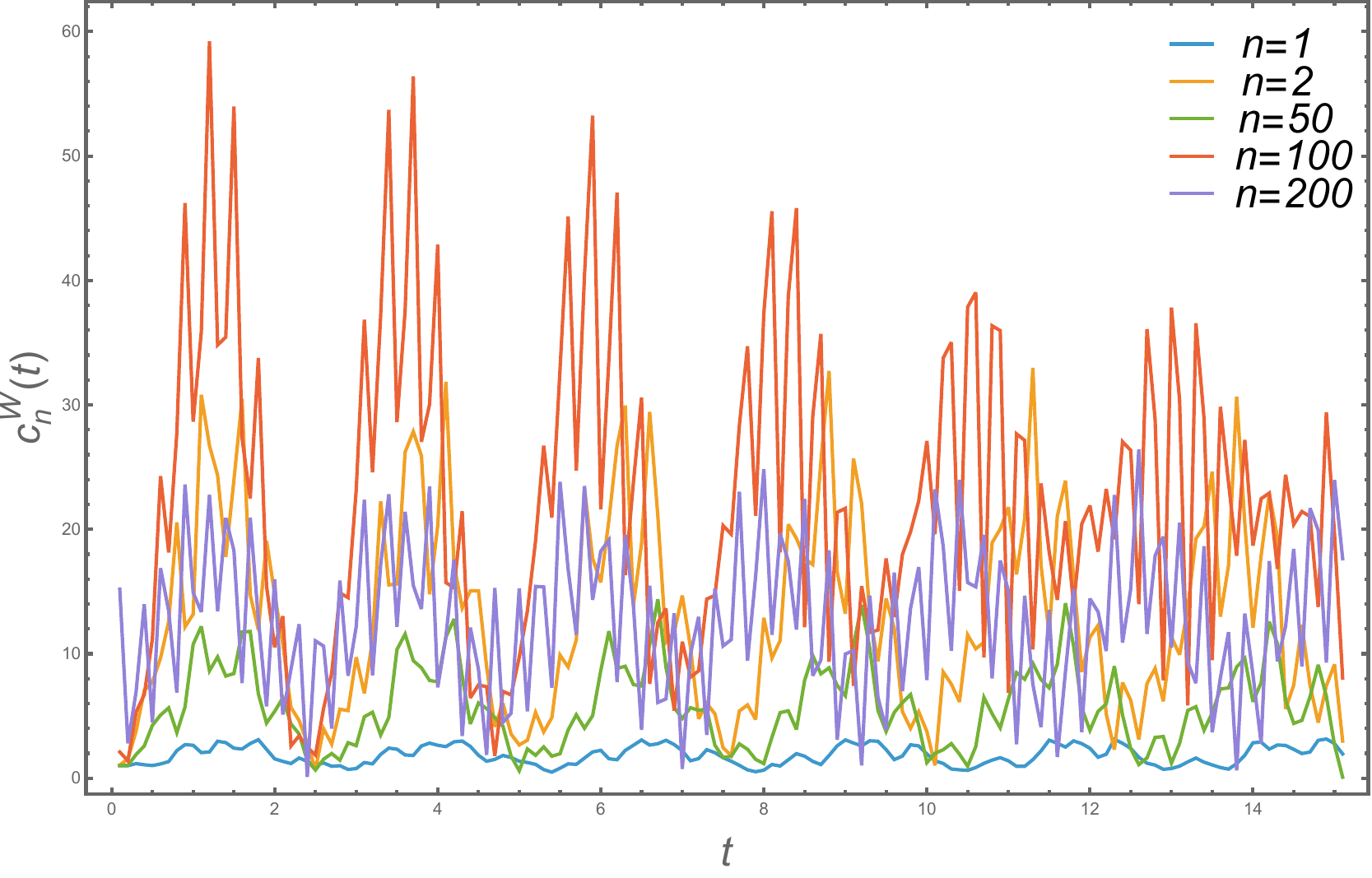}\includegraphics[scale=0.25]{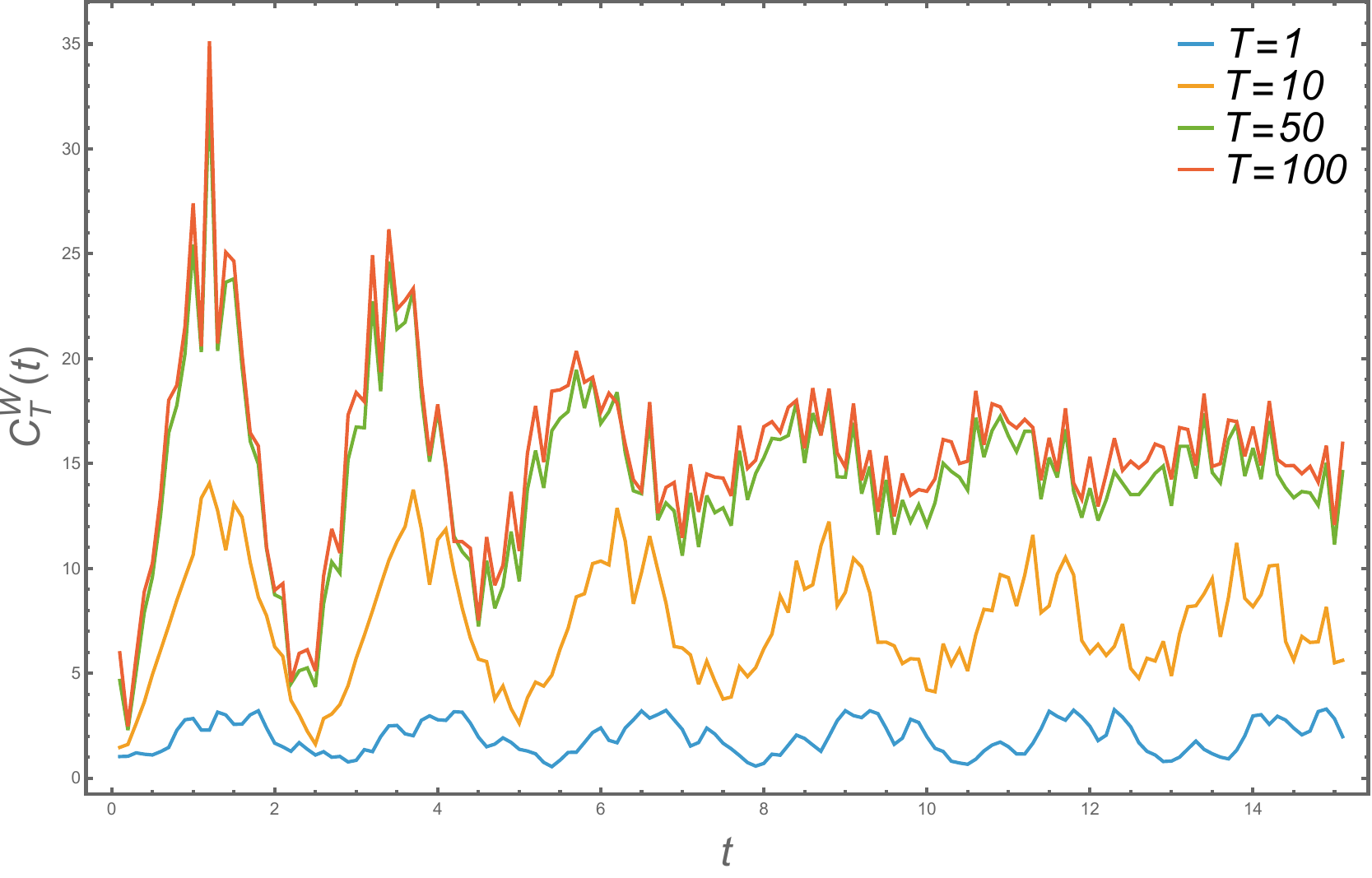}
\par\end{centering}
\caption{\label{fig:6}The microcanonical and thermal OTOCs of the baryonic
matrix model at $N_{c}=10$. The index $X,W$ refers respectively
to the OTOCs form the commutator of $\left[x\left(t\right),p_{x}\right]^{2}$
and $\left[w\left(t\right),p_{w}\right]^{2}$.\textbf{ Upper:} The
microcanonical OTOC $c_{n}^{X}\left(t\right)$ and thermal OTOC $C_{T}^{X}\left(t\right)$
as functions of time $t$ with various temperatures. \textbf{Lower:}
The microcanonical OTOC $c_{n}^{W}\left(t\right)$ and thermal OTOC
$C_{T}^{W}\left(t\right)$ functions of time $t$ with various temperatures.}

\end{figure}

\begin{figure}
\begin{centering}
\includegraphics[scale=0.25]{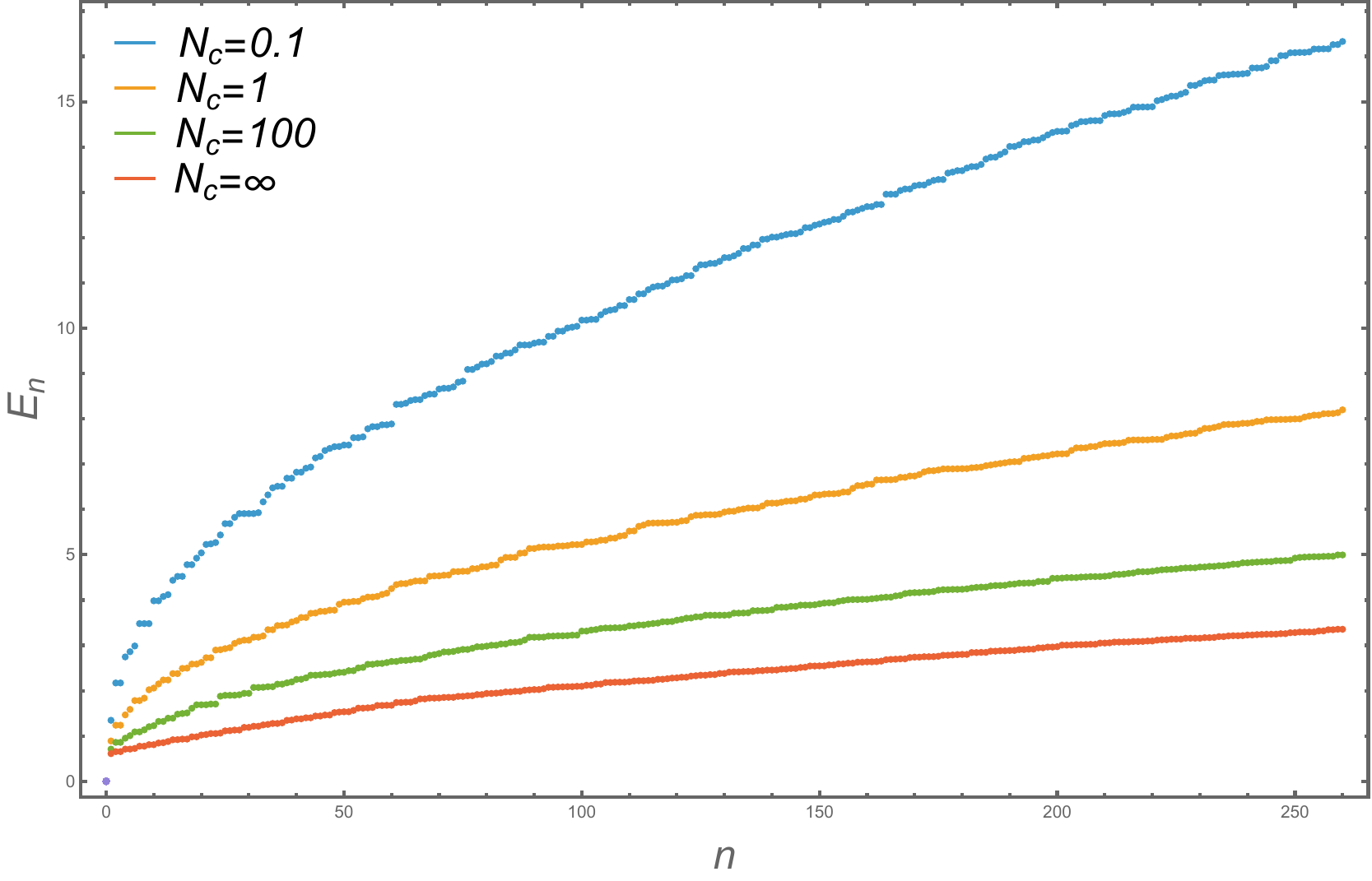}\includegraphics[scale=0.26]{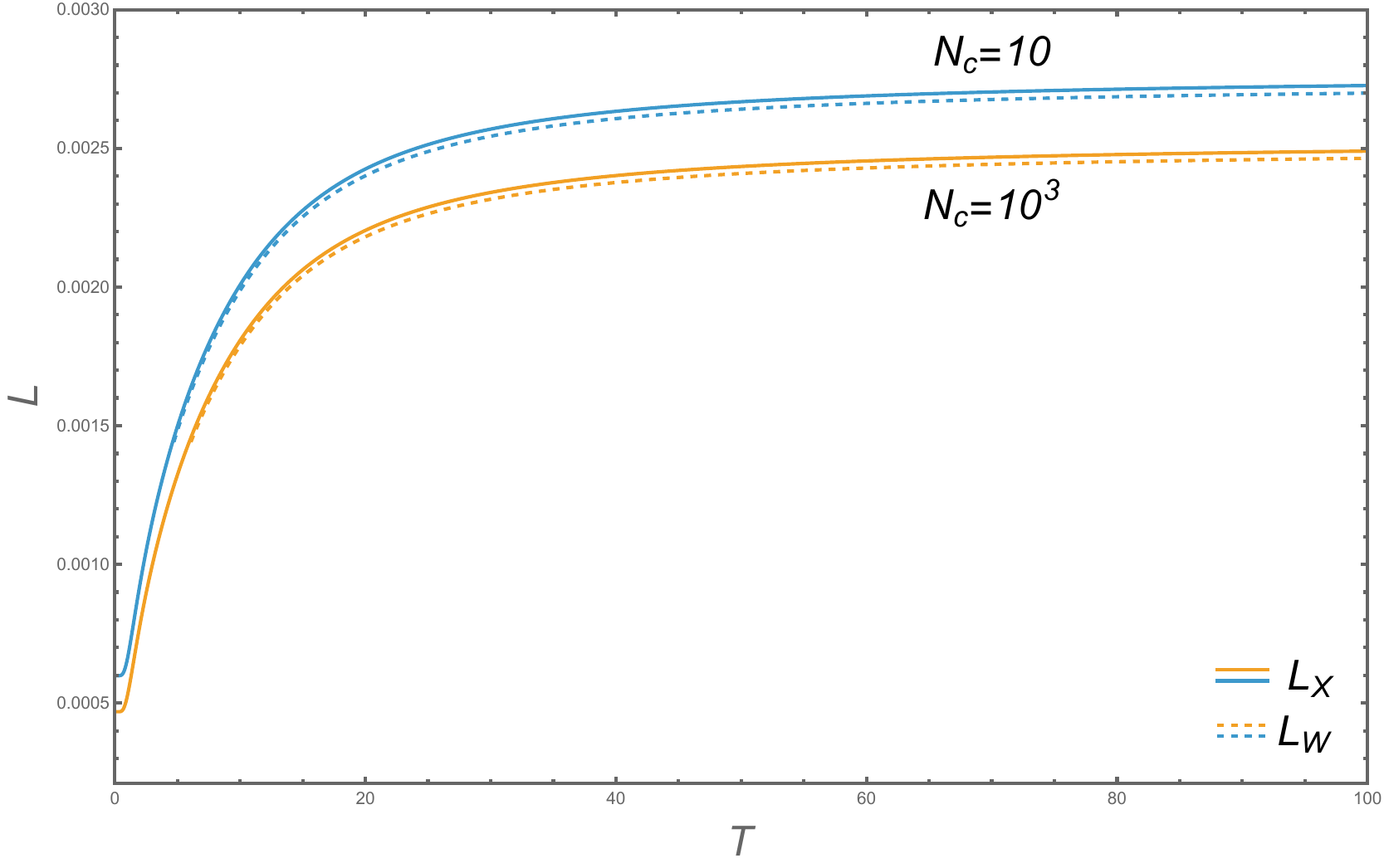}
\par\end{centering}
\caption{\label{fig:7}Energy spectrum $E_{n}$ and quantum Lyapunov coefficient
$L$ of the baryonic matrix model with various $N_{c}$. The index
$X,W$ refers respectively to the Lyapunov coefficient from the OTOC
$C_{T}^{X}\left(t\right)$ and $C_{T}^{W}\left(t\right)$. \textbf{Left:}
Energy spectrum $E_{n}$ as a function of the quantum number $n$.
\textbf{Right:} Lyapunov coefficient $L$ as a function of temperature
$T$.}

\end{figure}

As the actions (\ref{eq:3.3}) and (\ref{eq:3.8}) are our concerns
to investigate the chaos in the classical analysis, the quantum version
of them can be obtained by replacing momentum in Hamiltonians with
the associated derivative operator. Then it would be possible to investigate
the quantum OTOCs by using the above numerical method. Specifically,
for the mesonic matrix model (\ref{eq:3.3}), the eigen equation of
Hamiltonian is,

\begin{equation}
\left[-\frac{1}{2}\left(\frac{\partial^{2}}{\partial x^{2}}+\frac{\partial^{2}}{\partial y^{2}}\right)+\frac{1}{2}m^{2}\left(x^{2}+y^{2}\right)+gx^{2}y^{2}\right]\psi_{n}=E_{n}\psi_{n},\label{eq:5.12}
\end{equation}
and for the baryonic matrix model (\ref{eq:3.8}), the eigen equation
of Hamiltonian is,

\begin{equation}
\left[\frac{1}{2}\left(\frac{\partial^{2}}{\partial x^{2}}+\frac{\partial^{2}}{\partial y^{2}}+\frac{\partial^{2}}{\partial w^{2}}\right)+\frac{1}{2}m^{2}w^{2}+g\left(x^{2}y^{2}+x^{2}w^{2}+y^{2}w^{2}\right)\right]\psi_{n}=E_{n}\psi_{n}.\label{eq:5.13}
\end{equation}
So the eigenvalues $E_{n}$ and eigenfunctions $\psi_{n}$ presented
in (\ref{eq:5.12}) and (\ref{eq:5.13}) are our inputs for the numerical
calculation. Keeping these in hand, we solve the equations (\ref{eq:5.12})
and (\ref{eq:5.13}) numerically, then plot out the microcanonical
and thermal OTOC, energy spectrum and average of quantum Lyapunov
coefficient with respect to the mesonic matrix model in Figure \ref{fig:5}
and with respect to the baryonic matrix model in Figure \ref{fig:6},
\ref{fig:7}.
For the mesonic matrix model, since the action (\ref{eq:5.12}) is
symmetric under the interchange of $x,y$, we only plot out the OTOCs
and quantum Lyapunov coefficient from $\left\langle \left[x\left(t\right),p_{x}\right]^{2}\right\rangle $
and omit the index $X$. For the baryonic model, we plot out OTOCs
and quantum Lyapunov coefficients from $\left\langle \left[x\left(t\right),p_{x}\right]^{2}\right\rangle $
and $\left\langle \left[w\left(t\right),p_{w}\right]^{2}\right\rangle $
respectively denoting by the indices $X,W$, since the action (\ref{eq:5.13})
is symmetric under interchanging $x,y$ instead of interchanging $x,w$.
Our numerically results illustrate the microcanonical OTOCs of the
lowest modes oscillate basically periodically in time while the high
modes do not. This is expected since the potentials presented in the
time-independent Schr\"odinger equation (\ref{eq:5.12}) and (\ref{eq:5.13})
include a structure of the oscillation. However, there is not any
obviously exponential growth in the OTOC. Besides, the energy spectrum
decreases when the $N_{c}$ becomes large, since, as we will discuss
it in the following sections, the energy spectrum depends on $N_{c}^{-1}$
at large $N_{c}$. The average of quantum Lyapunov coefficient is
evaluated by using the generalization of (\ref{eq:4.4}) as,

\begin{equation}
L\left(T\right)=\frac{1}{2t_{max}}\log\left[\frac{1}{t_{max}}\int_{0}^{t_{max}}dtC_{T}\left(t\right)\right],
\end{equation}
which depends on the temperature. At very low temperature, the ground
state ($n=1$) dominates the thermal OTOC due to the factor $e^{-\frac{E_{n}}{T}}$
presented in (\ref{eq:5.5}), thus the system includes few degree
of freedoms characterized by the small quantum Lyapunov coefficient.
At high temperature, more modes contribute to the thermal OTOC so
that it is indicated by a larger Lyapunov coefficient as it is displayed
in Figure \ref{fig:5} and \ref{fig:7}. There is a critical temperature
that the Lyapunov coefficient begin to saturate and this behavior
covers qualitatively the classical analysis of the Lyapunov coefficient.
Notice that the average of Lyapunov coefficient also decreases when
$N_{c}$ increases. This is also expected because, at $N_{c}\rightarrow\infty$,
equation (\ref{eq:5.12}) becomes decoupled harmonic oscillator whose
Lyapunov coefficient is negative as it is given in Appendix C, and
(\ref{eq:5.13}) becomes free moving system along $x,y$ with harmonic
oscillator along $w$. For a free moving particle, we have $x\left(t\right)=x\left(0\right)+vt$,
its OTOC is $C_{T}\left(t\right)=-\left\langle \left[x\left(t\right),p\right]^{2}\right\rangle _{T}=1=e^{2Lt}$
leading to vanished Lyapunov coefficient. Therefore the Lyapunov coefficient
will go gradually close to its value for a harmonic oscillated or
a free moving system when $N_{c}$ grows, and this reduces the value
of the Lyapunov coefficient at large $N_{c}$.

\subsection{Asymptotics of the thermal OTOC}

\begin{figure}
\begin{centering}
\includegraphics[scale=0.25]{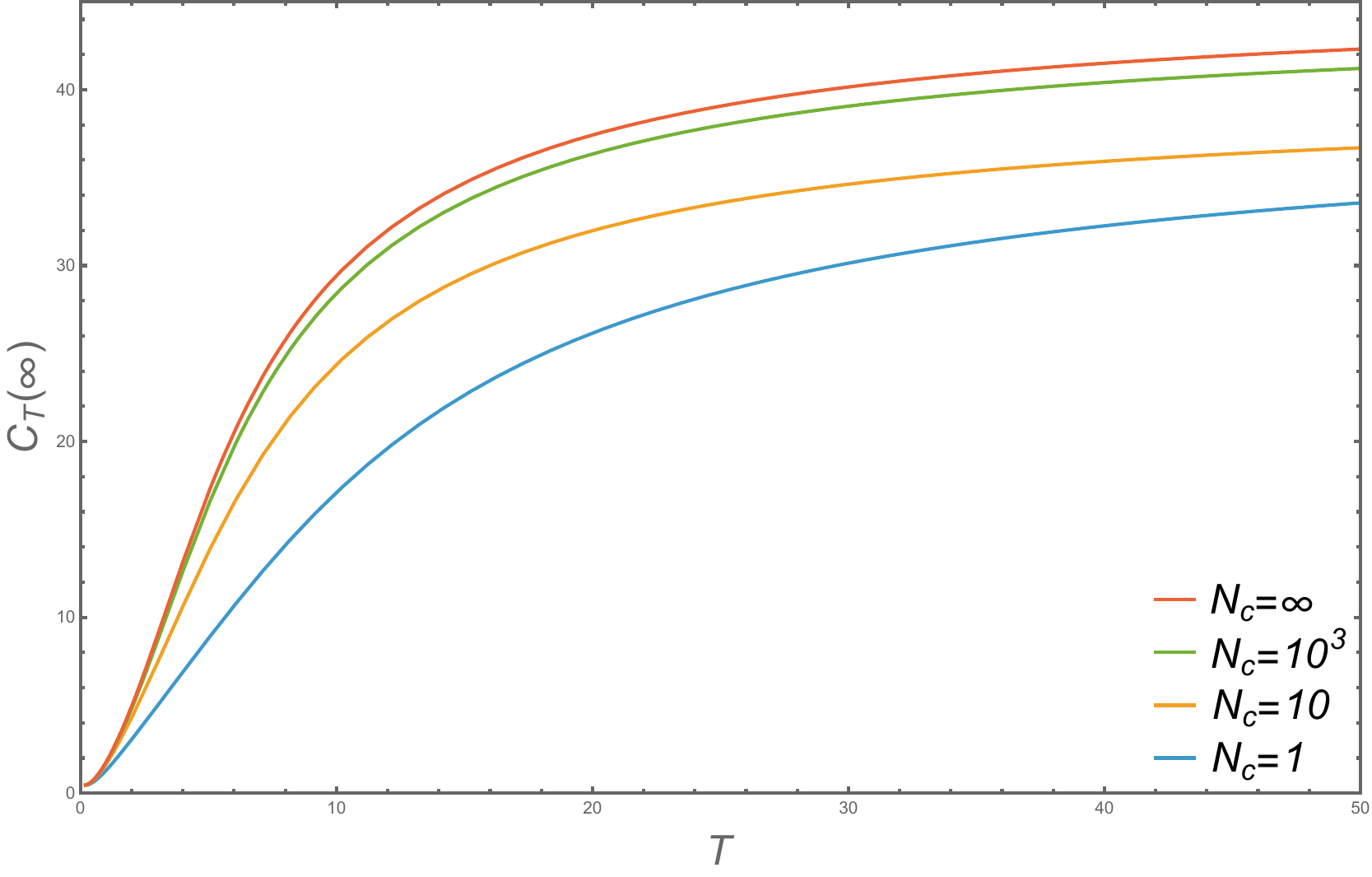}\includegraphics[scale=0.25]{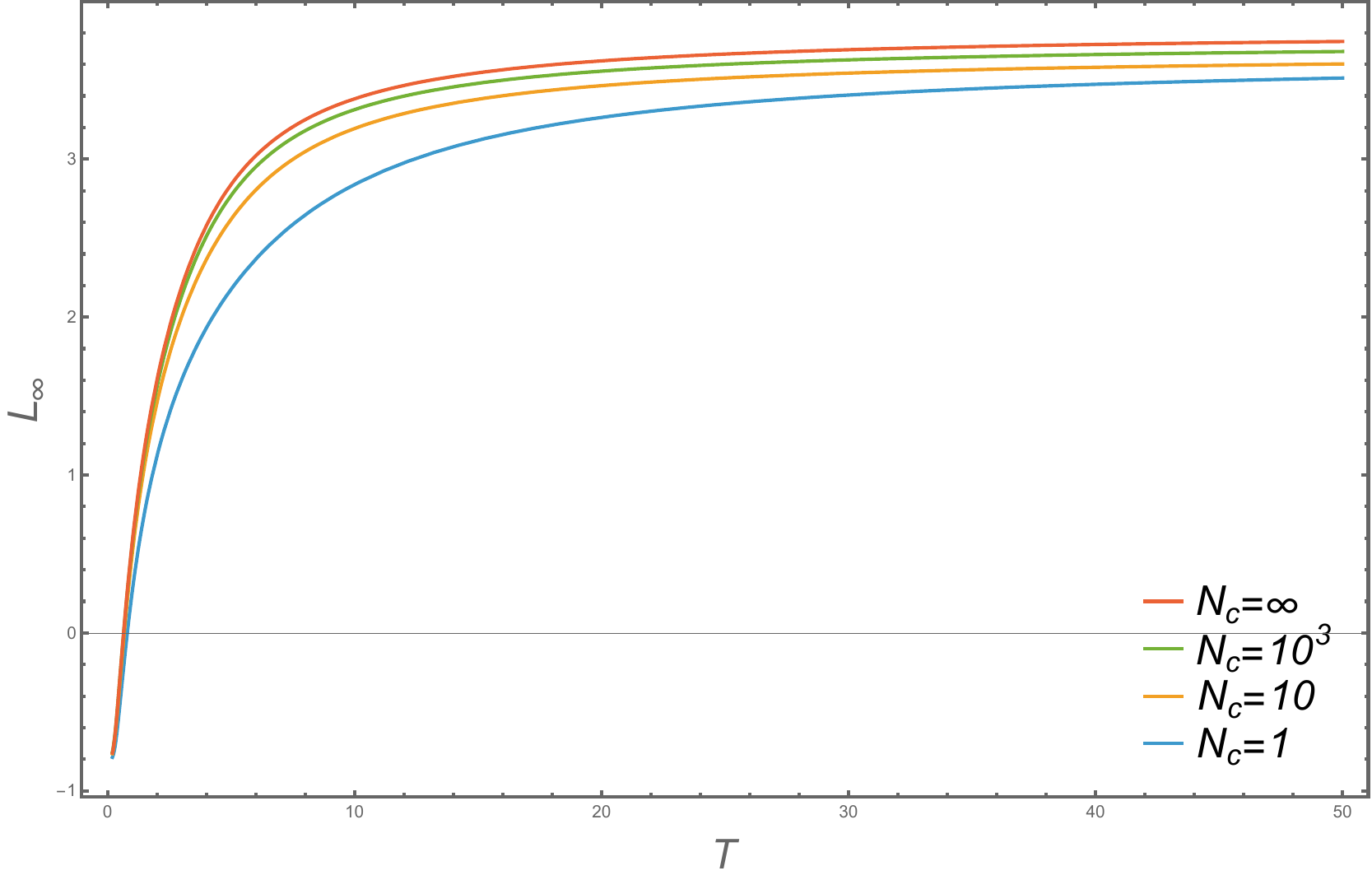}
\par\end{centering}
\begin{centering}
\includegraphics[scale=0.25]{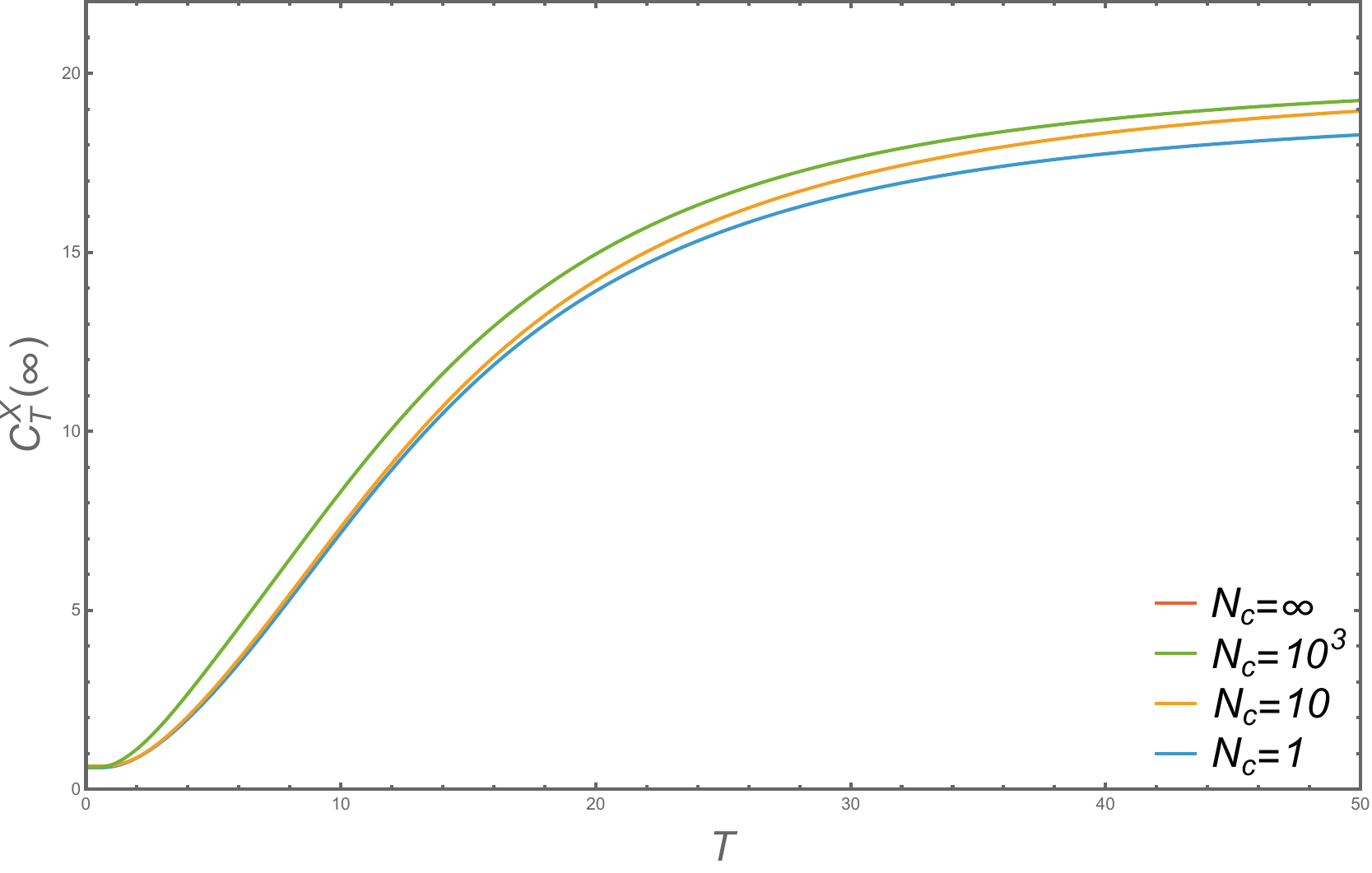}\includegraphics[scale=0.25]{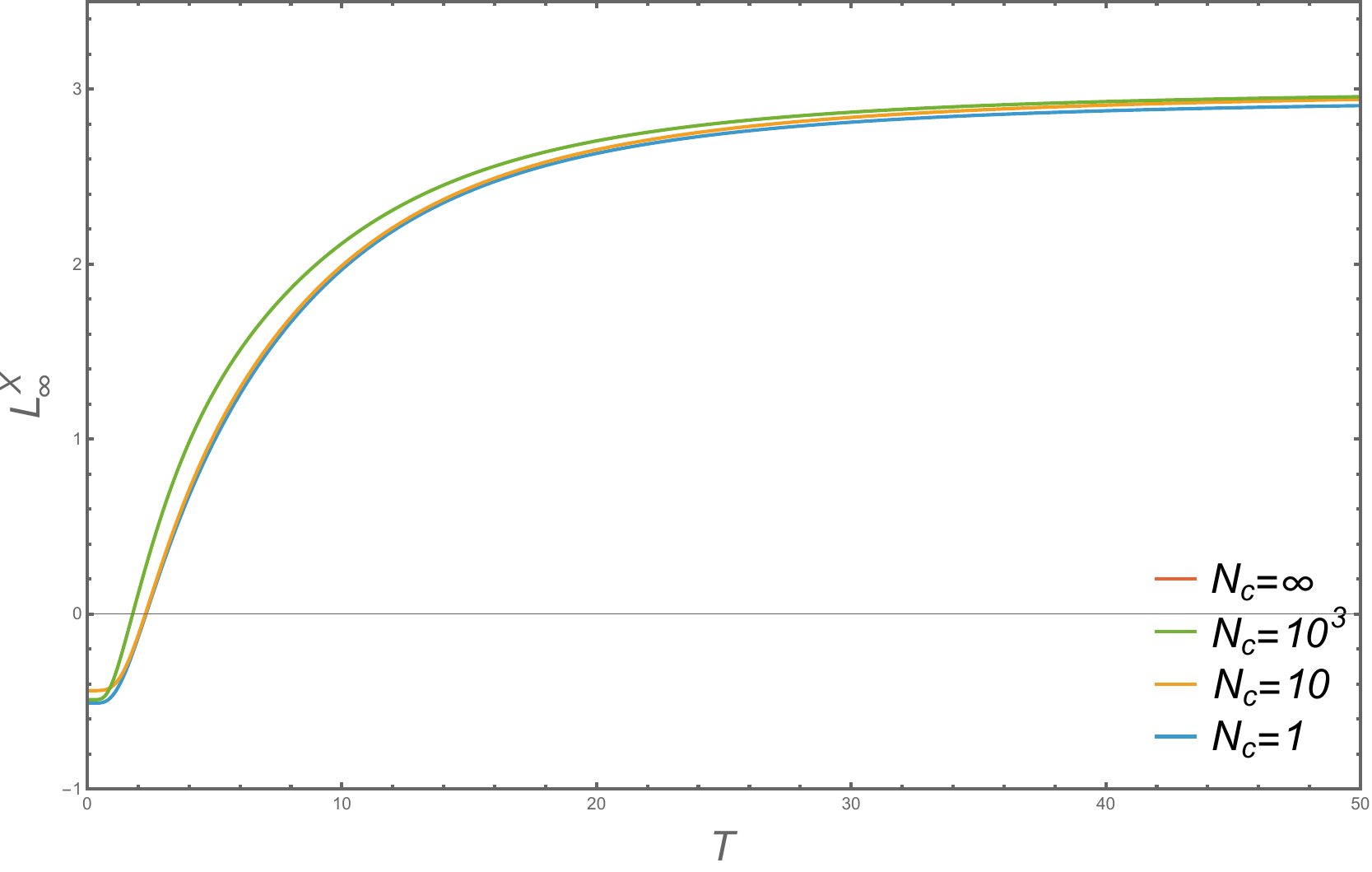}
\par\end{centering}
\caption{\label{fig:8}The asymptotics of OTOC $C_{T}\left(\infty\right)$
and the associated Lyapunov coefficient $L_{\infty}$ as functions
of $T$ with various $N_{c}$. \textbf{Upper:} Mesonic matrix model.
\textbf{Lower:} Baryonic matrix model. Only $C_{T}^{X},L_{\infty}^{X}$
are plotted since the behavior of $C_{T}^{W},L_{\infty}^{W}$ is quite
similar as $C_{T}^{X},L_{\infty}^{X}$.}

\end{figure}

In generic quantum chaotic systems, the correlation of operator $x\left(t\right)$
and $p$ will be lost as time goes by. After a critical point of time,
named as Ehrenfest time which refers to the timescale that the wave
function spreads over the whole system, the value of the thermal OTOC
approach asymptotically at $t\rightarrow\infty$ as \cite{key-8},

\begin{equation}
C_{T}\left(\infty\right)=2\left\langle x^{2}\right\rangle _{T}\left\langle p^{2}\right\rangle _{T},\label{eq:5.15}
\end{equation}
and we will use (\ref{eq:5.15}) to evaluate the asymptotic behavior
of the thermal OTOCs at large time. To further indicate the chaos,
we defined the asymptotic Lyapunov coefficient by using (\ref{eq:5.15})
as,

\begin{equation}
L_{\infty}=\frac{1}{2}\log C_{T}\left(\infty\right).
\end{equation}
The numerical results for the asymptotics of OTOCs and Lyapunov coefficient
are illustrated in Figure \ref{fig:8}.
 We can see while $C_{T}\left(\infty\right)$ grows fast at small
temperature, it trends to be flat at high temperature. The dependence
on $T$ of $L_{\infty}$ is therefore similar as the average of Lyapunov
coefficient illustrated given in Figure \ref{fig:5} and \ref{fig:7}.
However, the large $N_{c}$ behavior of the asymptotic Lyapunov coefficient
and the average value of Lyapunov coefficient is opposite. So while
$C_{T}\left(\infty\right)$ could indeed be the asymptotics of the
thermal OTOC with various $N_{c}$, we comment that the asymptotics
$C_{T}\left(\infty\right)$ does not have the same dependence of $N_{c}$
as $C_{T}\left(t\right)$. And we will further confirm this in the
following section.

\subsection{The large $N_{c}$ analytics}

In the large $N_{c}$ limit, as the coupling constant $g$ presented
in the Schr\"odinger equation (\ref{eq:5.12}) and (\ref{eq:5.13})
goes to zero i.e. $g\propto N_{c}^{-1}\rightarrow0$, it is possible
to calculate the OTOCs analytically by using the standard method of
perturbation in quantum mechanics. So let us discuss the leading order
approximation to the case of mesonic matrix model and baryonic matrix
model respectively.

\subsubsection*{The mesonic matrix model}

Let us start with the Schr\"odinger equation (\ref{eq:5.12}). As $g\propto N_{c}^{-1}\rightarrow0$
at large $N_{c}$, we decompose the Hamiltonian associated to equation
(\ref{eq:5.12}) as,

\begin{align}
H & =H_{0}+H_{1},\nonumber \\
H_{0} & =-\frac{1}{2}\left(\frac{\partial^{2}}{\partial x^{2}}+\frac{\partial^{2}}{\partial y^{2}}\right)+\frac{1}{2}m^{2}\left(x^{2}+y^{2}\right),\nonumber \\
H_{1} & =gx^{2}y^{2}.
\end{align}
The eigenstates and eigenvalues of $H_{0},H$ are denoted by the quantum
number $n_{x},n_{y}$ as 
\begin{equation}
H_{0}\left|n_{x},n_{y}\right\rangle _{\left(0\right)}=E_{n_{x},n_{y}}^{\left(0\right)}\left|n_{x},n_{y}\right\rangle _{\left(0\right)},H\left|n_{x},n_{y}\right\rangle =E_{n_{x},n_{y}}\left|n_{x},n_{y}\right\rangle .
\end{equation}
Since $H_{0}$ represents nothing but the two-dimensional harmonic
oscillator, its eigenstates $\left|n_{x},n_{y}\right\rangle _{\left(0\right)}$
can be constructed by the creation operators and its eigenvalue is
known as

\begin{equation}
E_{n_{x},n_{y}}^{\left(0\right)}=\left(n_{x}+n_{y}+1\right)m,\ n_{x},n_{y}=0,1,2...
\end{equation}
Besides, the eigenstates $\left|s_{x},s_{y}\right\rangle $ of $H$
can be decomposed by the combination of $\left|n_{x},n_{y}\right\rangle _{\left(0\right)}$
due to its completeness, so we can write down,

\begin{equation}
\left|s_{x},s_{y}\right\rangle =\sum_{n_{x},n_{y}}\mathcal{C}_{n_{x},n_{y}}^{s_{x},s_{y}}\left|n_{x},n_{y}\right\rangle _{\left(0\right)},
\end{equation}
where the coefficient $\mathcal{C}_{n_{x},n_{y}}^{s_{x},s_{y}}$ can
be expanded as a series of coupling constant $g$. Up to leading order
of $g$, we have

\begin{equation}
\mathcal{C}_{n_{x},n_{y}}^{s_{x},s_{y}}=\delta_{n_{x}}^{s_{x}}\delta_{n_{y}}^{s_{y}}+g\mathcal{B}_{n_{x},n_{y}}^{s_{x},s_{y}}+\mathcal{O}\left(g^{2}\right),
\end{equation}
and $\mathcal{B}_{n_{x},n_{y}}^{s_{x},s_{y}}$ is evaluated as,

\begin{equation}
\mathcal{B}_{n_{x},n_{y}}^{s_{x},s_{y}}=\begin{cases}
\frac{\ _{0}\left\langle s_{x},s_{y}\left|x^{2}y^{2}\right|n_{x},n_{y}\right\rangle _{0}}{E_{s_{x},s_{y}}^{\left(0\right)}-E_{n_{x},n_{y}}^{\left(0\right)}} & ,E_{s_{x},s_{y}}^{\left(0\right)}\neq E_{n_{x},n_{y}}^{\left(0\right)},\\
0, & E_{s_{x},s_{y}}^{\left(0\right)}=E_{n_{x},n_{y}}^{\left(0\right)},
\end{cases}\label{eq:5.22}
\end{equation}
according to the Schr\"odinger equation. Since $\left|n_{x},n_{y}\right\rangle _{\left(0\right)}$
is the eigenstate of harmonic oscillator, we can obtain the matrix
elements of $H_{1}$ in the representation of $H_{0}$ as,

\begin{align}
 & \ _{0}\left\langle s_{x},s_{y}\left|x^{2}y^{2}\right|n_{x},n_{y}\right\rangle _{0}\nonumber \\
 & =\frac{1}{4m^{2}}\left[\sqrt{n_{x}\left(n_{x}-1\right)}\delta_{s_{x},n_{x}-2}+\left(2n_{x}+1\right)\delta_{s_{x},n_{x}}+\sqrt{\left(n_{x}+1\right)\left(n_{x}+2\right)}\delta_{s_{x},n_{x}+2}\right]\nonumber \\
 & \times\left[\sqrt{n_{y}\left(n_{y}-1\right)}\delta_{s_{y},n_{y}-2}+\left(2n_{y}+1\right)\delta_{s_{y},n_{y}}+\sqrt{\left(n_{y}+1\right)\left(n_{y}+2\right)}\delta_{s_{y},n_{y}+2}\right],\label{eq:5.23}
\end{align}
so that $\mathcal{B}_{n_{x},n_{y}}^{s_{x},s_{y}}$ is completely determined
by $H_{0}$. Accordingly, the matrix element of $x$ is obtained as\footnote{We note here the coefficient $\mathcal{B}_{n_{x},n_{y}}^{s_{x},s_{y}}$
is completely real according to (\ref{eq:5.22}) and (\ref{eq:5.23}). },

\begin{align}
\left\langle t_{x},t_{y}\left|x\right|s_{x},s_{y}\right\rangle  & =\sum_{m_{x},m_{y},n_{x},n_{y}}\left(\delta_{m_{x}}^{t_{x}}\delta_{m_{y}}^{t_{y}}+g\mathcal{B}_{m_{x},m_{y}}^{t_{x},t_{y}}\right)\left(\delta_{n_{x}}^{s_{x}}\delta_{n_{y}}^{s_{y}}+g\mathcal{B}_{n_{x},n_{y}}^{s_{x},s_{y}}\right)\nonumber \\
 & \times\ _{0}\left\langle m_{x},m_{y}\left|x\right|n_{x},n_{y}\right\rangle _{0}\nonumber \\
 & =\ _{0}\left\langle t_{x},t_{y}\left|x\right|s_{x},s_{y}\right\rangle _{0}+g\sum_{m_{x},m_{y}}\mathcal{B}_{m_{x},m_{y}}^{t_{x},t_{y}}\ _{0}\left\langle m_{x},m_{y}\left|x\right|s_{x},s_{y}\right\rangle _{0}\nonumber \\
 & +g\sum_{n_{x},n_{y}}\mathcal{B}_{n_{x},n_{y}}^{s_{x},s_{y}}\ _{0}\left\langle t_{x},t_{y}\left|x\right|n_{x},n_{y}\right\rangle _{0}+\mathcal{O}\left(g^{2}\right),\label{eq:5.24}
\end{align}
where the zeroth order matrix element $\ _{0}\left\langle m_{x},m_{y}\left|x\right|n_{x},n_{y}\right\rangle _{0}$
can be calculated as,

\begin{equation}
\ _{0}\left\langle m_{x},m_{y}\left|x\right|n_{x},n_{y}\right\rangle _{0}=\frac{1}{\sqrt{2m}}\left(\sqrt{n_{x}}\delta_{m_{x},n_{x}-1}+\sqrt{n_{x}+1}\delta_{m_{x},n_{x}+1}\right)\delta_{m_{y},n_{y}},
\end{equation}
thus (\ref{eq:5.24}) is totally analytical. Besides, the energy spectrum
up to $\mathcal{O}\left(g\right)$ can be computed as,

\begin{align}
E_{n_{x},n_{y}} & =E_{n_{x},n_{y}}^{\left(0\right)}+g\ _{0}\left\langle n_{x},n_{y}\left|x^{2}y^{2}\right|n_{x},n_{y}\right\rangle _{0}\nonumber \\
 & =\left(n_{x}+n_{y}\right)m+\frac{1}{4}g\left(2n_{x}+1\right)\left(2n_{y}+1\right),\label{eq:5.25}
\end{align}
so we find the energy spectrum depends on $g\propto N_{c}^{-1}$ at
large $N_{c}$. 

\begin{figure}
\begin{centering}
\includegraphics[scale=0.25]{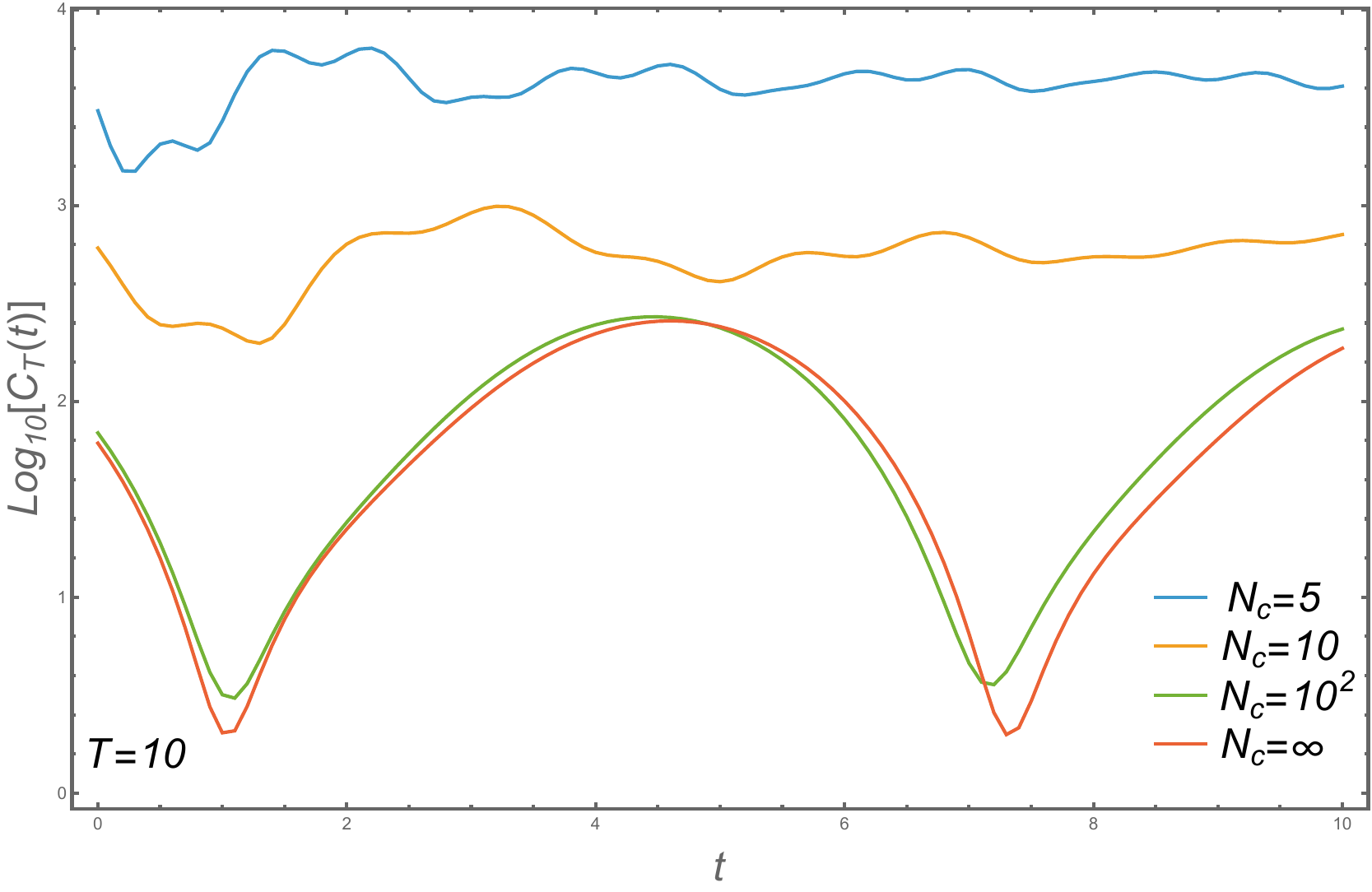}\includegraphics[scale=0.25]{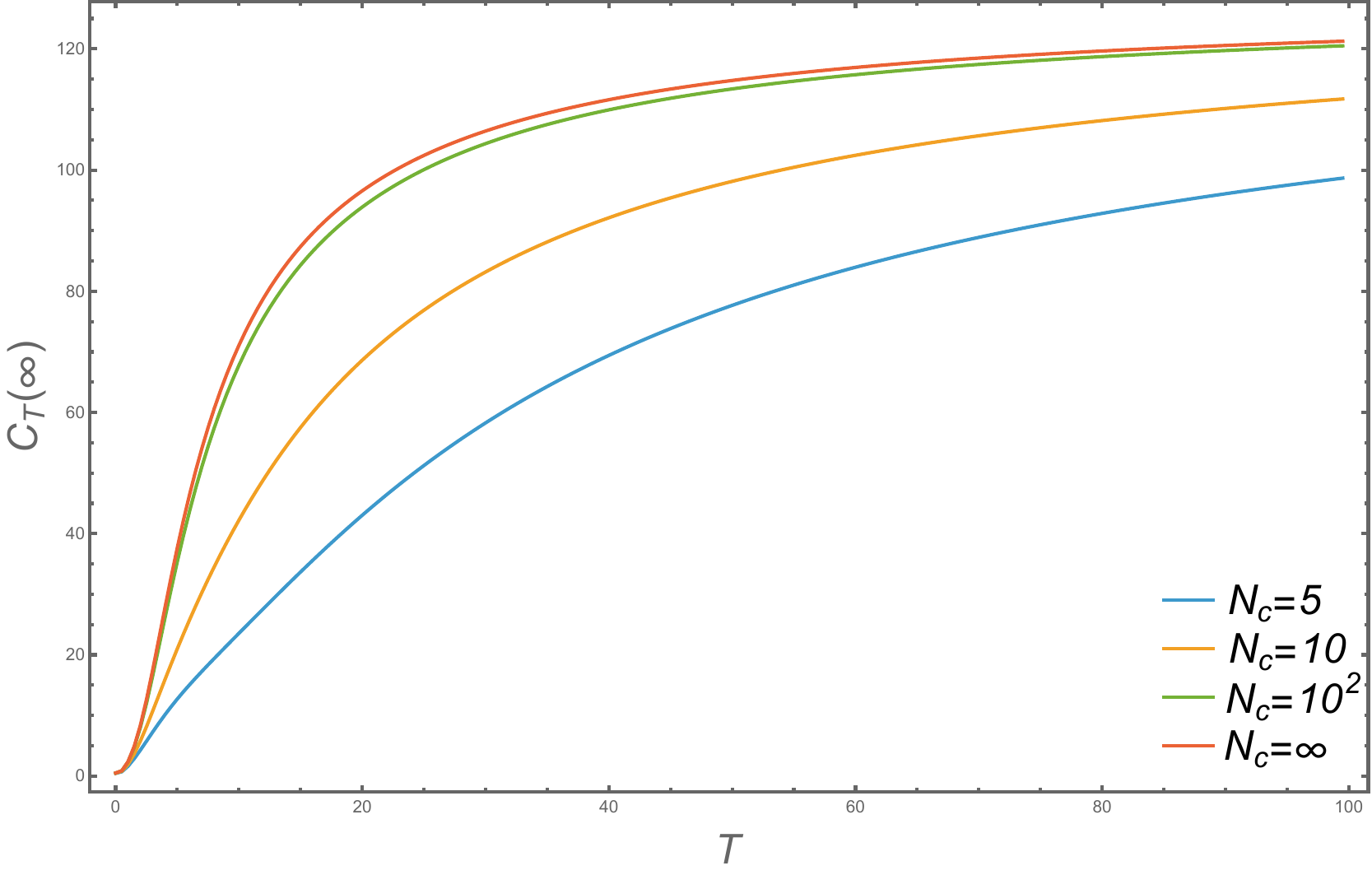}
\par\end{centering}
\caption{\label{fig:9}The OTOCs at large $N_{c}$ up to $\mathcal{O}\left(N_{c}^{-1}\right)$
in the mesonic matrix model. \textbf{Left:} The thermal OTOC as a
function of time $t$ with various $N_{c}$. \textbf{Right:} the asymptotics
of the thermal OTOC as a function of temperature $T$ with various
$N_{c}$.}

\end{figure}

In summary, by using the perturbative method up to leading order of
$g$, we have the energy spectrum given in (\ref{eq:5.25}), the matrix
element $\left\langle n\left|x\right|k\right\rangle $ given in (\ref{eq:5.24}).
Since the coefficients $\mathcal{C}_{n_{x},n_{y}}^{s_{x},s_{y}},\mathcal{B}_{n_{x},n_{y}}^{s_{x},s_{y}}$
are totally determined by $H_{0}$ and are completely analytical,
altogether, it is possible to calculate analytically the OTOCs with
these inputs up to the order of $N_{c}^{-1}$ through (\ref{eq:5.6})
and (\ref{eq:5.11}). Note that, although the leading order OTOC is
totally analytical, it is difficult to see obviously the dependence
of $N_{c}$ due to the lengthy sums in the formulas of the OTOC. Therefore,
we plot out the final results in Figure \ref{fig:9}, and it illustrates
that the thermal OTOC indeed decreases when the $N_{c}$ grows, agreeing
qualitatively with our numerical results presented in Section 5.2.
And at very large $N_{c}$, the thermal OTOC returns to be periodic
since the eigen equation of the Hamiltonian becomes the equation of
motion for a 2d harmonic oscillator. In this sense, we believe our
analysis is consistent.

In particular, the asymptotics of the thermal OTOC can also be evaluated
analytically up to the order of $N_{c}^{-1}$. Recall the definition
of $C_{T}\left(\infty\right)$ in (\ref{eq:5.15}) and the matrix
element given in (\ref{eq:5.23}), we find

\begin{align}
\ _{0}\left\langle s_{x},s_{y}\left|x^{2}\right|n_{x},n_{y}\right\rangle _{0} & =\frac{1}{2m}\bigg[\sqrt{n_{x}\left(n_{x}-1\right)}\delta_{s_{x},n_{x}-2}+\left(2n_{x}+1\right)\delta_{s_{x},n_{x}}\nonumber \\
 & +\sqrt{\left(n_{x}+1\right)\left(n_{x}+2\right)}\delta_{s_{x},n_{x}+2}\bigg]\delta_{s_{y},n_{y}},\label{eq:5.27}
\end{align}
and

\begin{align}
\ _{0}\left\langle s_{x},s_{y}\left|p^{2}\right|n_{x},n_{y}\right\rangle _{0} & =-\frac{m}{2}\bigg[\delta_{s_{x},n_{x}-2}\sqrt{n_{x}\left(n_{x}-1\right)}-\left(2n_{x}+1\right)\delta_{s_{x},n_{x}}\nonumber \\
 & +\sqrt{\left(n_{x}+1\right)\left(n_{x}+2\right)}\delta_{s_{x},n_{x}+2}\bigg]\delta_{s_{y},n_{y}}.\label{eq:5.28}
\end{align}
Therefore, the asymptotics of the thermal OTOC can be computed analytically
up to the leading order of $g$ as $C_{T}\left(\infty\right)=2\left\langle x^{2}\right\rangle _{T}\left\langle p^{2}\right\rangle _{T}$
with

\begin{align}
\left\langle n_{x},n_{y}\left|x^{2}\right|n_{x},n_{y}\right\rangle  & =\ _{0}\left\langle n_{x},n_{y}\left|x^{2}\right|n_{x},n_{y}\right\rangle _{0}+2g\sum_{m_{x},m_{y}}\mathcal{B}_{m_{x},m_{y}}^{n_{x},n_{y}}\ _{0}\left\langle m_{x},m_{y}\left|x^{2}\right|n_{x},n_{y}\right\rangle _{0}+\mathcal{O}\left(g^{2}\right),\nonumber \\
\left\langle n_{x},n_{y}\left|p^{2}\right|n_{x},n_{y}\right\rangle  & =\ _{0}\left\langle n_{x},n_{y}\left|p^{2}\right|n_{x},n_{y}\right\rangle _{0}+2g\sum_{m_{x},m_{y}}\mathcal{B}_{m_{x},m_{y}}^{n_{x},n_{y}}\ _{0}\left\langle m_{x},m_{y}\left|p^{2}\right|n_{x},n_{y}\right\rangle _{0}+\mathcal{O}\left(g^{2}\right),
\end{align}
and we also plot the final result in Figure \ref{fig:9}. The analytical
result basically covers our numerical results in Figure \ref{fig:8}
and it confirms that the large $N_{c}$ behavior of the thermal OTOC
$C_{T}\left(t\right)$ and its asymptotics $C_{T}\left(\infty\right)$
is indeed opposite.

\subsubsection*{The baryonic matrix model}

The analysis for the baryonic matrix model would be quite similar
as the case of the mesonic matrix model. The associated Hamiltonian
to the Schr\"odinger equation (\ref{eq:5.13}) can be written as,

\begin{align}
H & =H_{0}+H_{1},\nonumber \\
H_{0} & =\frac{1}{2}\left(\frac{\partial^{2}}{\partial x^{2}}+\frac{\partial^{2}}{\partial y^{2}}+\frac{\partial^{2}}{\partial w^{2}}\right)+\frac{1}{2}m^{2}w^{2},\nonumber \\
H_{1} & =g\left(x^{2}y^{2}+x^{2}w^{2}+y^{2}w^{2}\right),
\end{align}
where $H_{1}$ can be treated as the perturbation due to $g\propto N_{c}^{-1}\rightarrow0$
at large $N_{c}$. Here we use $\left|...\right\rangle _{0}$ to denote
the eigenstate of $H_{0}$. The eigenfunction of $H_{0}$ can be written
as,

\begin{equation}
\left\langle x,y,w|n_{x},n_{y},n_{w}\right\rangle _{0}=\left\langle x|n_{x}\right\rangle _{0}\left\langle y|n_{y}\right\rangle _{0}\left\langle y|n_{w}\right\rangle _{0},
\end{equation}
where $\left\langle x|n_{x}\right\rangle _{0},\left\langle y|n_{y}\right\rangle _{0}$
refers to the eigenfunctions of momentum describing a free moving
particle and $\left\langle y|n_{w}\right\rangle _{0}$ remains to
be the eigenfunction of a harmonic oscillator. However it is impossible
to obtain discrete eigenvalues if we chose $\left\langle x|n_{x}\right\rangle _{0},\left\langle y|n_{y}\right\rangle _{0}$
as the eigenfunctions of a free moving particle. To compare with our
numerical calculations in the previous sections and the discussion
of the classical case, we consider the particle moves in a square
box on the $x-y$ plane with length $2L$, hence the eigenfunctions
$\left\langle x|n_{x}\right\rangle _{0},\left\langle y|n_{y}\right\rangle _{0}$
can be chosen as,

\begin{equation}
\left\langle x|n_{x}\right\rangle _{0}\left\langle y|n_{y}\right\rangle _{0}=\left\langle x,y|n_{x},n_{y}\right\rangle _{0}=\begin{cases}
\frac{1}{L}\sin\left[\frac{n_{x}\pi}{2L}\left(x+L\right)\right]\sin\left[\frac{n_{y}\pi}{2L}\left(y+L\right)\right], & \left|x\right|,\left|y\right|\leq L,\\
0, & \left|x\right|,\left|y\right|>L,
\end{cases}\label{eq:5.32}
\end{equation}
and the energy spectrum of $H_{0}$ is,

\begin{equation}
E_{n_{x},n_{y},n_{w}}^{\left(0\right)}=\frac{\pi^{2}}{8L^{2}}\left(n_{x}^{2}+n_{y}^{2}\right)+\left(n_{w}+\frac{1}{2}\right)m.
\end{equation}
Then the eigenvalue $E_{n_{x},n_{y},n_{w}}$ of $H$ is evaluated
as,

\begin{equation}
E_{n_{x},n_{y},n_{w}}\simeq E_{n_{x},n_{y},n_{w}}^{\left(0\right)}+g\ _{0}\left\langle n_{x},n_{y},n_{w}\left|x^{2}y^{2}+x^{2}w^{2}+y^{2}w^{2}\right|n_{x},n_{y},n_{w}\right\rangle _{0},\label{eq:5.34}
\end{equation}
and the eigenstate $\left|s_{x},s_{y},s_{w}\right\rangle $ of $H$
can be written as the combination of $\left|n_{x},n_{y},n_{w}\right\rangle _{\left(0\right)}$
as, 

\begin{equation}
\left|s_{x},s_{y},s_{w}\right\rangle =\sum_{n_{x},n_{y}}\mathcal{C}_{n_{x},n_{y},n_{w}}^{s_{x},s_{y},s_{w}}\left|n_{x},n_{y},n_{w}\right\rangle _{\left(0\right)},
\end{equation}
where, up to the leading order of $g$, the coefficient $\mathcal{C}_{n_{x},n_{y},n_{w}}^{s_{x},s_{y},s_{w}}$
is given by,

\begin{equation}
\mathcal{C}_{n_{x},n_{y},n_{w}}^{s_{x},s_{y},s_{w}}=\delta_{n_{x}}^{s_{x}}\delta_{n_{y}}^{s_{y}}\delta_{n_{w}}^{s_{w}}+g\mathcal{B}_{n_{x},n_{y},n_{w}}^{s_{x},s_{y},s_{w}}+\mathcal{O}\left(g^{2}\right),
\end{equation}
and

\begin{equation}
\mathcal{B}_{n_{x},n_{y},n_{w}}^{s_{x},s_{y},s_{w}}=\begin{cases}
\frac{\ _{0}\left\langle s_{x},s_{y},s_{w}\left|x^{2}y^{2}+x^{2}w^{2}+y^{2}w^{2}\right|n_{x},n_{y},n_{w}\right\rangle _{0}}{E_{s_{x},s_{y},s_{w}}^{\left(0\right)}-E_{n_{x},n_{y},n_{w}}^{\left(0\right)}} & ,E_{s_{x},s_{y},s_{w}}^{\left(0\right)}\neq E_{n_{x},n_{y},n_{w}}^{\left(0\right)},\\
0, & E_{s_{x},s_{y},s_{w}}^{\left(0\right)}=E_{n_{x},n_{y},n_{w}}^{\left(0\right)}.
\end{cases}
\end{equation}
All the matrix elements $\ _{0}\left\langle s_{x}\left|x^{2}\right|n_{x}\right\rangle _{0},\ _{0}\left\langle s_{y}\left|y^{2}\right|n_{y}\right\rangle _{0},\ _{0}\left\langle s_{w}\left|w^{2}\right|n_{w}\right\rangle _{0}$
can be work out analytically due to the wave function given in (\ref{eq:5.32}),
leading to

\begin{equation}
\ _{0}\left\langle s_{x}\left|x^{2}\right|n_{x}\right\rangle _{0}=\begin{cases}
\frac{L^{2}}{3}\left(1-\frac{6}{n_{x}^{2}\pi^{2}}\right), & s_{x}=n_{x},\\
\frac{16\left[1+\left(-1\right)^{n_{x}+s_{x}}\right]n_{x}s_{x}L^{2}}{\left(s_{x}-n_{x}\right)^{2}\left(s_{x}+n_{x}\right)^{2}\pi^{2}}, & s_{x}\neq n_{x}.
\end{cases}\label{eq:5.36}
\end{equation}
The $\ _{0}\left\langle s_{y}\left|y^{2}\right|n_{y}\right\rangle _{0}$
takes the same form as (\ref{eq:5.36}) and the matrix elements $\ _{0}\left\langle s_{w}\left|w^{2}\right|n_{w}\right\rangle _{0}$
takes the same form as it is given in (\ref{eq:5.27}). Therefore
the coefficient $\mathcal{B}_{n_{x},n_{y},n_{w}}^{s_{x},s_{y},s_{w}}$
and the eigenvalue $E_{n_{x},n_{y},n_{w}}$ are totally determined
analytically by $H_{0}$. Besides, the matrix element $\left\langle t_{x},t_{y},t_{w}\left|x\right|s_{x},s_{y},s_{w}\right\rangle $
can be further computed as,

\begin{align}
\left\langle t_{x},t_{y},t_{w}\left|x\right|s_{x},s_{y},s_{w}\right\rangle  & =\ _{0}\left\langle t_{x},t_{y},t_{w}\left|x\right|s_{x},s_{y},s_{w}\right\rangle _{0}\nonumber \\
 & +g\sum_{m_{x},m_{y},m_{w}}\mathcal{B}_{m_{x},m_{y},m_{w}}^{t_{x},t_{y},t_{w}}\ _{0}\left\langle m_{x},m_{y},m_{w}\left|x\right|s_{x},s_{y},s_{w}\right\rangle _{0}\nonumber \\
 & +g\sum_{n_{x},n_{y},n_{w}}\mathcal{B}_{n_{x},n_{y},n_{w}}^{s_{x},s_{y},s_{w}}\ _{0}\left\langle t_{x},t_{y},t_{w}\left|x\right|n_{x},n_{y},n_{w}\right\rangle _{0}+\mathcal{O}\left(g^{2}\right),\label{eq:5.37}
\end{align}
where 

\begin{equation}
\ _{0}\left\langle t_{x},t_{y},t_{w}\left|x\right|s_{x},s_{y},s_{w}\right\rangle _{0}=\begin{cases}
0, & t_{x}=s_{x},\\
\frac{8\left[-1+\left(-1\right)^{n_{x}+t_{x}}\right]n_{x}t_{x}L}{\pi^{2}\left(n_{x}^{2}-t_{x}^{2}\right)}\delta_{t_{y},s_{y}}\delta_{t_{w},s_{w}}, & t_{x}\neq s_{x}.
\end{cases}
\end{equation}
\begin{figure}
\begin{centering}
\includegraphics[scale=0.25]{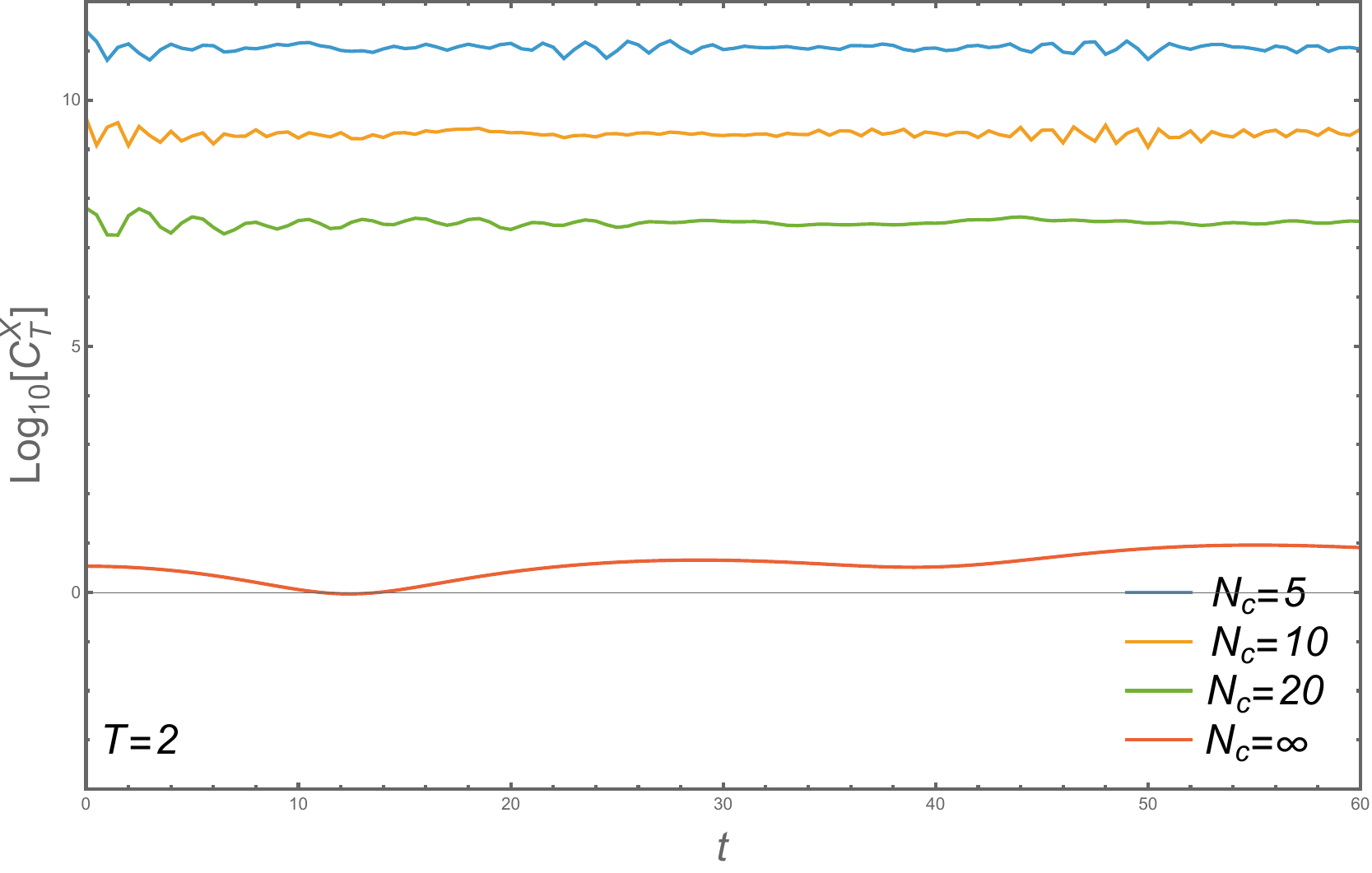}\includegraphics[scale=0.25]{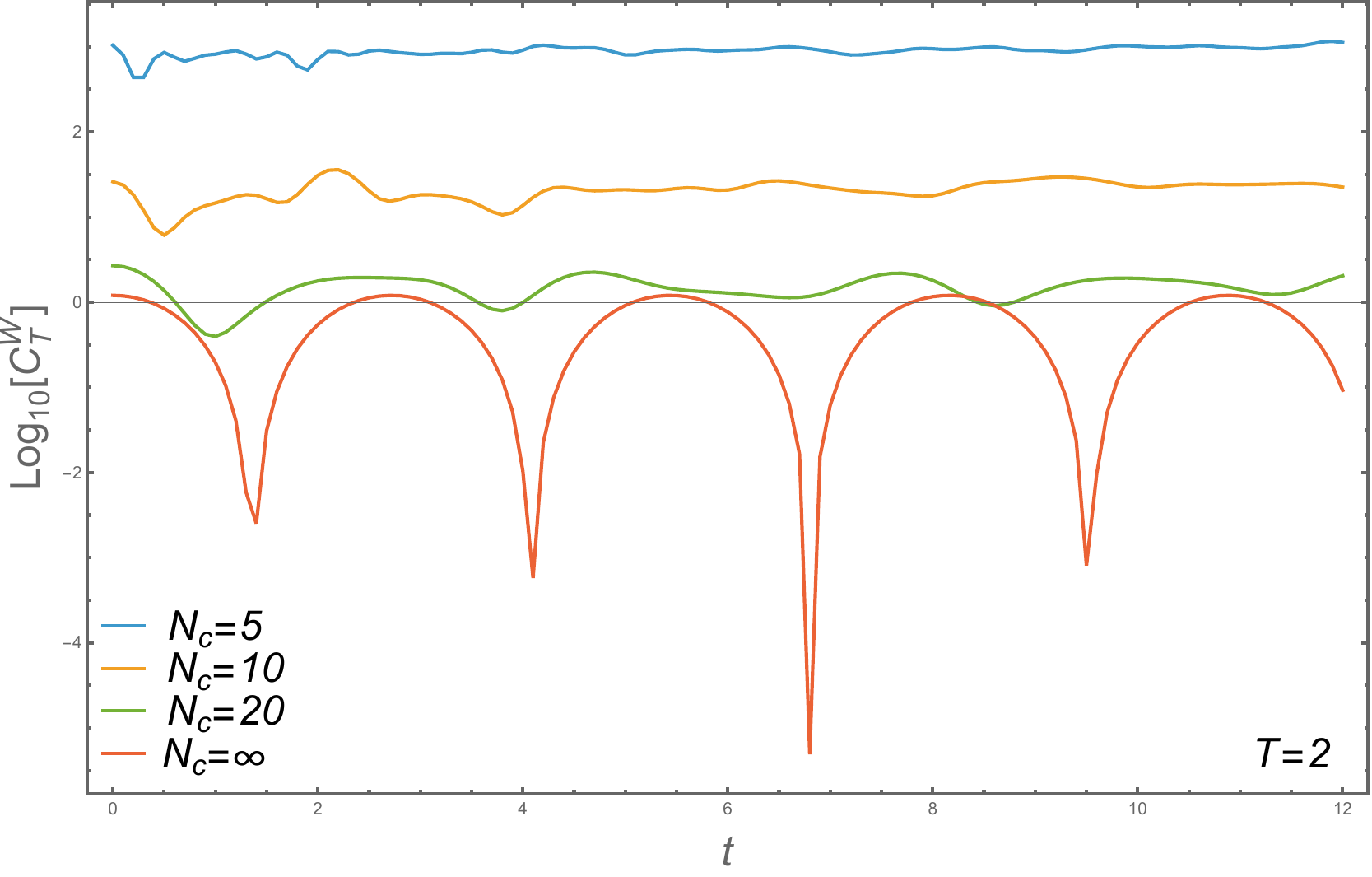}
\par\end{centering}
\begin{centering}
\includegraphics[scale=0.25]{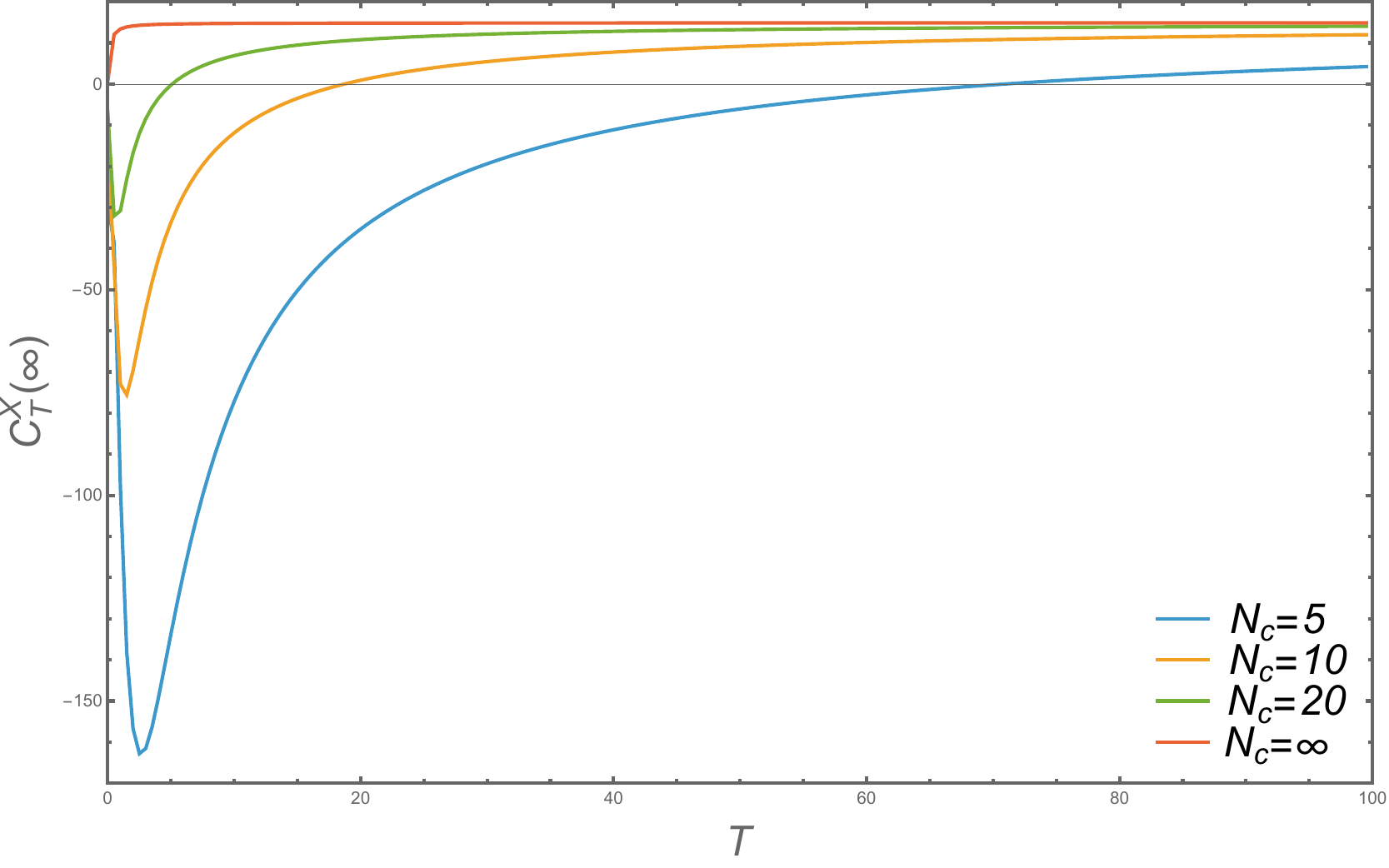}\includegraphics[scale=0.25]{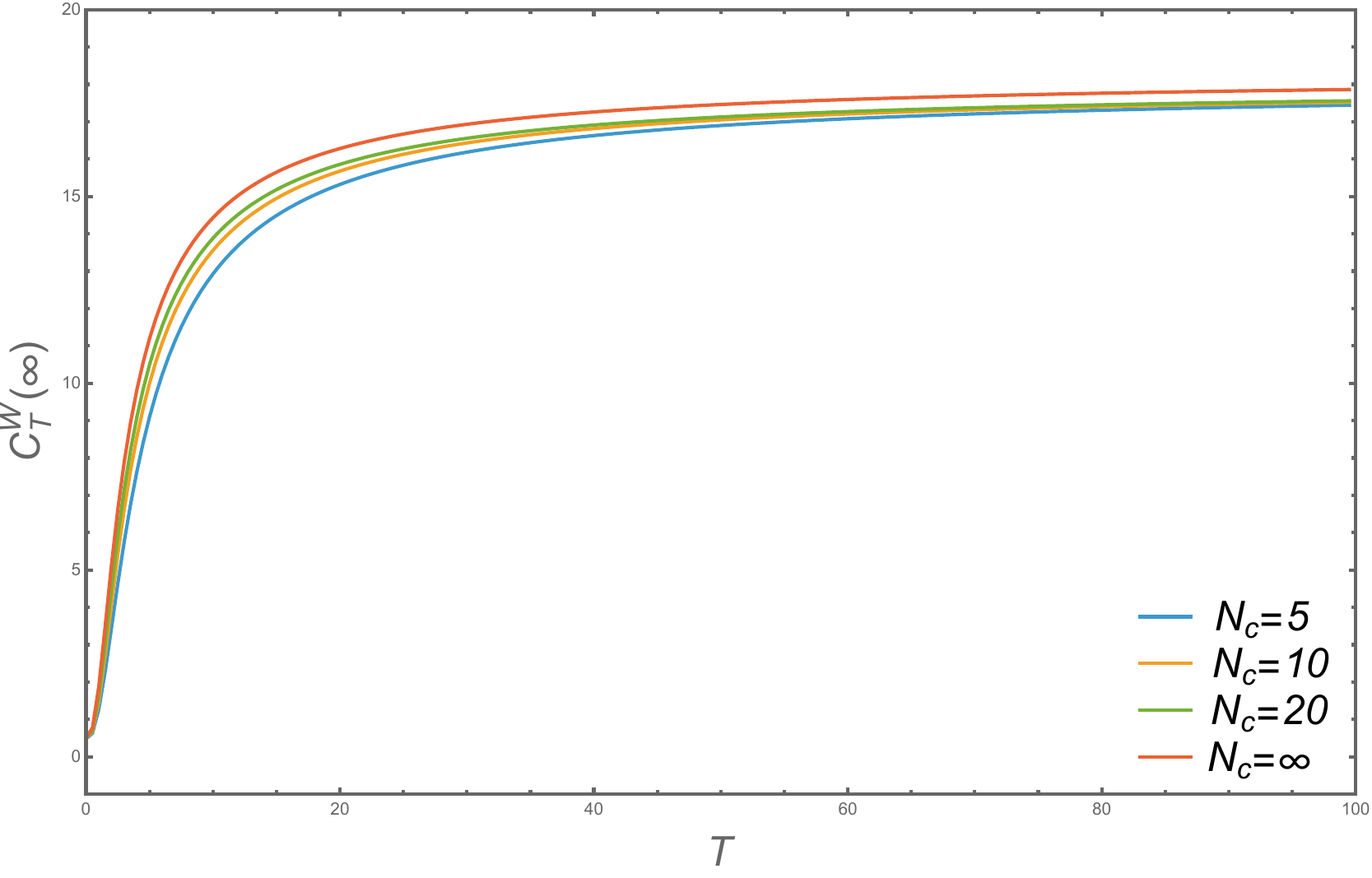}
\par\end{centering}
\caption{\label{fig:10}The OTOCs at large $N_{c}$ up to $\mathcal{O}\left(N_{c}^{-1}\right)$
in the baryonic matrix model. \textbf{Upper:} The thermal OTOC as
a function of time $t$ with various $N_{c}$. \textbf{Lower:} the
asymptotics of the thermal OTOC as a function of temperature $T$
with various $N_{c}$.}
\end{figure}
In addition, we can also obtain the matrix element $\left\langle t_{x},t_{y},t_{w}\left|w\right|s_{x},s_{y},s_{w}\right\rangle $
up to leading order of $g$ as,

\begin{align}
\left\langle t_{x},t_{y},t_{w}\left|w\right|s_{x},s_{y},s_{w}\right\rangle  & =\ _{0}\left\langle t_{x},t_{y},t_{w}\left|w\right|s_{x},s_{y},s_{w}\right\rangle _{0}\nonumber \\
 & +g\sum_{m_{x},m_{y},m_{w}}\mathcal{B}_{m_{x},m_{y},m_{w}}^{t_{x},t_{y},t_{w}}\ _{0}\left\langle m_{x},m_{y},m_{w}\left|w\right|s_{x},s_{y},s_{w}\right\rangle _{0}\nonumber \\
 & +g\sum_{n_{x},n_{y},n_{w}}\mathcal{B}_{n_{x},n_{y},n_{w}}^{s_{x},s_{y},s_{w}}\ _{0}\left\langle t_{x},t_{y},t_{w}\left|w\right|n_{x},n_{y},n_{w}\right\rangle _{0}+\mathcal{O}\left(g^{2}\right),\label{eq:5.39}
\end{align}
where 

\begin{equation}
\ _{0}\left\langle t_{x},t_{y},t_{w}\left|w\right|s_{x},s_{y},s_{w}\right\rangle _{0}=\frac{1}{\sqrt{2m}}\left(\sqrt{s_{w}}\delta_{t_{w},s_{w}-1}+\sqrt{s_{w}+1}\delta_{t_{w},s_{w}+1}\right)\delta_{t_{y},s_{y}}\delta_{t_{w},s_{w}.}
\end{equation}
In summary, by using the method of the perturbation up to $\mathcal{O}\left(g\right)$,
we have the matrix elements $\left\langle n\left|x\right|k\right\rangle ,\left\langle n\left|w\right|k\right\rangle $
given in (\ref{eq:5.37}) and (\ref{eq:5.39}), energy spectrum given
in (\ref{eq:5.34}). With the matrix elements $\left\langle n\left|x\right|k\right\rangle ,\left\langle n\left|w\right|k\right\rangle $
and the energy spectrum as the inputs, it is possible to evaluate
$C_{T}^{X}\left(t\right)=\left\langle \left[x\left(t\right),p_{x}\right]^{2}\right\rangle _{T}$
and $C_{T}^{W}\left(t\right)=\left\langle \left[w\left(t\right),p_{w}\right]^{2}\right\rangle _{T}$
analytically by using the (\ref{eq:5.6}) and (\ref{eq:5.11}). In
particular, we can further compute the asymptotics of the thermal
OTOC 

\begin{equation}
C_{T}^{X}\left(\infty\right)=2\left\langle x^{2}\right\rangle _{T}\left\langle p_{x}^{2}\right\rangle _{T},C_{T}^{W}\left(\infty\right)=2\left\langle w^{2}\right\rangle _{T}\left\langle p_{w}^{2}\right\rangle _{T}.
\end{equation}
Note that $\left\langle x^{2}\right\rangle _{T}$ can be obtained
by (\ref{eq:5.36}) and
\begin{equation}
\ _{0}\left\langle s_{x}\left|p_{x}^{2}\right|n_{x}\right\rangle _{0}=\begin{cases}
\frac{\pi^{2}s_{x}^{2}}{4L^{2}}, & s_{x}=n_{x},\\
0, & s_{x}\neq n_{x}.
\end{cases}
\end{equation}
$\left\langle w^{2}\right\rangle _{T}$ and $\left\langle p_{w}^{2}\right\rangle _{T}$
takes the same form as (\ref{eq:5.27}) and (\ref{eq:5.28}). Therefore,
both $C_{T}^{X}\left(\infty\right)$ and $C_{T}^{W}\left(\infty\right)$
are analytical in the leading order perturbation of $g$. 

The associated results are plotted in Figure \ref{fig:10}, we can
see both $C_{T}^{X}\left(t\right)$ and $C_{T}^{W}\left(t\right)$
are suppressed by the growth of $N_{c}$ which qualitatively agrees 
with our numerical calculation for the average value of the Lyapunov
coefficient presented in Figure \ref{fig:7}. Notably, $C_{T}^{W}\left(t\right)$
returns to be periodic at very large $N_{c}$, since in this limit,
$C_{T}^{W}\left(t\right)$ returns to the OTOC of a harmonic oscillator.
Moreover, both $C_{T}^{X}\left(\infty\right)$ and $C_{T}^{W}\left(\infty\right)$
increase when $N_{c}$ grows, so it confirms that the asymptotics
of the thermal OTOC defined in (\ref{eq:5.15}) has opposite large
$N_{c}$ behavior to the thermal OTOC. Here we need to clarify, although
we plot $C_{T}^{X,W}\left(t\right)$ and $C_{T}^{X,W}\left(\infty\right)$
with small $N_{c}$, it is not worthy of discussion. The reason is that the
perturbation method we employed becomes invalid if $g$ is not sufficiently
small. So, our concern is the case of $N_{c}\rightarrow\infty$ in
this section, and the current calculation in this limit displays consistent
evaluation.

\subsection{The truncation error}

\begin{figure}
\begin{centering}
\includegraphics[scale=0.25]{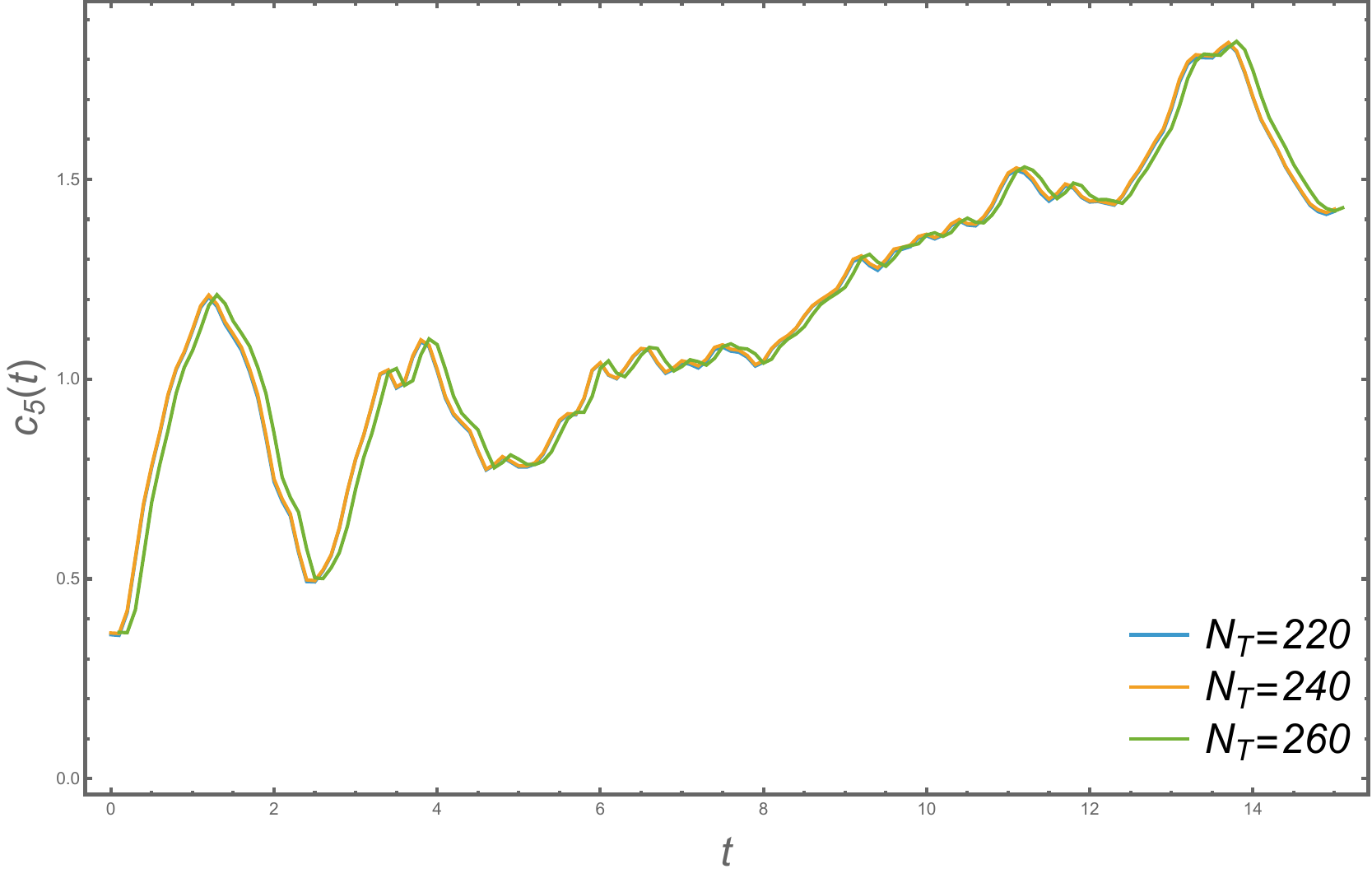}\includegraphics[scale=0.25]{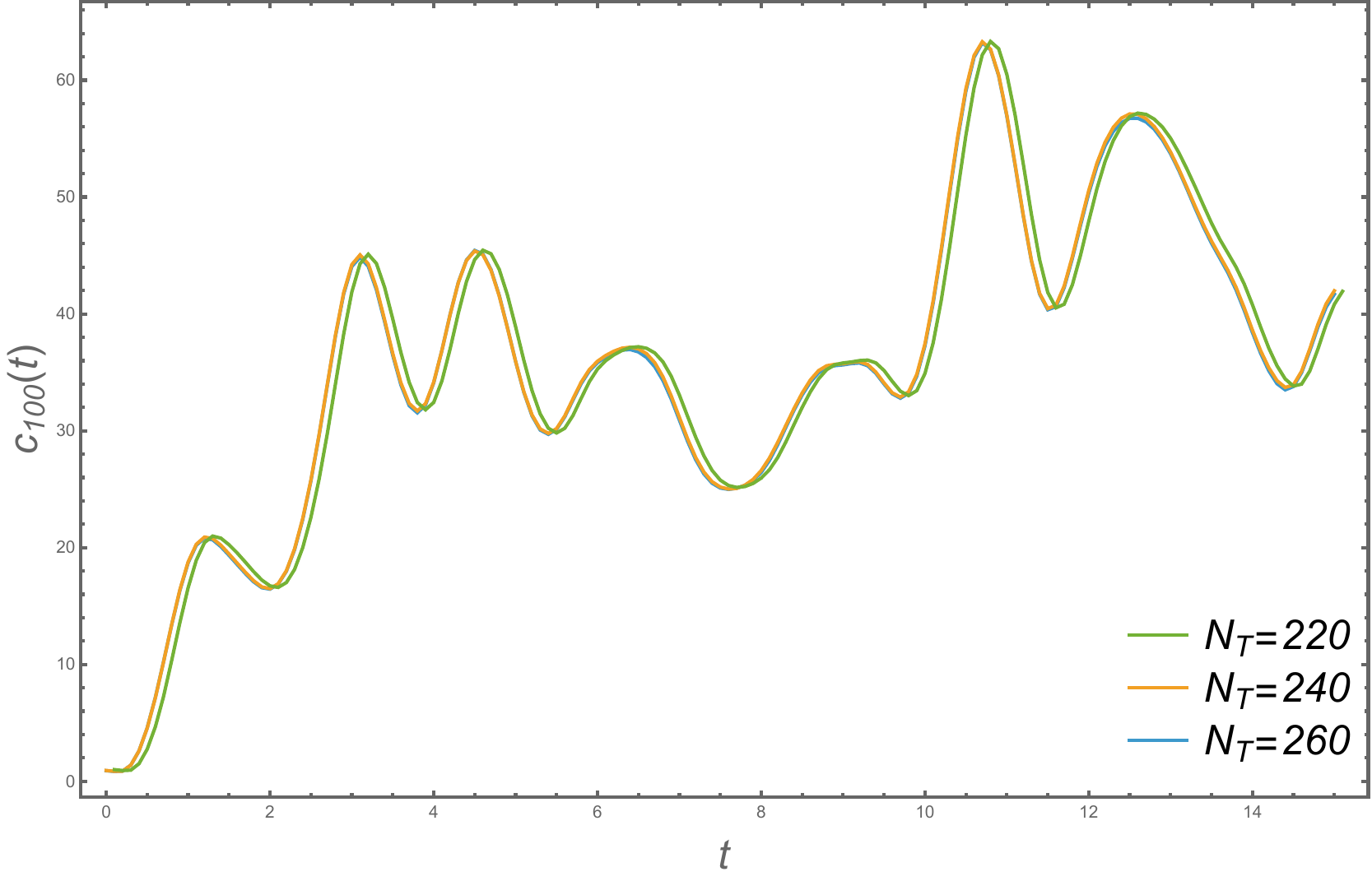}
\par\end{centering}
\caption{\label{fig:11}The microcanonical OTOC $c_{5}\left(t\right),c_{100}\left(t\right)$
of the mesonic matrix model with various truncation values $N_{T}$
at $N_{c}=10$.}

\end{figure}

\begin{figure}
\begin{centering}
\includegraphics[scale=0.25]{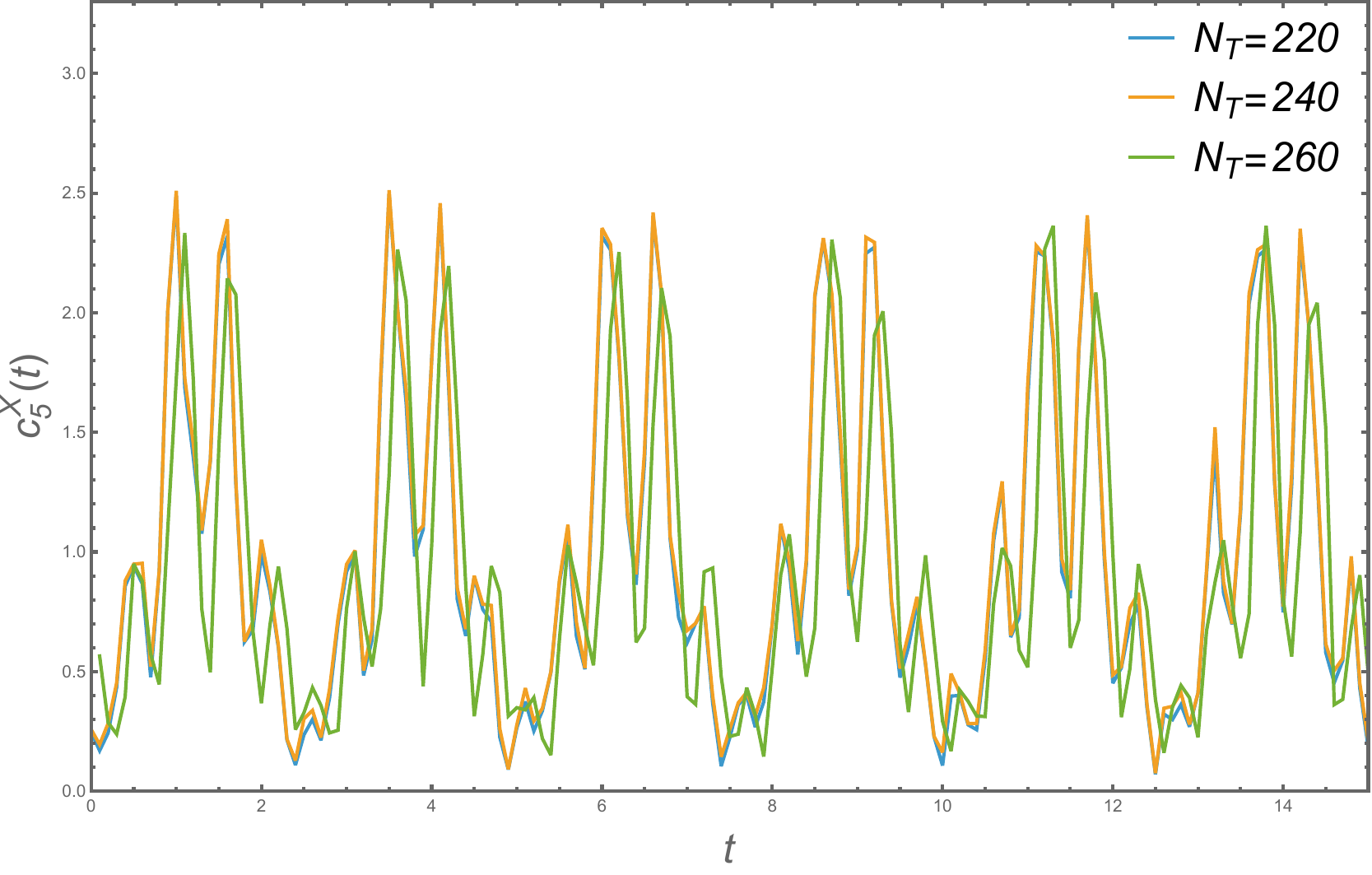}\includegraphics[scale=0.25]{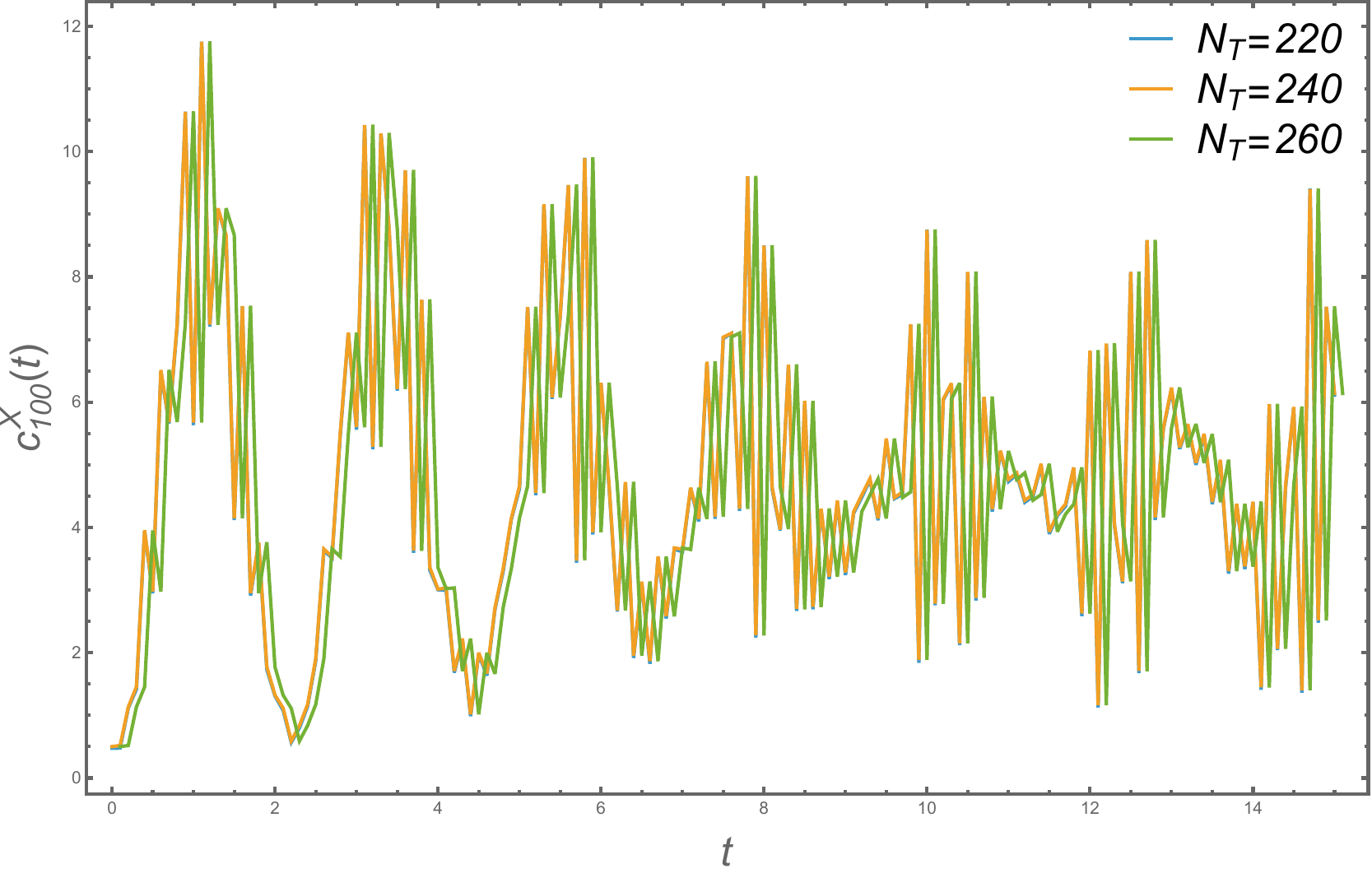}
\par\end{centering}
\begin{centering}
\includegraphics[scale=0.25]{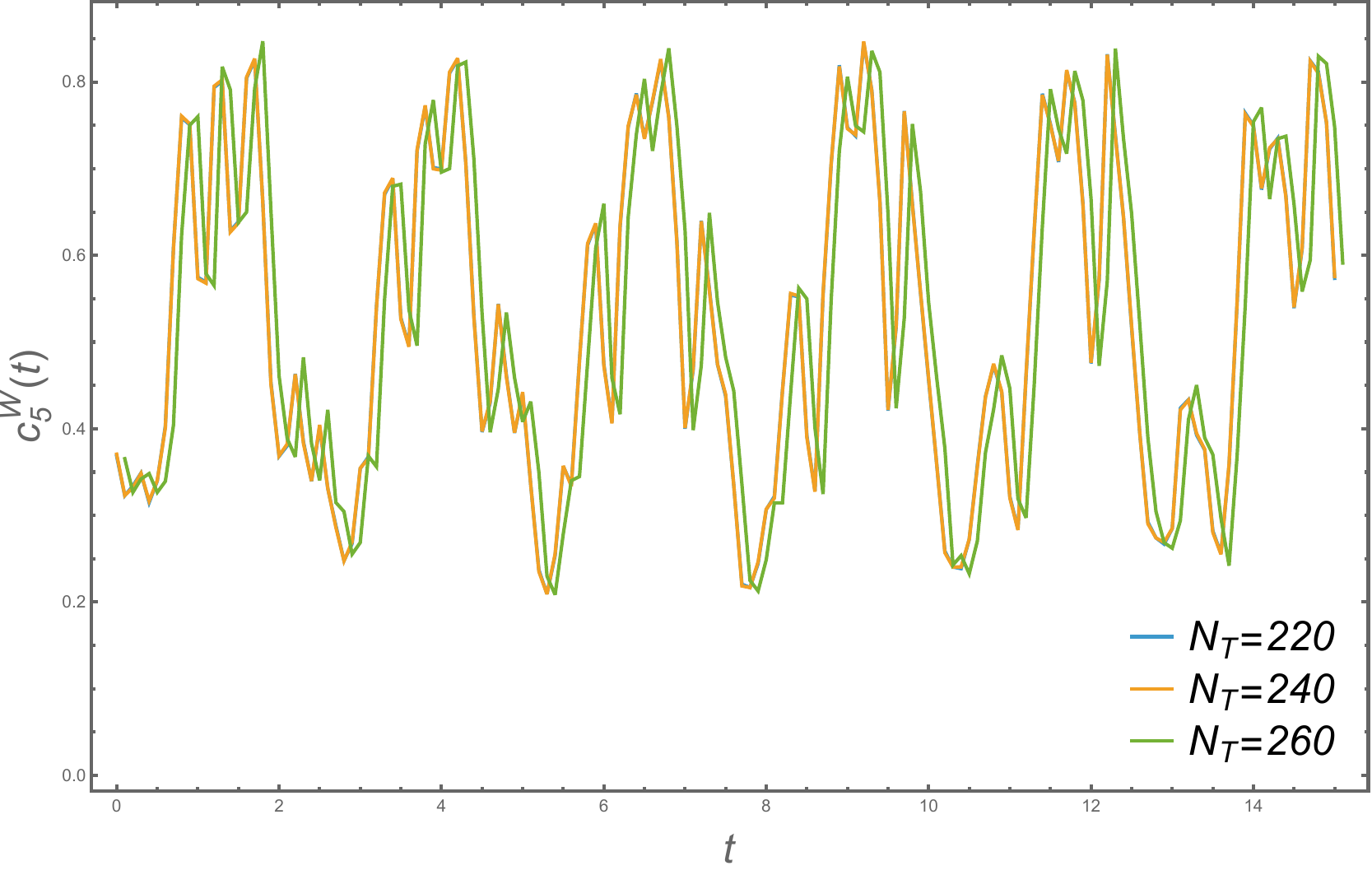}\includegraphics[scale=0.25]{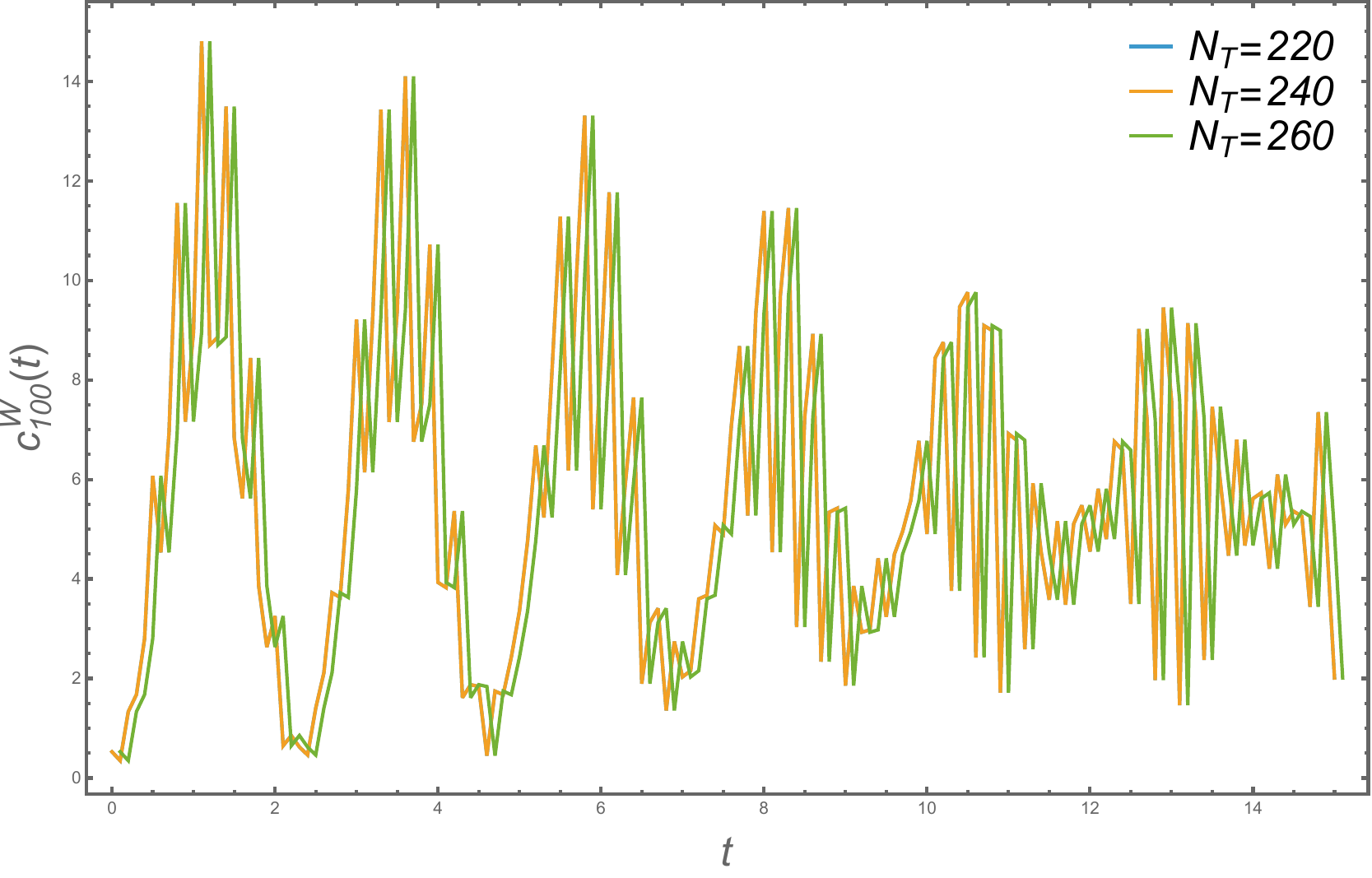}
\par\end{centering}
\caption{\label{fig:12}The microcanonical OTOC $c_{5}^{X,W}\left(t\right),c_{100}^{X,W}\left(t\right)$
of the baryonic matrix model with various truncation values $N_{T}$
at $N_{c}=10$.}

\end{figure}

In the actual numerical calculation, there must be a cut-off in the
sum presented in the calculations of the OTOC e.g. in (\ref{eq:5.5})
or in (\ref{eq:5.6}). Therefore, we need to confirm that our calculation
is less dependent on the truncation we chose, otherwise the numerical
calculation is not credible. In this work, we chose the truncation
$N_{T}=260$ to basically include about 260 states in our numerical
calculation. Here we plot the microcanonical OTOC of the mesonic
and baryonic matrix model with various truncations $N_{T}$ in Figure
\ref{fig:11} and \ref{fig:12}.
We can see the microcanonical OTOCs have converged when the truncation
is chosen to be $N_{T}=220,240,260$, so the lower bound for the OTOCs
to converge must be smaller than $N_{T}=220$. Accordingly, all the
calculations in this work with respect to the OTOCs should be less
sensitive to the truncation $N_{T}=260$ we chose. Since the thermal
OTOCs are based on the sum of microcanonical OTOCs, it also converges
with respect to the choice of $N_{T}=260$.

\section{Summary and discussion}

In this work, we study holographically the chaos in the mesonic and
baryonic system through the matrix models for meson and baryon. The
concerned matrix models are obtained from the $\mathrm{D4}/\mathrm{D6}/\overline{\mathrm{D6}}$
approach, which is a top-down holographic model of QCD. We derive
the Lagrangians for the concerned matrix models and simplify their
associated Hamiltonians to be the model of the coupled oscillators
given in (\ref{eq:1.5}) with special parameters of dimension, mass
and coupling constant, so the chaos in the mesonic and baryonic matrix
models can be evaluated equivalently in the model (\ref{eq:1.5})
with those special parameters. 

In the analysis of the classical chaos, we calculate numerically the
orbits on the Poincar\'e section, the relation between the classical
Lyapunov coefficient and the total energy. Our numerical results illustrate
the chaotic phase transition in the mesonic matrix model when the
total energy increases. While this conclusion agrees basically with
the analysis of the linear sigma models for mesons in QCD \cite{key-6,key-49},
there is not a definitely regular phase of the baryonic matrix model.
In addition, since the gauge theory with spontaneous breaking of symmetry
in general can also be simplified to be the model (\ref{eq:1.5}),
chaos in the model (\ref{eq:1.5}) may indicate the phase with breaking
or restoration of the gauge symmetry as an order parameter. For example,
when the gauge symmetry is spontaneously broken totally, it has a
regular phase at low-energy with respect to the analysis of chaos,
otherwise the gauge theory always takes a chaotic phase. Besides,
we also demonstrate the analytical calculations about the Lyapunov
coefficient by using the perturbation method up to $\mathcal{O}\left(N_{c}^{-1}\right)$,
since the coupling constant going to be vanished can be treated as
a perturbation in the large $N_{c}$ limit. The large $N_{c}$ analytics
indicates the Lyapunov coefficient is suppressed by the growth of
$N_{c}$ which agrees basically with our numerical evaluation and
with some large $N_{c}$ analyses \cite{key-6,key-10,key-11} in holography.
However, it leads to non-positive value of Lyapunov coefficient by
taking the limit $t\rightarrow\infty$. The reason is that, in the
perturbation up $\mathcal{O}\left(N_{c}^{-1}\right)$, both the mesonic
and baryonic matrix models are analytical thus they are integrable
models, and the Lyapunov coefficient is non-positive due the integrable
system.

In the analysis of the quantum chaos, we compute numerically the microcanonical
and thermal OTOCs with various variables and define the quantum Lyapunov
exponent as an analogue of the classical Lyapunov exponent. The numerical
results display the thermal OTOC oscillates periodically in time direction
at low temperature while it trends to be saturated at high temperature.
The reason is that, due to the factor $e^{-\frac{E_{n}}{T}}$ presented
in the definition (\ref{eq:5.5}) of the thermal OTOC, at low temperature,
only the lowest modes contribute to the thermal OTOC while more modes
contribute to it at high temperature. So, if we identify the temperature
as the energy scale, we can see the quantum Lyapunov coefficient begins
to saturate after a critical energy (as a critical temperature) and
this behavior covers qualitatively the analysis of the classical Lyapunov
coefficients given in (\ref{eq:4.5}) and (\ref{eq:4.6}). Moreover,
we evaluate the large $N_{c}$ dependence of the thermal OTOC and
of the thermal OTOCs' asymptotics. To further confirm our analysis
at large $N_{c}$, we derive the analytical forms for the OTOCs by
using the perturbative methods in quantum mechanics. These calculations
illustrate while the thermal OTOC is suppressed by the growth of $N_{c}$,
its asymptotics defined in (\ref{eq:5.15}) behaves oppositely. The
analytical derivation reveals that the discrepant terms between the
thermal OTOC and its asymptotics are significant to determine their
large $N_{c}$ behavior. 

Overall, since the simple model (\ref{eq:1.5}) we discussed in this
work connects to the gauge theory, hadronic system, matrix model,
quantum mechanics, gauge-gravity duality, and even quantum gravity,
the investigation of its chaos may be helpful to find the common features
of these theories and may provide us a novel way to study the underlying
physics.

\section*{Acknowledgements}

Si-wen Li is supported by the National Natural Science Foundation
of China (NSFC) under Grant No. 12005033, the Fundamental Research
Funds for the Central Universities under Grant No. 3132025192. Xun Chen is supported by the National Natural Science Foundation of China (NSFC) Grant Nos: 12405154, and the European Union---Next Generation EU through the research grant number P2022Z4P4B ``SOPHYA - Sustainable Optimised PHYsics Algorithms: fundamental physics to build an advanced society'' under the program PRIN 2022 PNRR of the Italian Ministero dell'Universit\`a e Ricerca (MUR).

\section{Appendixes}

\subsection*{Appendix A: The bosonic action of D-brane}

The action of D-branes with non-Abelian excitations describes the
dynamics of coincident $N$ D$p$-branes. It could be obtained by
T-duality which is the standard technique in string theory. Without
loss of generality, let us consider a stack of D$p$-branes in $D$
dimensional spacetime parametrized by $\left\{ X^{\mu}\right\} ,\mu=0,1...D-1$.
Then we use the indices $a,b=0,1...p$ and $i,j,k=p+1...D-1$ to denote
respectively the directions parallel and vertical to the D$p$-branes
in the spacetime. The complete bosonic action of such D$p$-branes
is therefore collected as,

\[
S_{\mathrm{D}_{p}-\mathrm{branes}}=S_{\mathrm{DBI}}+S_{\mathrm{CS}},\tag{A-1}
\]
where \cite{key-50}

\begin{align}
S_{\mathrm{DBI}}= & -T_{\mathrm{D}_{p}}\mathrm{STr}\int d^{p+1}\xi e^{-\phi}\sqrt{\det\left[Q_{\ j}^{i}\right]\det\left\{ -\left[E_{ab}+E_{ai}\left(Q^{-1}-\delta\right)^{ij}E_{jb}+2\pi\alpha^{\prime}F_{ab}\right]\right\} },\nonumber \\
S_{\mathrm{CS}}= & \mu_{p}\sum_{n=0,1}\int_{\mathrm{D}_{p}\mathrm{-branes}}C_{p-2n+1}\wedge\frac{\left(B+2\pi\alpha^{\prime}F\right)^{n}}{n!},\nonumber \\
Q_{\ j}^{i}= & \delta^{i}\ j+2\pi\alpha^{\prime}\left[\varphi^{i},\varphi^{k}\right]E_{kj},\ E_{\mu\nu}=g_{\mu\nu}+B_{\mu\nu}.\tag{A-2}\label{eq:A-2}
\end{align}
and $T_{\mathrm{D}_{p}}=\frac{1}{\left(2\pi\right)^{p}l_{s}^{p+1}}$
is the tension of the D$p$-brane, $\phi$ refers to the dilaton.
Here $g_{\mu\nu},B_{\mu\nu}$ is the metric of the $D$ dimensional
spacetime and the 2-form field respectively. The ``STr'' refers
to the ``symmetric trace'' and $F$ is the non-Abelian gauge field
strength defined as $F_{ab}=\partial_{a}A_{b}-\partial_{b}A_{a}-i\left[A_{a},A_{b}\right]$.
The transverse modes of the D$p$-branes are denoted as $\varphi^{i}$
's which are in fact the ``T-dualitized'' coordinates given by $2\pi\alpha^{\prime}\varphi^{i}=X^{i}$.
Note that we have chosen the ``static gauge'' throughout the manuscript
in order to gauge away the 2-form field $B$ i.e. $B_{\mu\nu}=0$.
By keeping these in mind, the DBI action in (\ref{eq:A-2}) could
be expanded as,

\begin{equation}
S_{\mathrm{DBI}}=-T_{p}\left(2\pi\alpha^{\prime}\right)^{2}\mathrm{Tr}\int d^{p+1}\xi e^{-\phi}\sqrt{-g}\left[1+\frac{1}{4}F_{ab}F^{ab}+\frac{1}{2}D_{a}\varphi^{i}D^{a}\varphi^{i}-\frac{1}{4}\left[\varphi^{i},\varphi^{j}\right]^{2}\right]+\mathrm{high\ orders}.\tag{A-3}\label{eq:A-3}
\end{equation}
Note the gauge field $A_{a}$ and scalar field $\varphi^{i}$ 's are
all in the adjoint representation of $U\left(N\right)$. The covariant
derivative is given as $D_{a}\varphi^{i}=\partial_{a}\varphi^{i}-i\left[A_{a},\varphi^{i}\right]$.
There is in addition a fermionic part of the action for a D-brane
which can be reviewed in \cite{key-51,key-52}. However, we do not
attempt to discuss it in this work since the fermionic part of the
D-brane action is less relevant in our concern.

\subsection*{Appendix B: Derivation of the baryonic matrix model}

Let us derive the conjectured action for the baryonic matrix model
(\ref{eq:3.5}) by consider $n_{b}$ coincident D4-branes wrapped
on $S^{4}$ located on a fixed point in the worldvolume of $N_{c}$
D4-branes. The bosonic action for a baryon vertex takes the form as,

\begin{align}
S_{\mathrm{D4}} & =S_{\mathrm{DBI}}+S_{\mathrm{CS}}\nonumber \\
S_{\mathrm{DBI}} & =-T_{\mathrm{D4}}\int d^{5}xe^{-\phi}\mathrm{Tr}\sqrt{-\det\left[g_{ab}+2\pi\alpha^{\prime}F_{ab}\right]}\nonumber \\
S_{\mathrm{CS}} & =-g_{s}T_{\mathrm{D4}}\left(2\pi\alpha^{\prime}\right)\mathrm{Tr}\int F\wedge C_{3},\tag{B-1}\label{eq:B-1}
\end{align}
In the gauge-gravity duality, we need $\alpha^{\prime}\rightarrow0$,
so the DBI action can be expanded as,

\begin{equation}
S_{\mathrm{DBI}}=-T_{\mathrm{D4}}\int d^{5}xe^{-\phi}\sqrt{-g}\left[1+\frac{1}{4}\left(2\pi\alpha^{\prime}\right)^{2}\mathrm{Tr}F_{ab}F^{ab}+...\right],\tag{B-2}\label{eq:B-2}
\end{equation}
where $a,b=0,1...4$ and

\begin{equation}
F_{ab}F^{ab}=2F_{0M}F^{0M}+F_{MN}F^{MN},\ M,N=1,2,3,4.\tag{B-3}
\end{equation}
Imposing the rules of the T-duality $G_{MN}X^{N}=\left(2\pi\alpha^{\prime}\right)A_{M}$
on the direction of $S^{4}$, we can obtain

\begin{align*}
F_{0M} & =\nabla_{0}A_{M}-\nabla_{M}A_{0}-i\left[A_{0},A_{M}\right]\\
 & =\frac{1}{\left(2\pi\alpha^{\prime}\right)}G_{MN}\partial_{0}X^{N}-\frac{i}{\left(2\pi\alpha^{\prime}\right)}G_{MN}\left[A_{0},X^{N}\right],\\
 & =\frac{1}{\left(2\pi\alpha^{\prime}\right)}G_{MN}D_{0}X^{M},\tag{B-4}
\end{align*}
and

\begin{align}
F_{MN} & =\nabla_{M}A_{N}-\nabla_{N}A_{M}-i\left[A_{M},A_{N}\right]\nonumber \\
 & =\frac{1}{\left(2\pi\alpha^{\prime}\right)}\left(G_{NK}\partial_{M}X^{K}-G_{MK}\partial_{N}X^{K}\right)-\frac{i}{\left(2\pi\alpha^{\prime}\right)^{2}}G_{MK}G_{NL}\left[X^{K},X^{L}\right]\nonumber \\
 & =-\frac{i}{\left(2\pi\alpha^{\prime}\right)^{2}}G_{MK}G_{NL}\left[X^{K},X^{L}\right],\tag{B-5}\label{eq:B-5}
\end{align}
so that it leads to

\begin{align}
F_{ab}F^{ab} & =-\frac{2}{\left(2\pi\alpha^{\prime}\right)^{2}}\left(D_{0}X^{i}D_{0}X^{i}+D_{0}X^{4}D_{0}X^{4}\right)\nonumber \\
 & -\frac{1}{\left(2\pi\alpha^{\prime}\right)^{4}}\left\{ \left(\frac{U_{KK}}{R}\right)^{3}\left[X^{i},X^{j}\right]\left[X^{i},X^{j}\right]+2\left(\frac{U_{KK}}{R}\right)^{3/2}\frac{U_{KK}^{3/2}}{R^{3/2}}\left[X^{i},X^{4}\right]\left[X^{i},X^{4}\right]\right\} ,\tag{B-6}
\end{align}
where only the leading order terms of $X^{M}$ are kept and we have
defined

\begin{equation}
\frac{3}{2}\left(\frac{U_{KK}}{R}\right)^{3/2}X^{4}=Z.\tag{B-7}
\end{equation}
Since the direction $\tau$ is periodic, the convenient coordinates
can be defined as \cite{key-34,key-36,key-40,key-42},

\begin{equation}
U^{3}=U_{KK}^{3}+U_{KK}r^{2},\ \theta=\frac{3U_{KK}^{1/2}}{2R^{3/2}}\tau,\tag{B-8}
\end{equation}
so that 
\begin{equation}
Z=r\cos\theta,Y=r\sin\theta.\tag{B-9}
\end{equation}
Assuming the baryonic D4-brane is embedded at $\theta=\frac{\pi}{2}$
sharing the same angular with D6-brane, under T-duality, the dynamic
of the baryonic D4-brane become effectively the dynamic of the D0-brane.
And the induced metric on the D0-brane is

\begin{align}
G_{00} & =\left(\frac{U}{R}\right)^{3/2}\left(-1+\delta_{ij}\partial_{0}x^{i}\partial_{0}x^{j}\right)+\frac{4}{9}\left(\frac{R}{U}\right)^{3/2}\frac{U_{KK}}{U}\left(\partial_{0}Z\right)^{2}\nonumber \\
 & =\left(\frac{U}{R}\right)^{3/2}\left[-1+\delta_{ij}\partial_{0}x^{i}\partial_{0}x^{j}+\frac{4}{9}\frac{R^{3}U_{KK}}{U^{4}}\left(\partial_{0}Z\right)^{2}\right],\tag{B-10}\label{eq:B-10}
\end{align}
where $U^{3}=U_{KK}^{3}+U_{KK}Z^{2}$. Afterwards, impose (\ref{eq:B-5})
and (\ref{eq:B-10}) into the DBI action (\ref{eq:B-2}), we find

\begin{align}
S_{\mathrm{DBI}} & =-T_{\mathrm{D4}}\int d^{5}xe^{-\phi}\sqrt{-g}\left[1+\frac{1}{4}\left(2\pi\alpha^{\prime}\right)^{2}\mathrm{Tr}F_{ab}F^{ab}+...\right]\nonumber \\
 & =-T_{\mathrm{D4}}\Omega_{4}\int dt\left(\frac{R}{U}\right)^{3/4}R^{3}U\sqrt{-G_{00}}\left[1+\frac{1}{4}\left(2\pi\alpha^{\prime}\right)^{2}\mathrm{Tr}F_{ab}F^{ab}+...\right]\nonumber \\
 & =\frac{2}{27\pi}\lambda M_{KK}N_{c}\int dt\mathrm{Tr}\left\{ \frac{1}{2}D_{0}X^{M}D_{0}X^{M}-\frac{2}{3}M_{KK}^{2}\left(X^{4}\right)^{2}+\frac{2}{3^{6}\pi^{2}}\lambda^{2}M_{KK}^{4}\left[X^{M},X^{N}\right]^{2}\right\} ,\tag{B-11}\label{eq:B-11}
\end{align}
where all the high order terms of $DX,X$ are dropped off. Finally,
the CS term given in (\ref{eq:B-1}) becomes,

\begin{align}
S_{\mathrm{CS}} & =-g_{s}T_{\mathrm{D4}}\left(2\pi\alpha^{\prime}\right)\mathrm{Tr}\int_{\mathbb{R}\times S^{4}}F\wedge C_{3}\nonumber \\
 & =-g_{s}T_{\mathrm{D4}}\left(2\pi\alpha^{\prime}\right)\mathrm{Tr}\int_{\mathbb{R}\times S^{4}}A\wedge F_{4}\nonumber \\
 & =N_{c}\int dt\mathrm{Tr}A_{0}.\tag{B-12}\label{eq:B-12}
\end{align}
Picking up (\ref{eq:B-11}) and (\ref{eq:B-12}), we can reach to
the conjectured action (\ref{eq:3.5}).

\subsection*{Appendix C: OTOC and Lyapunov coefficient for 1d harmonic oscillator}

Let us consider the Hamiltonian for a 1d harmonic oscillator as

\begin{equation}
H=\frac{p^{2}}{2m}+\frac{1}{2}m\omega^{2}x^{2}.\tag{C-1}\label{eq:C-1}
\end{equation}
Its classical solution takes the form as

\begin{align}
x\left(t\right) & =x\cos\omega t+\frac{p}{m}\sin\omega t,\nonumber \\
p\left(t\right) & =\omega p\cos\omega t-m\omega x\sin\omega t,\tag{C-2}\label{eq:C-2}
\end{align}
where $x=x\left(0\right),p=p\left(0\right)$ are initial constants.
Using the definition of the function $C\left(t\right)$, we have

\begin{equation}
C\left(t\right)=\left[\frac{\delta x\left(t\right)}{\delta x}\right]^{2}=\cos^{2}\omega t,\tag{C-3}\label{eq:C-3}
\end{equation}
which is not exponential since 1d harmonic oscillator is totally integrable
system. Nonetheless, we can evaluate its classical Lyapunov coefficient
by taking the average value of (\ref{eq:C-3}) as,

\begin{equation}
L=\frac{1}{2T}\log\left(\frac{1}{T}\int_{0}^{T}\cos^{2}\omega t\right)=-\frac{\omega}{2\pi}\log2,\tag{C-4}\label{eq:C-4}
\end{equation}
where $T=\pi/\omega$ is the period of function $C\left(t\right)$.
Therefore we can see the Lyapunov coefficient is negative which implies
the system is totally regular as it is expected. 

\begin{figure}
\begin{centering}
\includegraphics[scale=0.25]{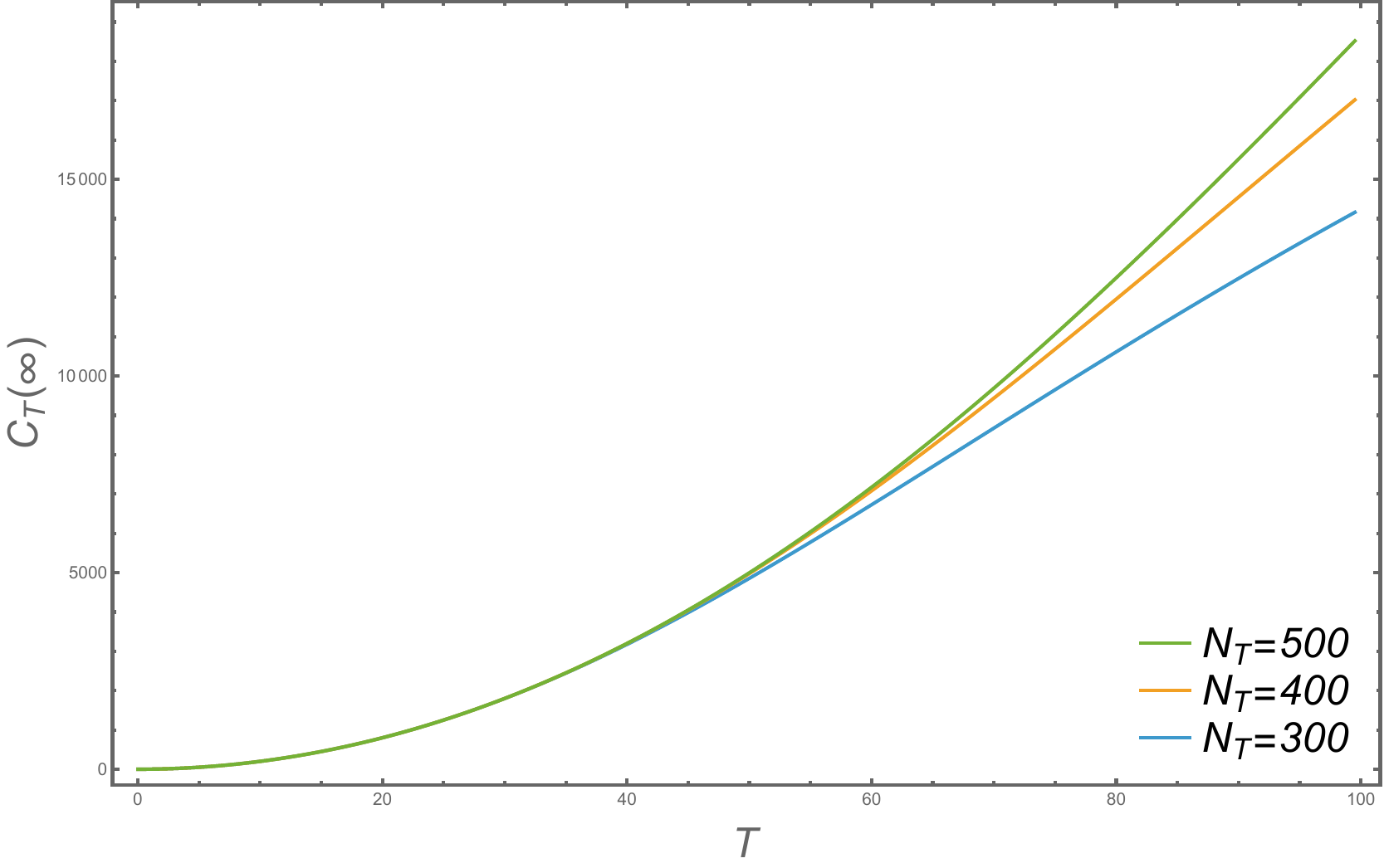}\includegraphics[scale=0.25]{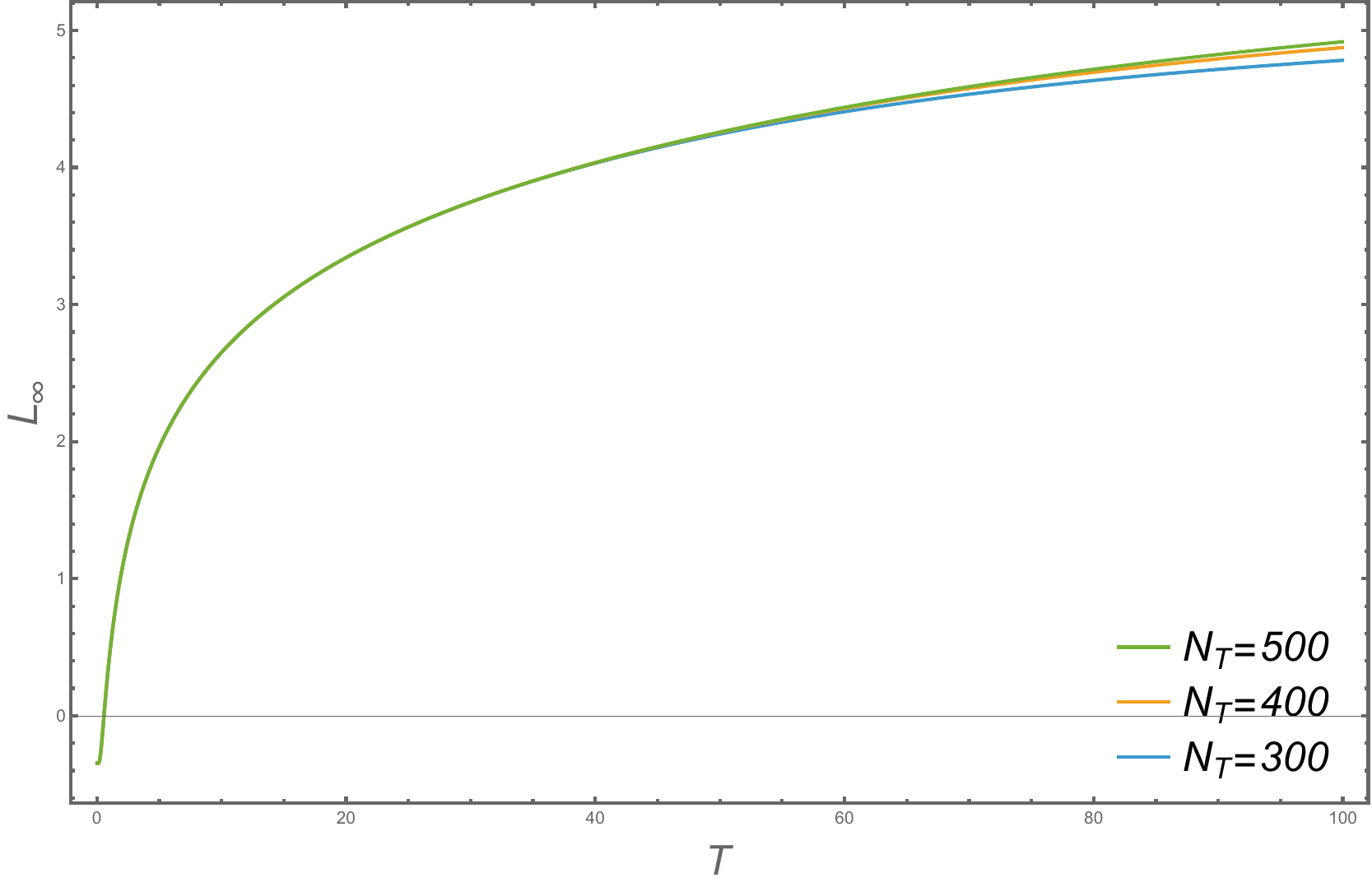}
\par\end{centering}
\caption{\label{fig:C}The late-time OTOC and the associated Lyapunov coefficient
of 1d quantum harmonic oscillator with various truncations.}

\end{figure}

The quantum OTOC leads to the same Lyapunov coefficient as its classical
version. As the equation of motion for the operators in the Heisenberg
picture takes form as,

\begin{align}
\frac{\partial x\left(t\right)}{\partial t} & =-i\left[x,H\right]=\frac{p}{m},\nonumber \\
\frac{\partial p\left(t\right)}{\partial t} & =-i\left[p,H\right]=-m\omega^{2}x,\tag{C-5}\label{eq:C-5}
\end{align}
which is same as the classical equation of motion, therefore its solution
takes the same form as (\ref{eq:C-2}). Keeping this in hand, the
microcanonical OTOC can be computed as,

\begin{equation}
c_{n}\left(t\right)=-\left\langle \left[x\left(t\right),p\right]\right\rangle ^{2}=\cos^{2}\omega t,\tag{C-6}
\end{equation}
and the thermal OTOC becomes,

\begin{equation}
C_{T}\left(t\right)=\frac{1}{\mathcal{Z}}\sum_{n}c_{n}\left(t\right)e^{-\frac{E_{n}}{T}}=\cos^{2}\omega t,\tag{C-7}\label{eq:C-7}
\end{equation}
which is same as its classical version. And we can further take the
average value of (\ref{eq:C-7}) to define the Lyapunov coefficient
leading to the same value as (\ref{eq:C-4}). Then let us further
investigate the approximation given in \cite{key-7,key-8} of the
late-time OTOC at large time as,

\begin{equation}
\lim_{t\rightarrow\infty}C_{T}\left(t\right)=2\left\langle x^{2}\right\rangle _{T}\left\langle p^{2}\right\rangle _{T}.\tag{C-8}
\end{equation}
Using the Virial theorem, the average value of kinetic energy and
potential energy for a harmonic oscillator satisfies,

\begin{equation}
\left\langle p^{2}\right\rangle =m^{2}\omega^{2}\left\langle x^{2}\right\rangle ,\tag{C-9}
\end{equation}
One the other hand, we have

\begin{equation}
\left\langle H\right\rangle =\left\langle n\left|\frac{p^{2}}{2m}+\frac{1}{2}m\omega^{2}x^{2}\right|n\right\rangle =E_{n}=\left(n+\frac{1}{2}\right)\hbar\omega,\tag{C-10}
\end{equation}
which means

\begin{equation}
\left\langle n\left|x^{2}\right|n\right\rangle =\left(n+\frac{1}{2}\right)\frac{\hbar}{m\omega}.\tag{C-11}
\end{equation}
Therefore the late-time OTOC is given by the thermal average value
of $x^{2}$ as,

\begin{align}
C_{T}\left(\infty\right)=\lim_{t\rightarrow\infty}C_{T}\left(t\right) & =2m^{2}\omega^{2}\left\langle x^{2}\right\rangle _{T}^{2},\tag{C-12}\label{eq:C-12}
\end{align}
with

\begin{equation}
\left\langle x^{2}\right\rangle _{T}=\frac{1}{\mathcal{Z}}\sum_{n=0}^{N_{T}}\left\langle n\left|x^{2}\right|n\right\rangle e^{-\frac{E_{n}}{T}}.\tag{C-13}\label{eq:C-13}
\end{equation}
Since (\ref{eq:C-12}) does not lead to a compact result, we plot
out the evaluation of (\ref{eq:C-12}) with a truncation $N_{T}$
which is illustrated in Figure \ref{fig:C}.
 According to (\ref{eq:C-12}) and (\ref{eq:C-13}), when the truncation
$N_{T}$ increases, the late-time OTOC is impossible to converge at
large $T$. Nevertheless, the behavior of the asymptotic Lyapunov
coefficient $L_{\infty}\equiv Lt_{\infty}$,
\begin{equation}
L_{\infty}=\frac{1}{2}\log\left[C_{T}\left(\infty\right)\right]\tag{C-14}
\end{equation}
is less dependent on the truncation $N_{T}$.

\end{document}